\documentclass[superscriptaddress, amsmath, amssymb,  prd, showpacs, floatfix ,preprintnumbers, natbib, longbibliography]{revtex4-1}
\usepackage{graphicx}
\usepackage{float}
\usepackage{dcolumn}
\usepackage[usenames]{color}
\usepackage[colorlinks=true,linkcolor=blue,citecolor=blue,urlcolor=blue]{hyperref}
\usepackage{natbib}
\usepackage{subcaption}
\usepackage{mathtools}

\usepackage{dcolumn}
\usepackage{bm}
\usepackage{epsfig}
\usepackage{xcolor}
\usepackage[normalem]{ulem}
\usepackage[makeroom]{cancel}

\usepackage{ulem}
\usepackage{color}
\definecolor{amethyst}{rgb}{0.6, 0.4, 0.8}

\newlength{\colw}
\newcommand{\trd}{\operatorname{Tr}_4}

\newcommand{\ga}{\gamma_{\alpha}}

\newcommand{\gm}{\gamma_{\mu}}

\newcommand{\ksp}{K^2(p)}

\newcommand{\Kt}{\widetilde{K}}

\newcommand{\qsp}{Q^2(p)}

\newcommand{\half}{\frac{1}{2}}
\newcommand{\bra}{\langle}
\newcommand{\ket}{\rangle}
\newcommand{\braket}[1]{\langle#1\rangle}
\newcommand{\bbra}{\left<\!\left<}
\newcommand{\kket}{\right>\!\right>}
\newcommand{\tr}{\operatorname{Tr}}
\renewcommand{\Re}{\operatorname{Re}}
\newcommand{\order}{{\cal O}}
\newcommand{\One}{1\kern-4.5pt1}

\newcommand{\psibar}{\bar{\psi}}

\newcommand{\ncfg}{N_{\text{cfg}}}

\newcommand{\Dslash}{/\hspace{-1.5ex}D}
\newcommand{\Dslashfwd}{\overrightarrow{\Dslash}}
\newcommand{\Dslashbak}{\overleftarrow{\Dslash}}
\newcommand{\pslash}{\not\!p}
\newcommand{\kslash}{\not\!k}
\newcommand{\qslash}{\not\!q}
\newcommand{\Kslash}{\not\!\!K}

\newcommand{\z}{{(0)}}
\newcommand{\csw}{c_{SW}}

\definecolor{DarkGreen}{rgb}{0,0.5,0}

\newcommand{\eqqref}[1]{Eq.~\eqref{#1}}

\newcommand{\cquark}{\cite{Oliveira:2018lln}}

\begin{document}
\preprint{ADP-20-32/T1142}
\title{Quark-gluon vertex from $N_f=2$ lattice QCD}

\author{Ay{\c s}e K{\i}z{\i}lers\"u}
\affiliation{CSSM, Department of Physics, Faculty of Sciences, School of Physical Sciences, University of Adelaide, 5005,
Adelaide, Australia.} 

\author{Orlando Oliveira}
\affiliation{CFisUC, Department of Physics, University of Coimbra, 
3004--516 Coimbra, Portugal.} 

\author{Paulo J.\ Silva}
\affiliation{CFisUC, Department of Physics, University of Coimbra, 
3004--516 Coimbra, Portugal.} 

\author{Jon-Ivar Skullerud\thanks{jonivar.skullerud@mu.ie}}

\affiliation{
Department of Theoretical Physics, National University of Ireland Maynooth,
Maynooth, County Kildare, Ireland.}
\affiliation{
School of Mathematics, Trinity College, Dublin 2, Ireland}

\author{Andr\'e Sternbeck}

\affiliation{
  Theoretisch-Physikalisches Institut and Universit\"atsrechenzentrum,
  Friedrich-Schiller-Universit\"at
Jena, 07743 Jena, Germany}

\begin{abstract}
We study the quark--gluon vertex in the limit of vanishing gluon
momentum using lattice QCD with 2 flavors of $\order(a)$ improved Wilson
fermions, for several lattice spacings and quark masses.  We find that all three form factors in this kinematics have a significant infrared strength, and that
both the leading form factor $\lambda_1$, multiplying the tree-level vertex structure, and the scalar, chiral symmetry breaking form factor
$\lambda_3$ are significantly enhanced in the infrared compared to the quenched ($N_f=0$) case.  These enhancements are orders of magnitude
larger than predicted by one-loop perturbation theory.  We find only a weak dependence on the lattice spacing and quark mass.
\end{abstract}

         
\maketitle

\section{Introduction}
\label{sec:intro}
The quark--gluon vertex is one of the basic building blocks of the
strong interaction.  It encodes the fundamental interactions of the
quarks and gluons, and can be used to define a nonperturbative running coupling.
In addition, it is an essential ingredient in functional approaches to
nonperturbative quantum chromodynamics (QCD), such as Dyson--Schwinger
equations (DSEs) and the functional renormalisation group (FRG).  In
particular, the DSE for the quark propagator contains the quark--gluon
vertex, and the amount of dynamical chiral symmetry breaking is
highly sensitive to the details of this vertex, see
Ref.~\cite{Kizilersu:2013ksw} for the case of QED.

Traditionally, many studies of hadron phenomenology in the DSE
framework have been carried out using the rainbow--ladder truncation,
where the quark--gluon vertex is approximated by its tree-level
structure, multiplied by an effective coupling which is taken to
depend only on the gluon momentum.  While this approach has been
successful in describing a range of properties of pseudoscalar and
vector mesons \cite{Maris:1997hd,Maris:1999nt,Maris:2000sk,Chang:2013nia}, it has failed to provide a satisfactory description of
other quantities including scalar and axial-vector mesons and the
chiral transition temperature \cite{Bashir:2011dp,Chang:2013pq,Chang:2009zb}.  It should also be noted that this
approach fails to satisfy the Slavnov--Taylor identities which encode
the gauge invariance of QCD.

Through a combination of lattice, DSE and FRG studies, the elementary
(gluon, quark and ghost) propagators in Landau-gauge QCD are now very
well known \cite{Maas:2011se,Sternbeck:2008mv,Sternbeck:2012mf,Oliveira:2016stx,Duarte:2016iko,Parappilly:2005ei,Kamleh:2007ud,Oliveira:2018lln,Fischer:2006ub,Binosi:2009qm,Braun:2014ata,Cyrol:2017ewj,Aguilar:2019uob}
although
a continuum extrapolation of lattice results is still outstanding.  It
is thus natural that attention has in recent years turned to the
structure of the 3- and 4-point vertices of QCD, including the
quark--gluon vertex.  There have been a number of recent studies
within the DSE framework exploring the full structure of this vertex
\cite{Bhagwat:2004kj,Kizilersu:2009kg,Chang:2010hb,Mitter:2014wpa,Aguilar:2014lha,Aguilar:2016lbe,Binosi:2016wcx,Bermudez:2017bpx,Cyrol:2017ewj,Kizilersu:2014ela,Kizilersu:2009kg, Williams:2014iea,Oliveira:2018fkj,Oliveira:2020yac,Serna:2018dwk,Gao:2021wun}
and the potential impact of the various non-leading form factors on
chiral symmetry breaking and hadron phenomenology
\cite{Chang:2010hb,Williams:2015cvx,Sultan:2018qpx,Aguilar:2018epe}.  Two critical
ingredients in determining the full structure of the vertex have been
the Slavnov--Taylor identity, which relates the longitudinal part of
the vertex to the quark propagator, and the transverse Ward--Takahashi
identities
\cite{He:2000we,Kondo:1996xn,Pennington:2005mw,He:2009sj} which
constrain the purely transverse part of the vertex.

The quark--gluon vertex has previously been studied on the lattice in
the {\it quenched approximation} in a series of papers \cite{Skullerud:2002ge,Skullerud:2003qu, Kizilersu:2006et, Pelaez:2015tba}.  Apart from
the uncontrolled systematic uncertainty of the quenched approximation,
these studies have only been carried out at a single lattice spacing
and volume and hence do not allow for a controlled approach to the
physical (continuum and infinite-volume) limit.  In this paper we take
the first steps towards rectifying this by computing the quark--gluon
vertex on state-of-the-art lattices with $N_f=2$ light dynamical
quarks with different masses and for several lattice spacings, and
compare with equivalent results in the quenched approximation
($N_f=0$).  Some preliminary findings were presented in
\cite{Oliveira:2016muq,Sternbeck:2017ntv}. We will conduct this study
in the soft gluon kinematics where the gluon momentum is vanishing,
and hence both quark momenta are equal.

The structure of this paper is as follows.  In
Section~\ref{sec:methods} we present our methods for
computing the vertex, including the notation and vertex decomposition
(Sec.~\ref{sec:vertex-notation}), lattice simulation details
(Sec.~\ref{sec:simulation}), procedure for extracting form factors
(Sec.~\ref{sec:extract-formfactors}) and tree-level correction of
lattice data (Sec.~\ref{sec:treelevel}).  Our results are presented in
Section~\ref{sec:results}, while in Section~\ref{sec:conclude} we
summarise our findings and outline prospects for future work.

\section{Continuum and lattice vertex}
\label{sec:methods}
\subsection{Quark-Gluon Vertex}
\label{sec:vertex-notation}
Our description of the quark-gluon vertex follows the notation used in 
\cite{Skullerud:2000un,
  Skullerud:2001aw,Skullerud:2002ge,Skullerud:2003qu,Kizilersu:2006et},
which we briefly summarise here.

\begin{figure}[hpbt]
\includegraphics*[width=0.3\textwidth]{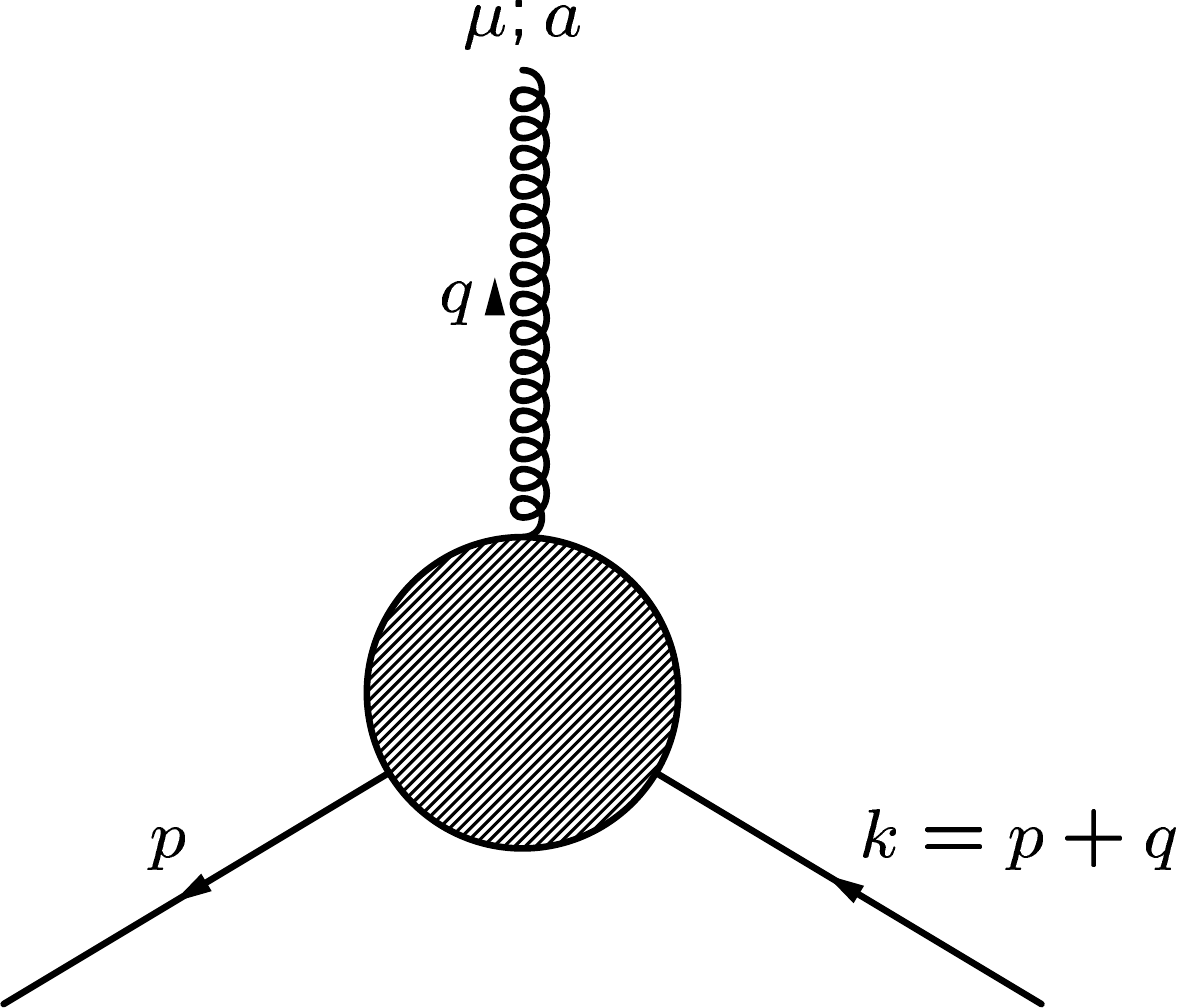}
\caption{The proper quark-gluon vertex.}
\label{fig:vertex}
\end{figure}

The proper quark-gluon vertex, $(\Lambda_\mu^a)_{\beta\rho}^{ij} = t_{ij}^a\,(\Lambda_\mu)_{\beta\rho} = t_{ij}^a\,( -ig_0\,\Gamma_\mu)_{\beta\rho} $, is depicted in
Fig.~\ref{fig:vertex} with $p$ 
the outgoing quark momentum, $q$ the outgoing gluon momentum and $k =p + q$ the incoming quark momentum where $t_{ij}^a$ is the group generator. 
The quark-gluon vertex can be described in terms of 12 spin amplitudes
which would be formed using the three vectors
$p_{\mu},q_{\mu},\gamma_{\mu}$, and four Dirac scalars, $I, \pslash,
\qslash$ and $\pslash\qslash$. 
Reorganising these spin amplitudes 
to form a tensor basis, the full one-particle irreducible function
$\Lambda_F^\mu (p,q, k )$
can be decomposed into non-transverse and transverse parts, 
\begin{align}
\begin{split}
 \Lambda^\mu_F(p,q,k)
 &= \Lambda^\mu_{NT}(p,q,k)  + \Lambda^\mu_T(p,q,k), \\
 &= \sum_{i=1}^{4} \lambda^i(p^2,q^2,k^2) L^{\mu}_i(p,q,k)
 + \sum_{i=1}^{8}\tau^i (p^2,q^2,k^2)T^{\mu}_i(p,q,k),
\end{split}
\end{align}
 where $\lambda^i$ and $L^{\mu}_i(p,k,q)$ are non-transverse form factors
 and basis vectors \cite{Ball:1980ay}, and $\tau^i$ and $T^{\mu}_i(p,k,q)$ are transverse form 
 factors and basis vectors \cite{Kizilersu:1995iz} respectively. By
 definition the transverse part is such that $q \cdot T _i(p,k,q)=0$.
 In terms of the incoming and outgoing quark momenta the longitudinal
 basis can be written as
 \begin{align} 
  {L_1}_{\mu} &= \gamma_{\mu}\,, \nonumber\\
  {L_2}_{\mu} &= - (\pslash+\kslash)(p_{\mu}+k_\mu)\,,\nonumber\\
  {L_3}_{\mu} &= -i(p_{\mu} + k_\mu)\,, \nonumber\\
  {L_4}_{\mu} &= -i \sigma_{\mu\nu}\,(p_{\nu}+k_\nu)\,,
 \end{align}
The non-transverse part of the vertex satisfies the Slavnov--Taylor identity,
\begin{equation}
q_{\mu} \Lambda^{\mu}(p,q,k)  = q_{\mu} \Lambda^{\mu}_{NT}(p,q,k)
 = G_h(q^2)\,[\bar{H}(k,-p,-q)S^{-1}(k)-S^{-1}(p)H(-p, k,-q)]\,,
 \label{Eq:STI}
\end{equation}
which encodes the gauge invariance of the system. Through these Slavnov-Taylor identities the non-transverse part of the vertex is related to the inverse quark propagator, $S^{-1}$, 
the ghost--quark scattering kernel, $H$, and the ghost dressing function,
$G_h$. However the transverse part remains unconstrained by the
Slavnov--Taylor identity, i.e.,
$q_{\mu} \Lambda^{\mu}_T (p,q)=0$ and
$\Lambda^{\mu}_T(p,p)=0$.  The transverse basis tensors
\cite{Kizilersu:1995iz} are given by
 \begin{align} 
 {T_1}_{\mu} &= i \left[(k\cdot q)p_\mu - (p\cdot q)k_\mu\right]\,,   \nonumber\\
 {T_2}_{\mu} &= (\pslash+\kslash)\, \left[(k\cdot q)p_\mu - (p\cdot q)k_\mu\right]\,,   \nonumber\\
 {T_3}_{\mu} &= \qslash\, q_{\mu}-q^2 \gamma_{\mu}\,, \nonumber\\
 {T_4}_{\mu} &= -i\left[ q^2\, \sigma_{\mu\nu} (p_{\nu}+k_\nu) + 2q_{\mu}\sigma_{\nu\lambda}p_{\nu}k_{\lambda} \right]\,,  \nonumber\\
 {T_5}_{\mu} &= -i\sigma_{\mu\nu}\,q_{\nu}  \nonumber\\
 {T_6}_{\mu} &= q \cdot (p+k)\,\gamma_{\mu} - {\not\! q}\,(p_{\mu}+k_\mu)\,, \nonumber\\
 {T_7}_{\mu} &= \frac{i}{2}\,q\cdot(p+k)\,\sigma_{\mu\nu}(p_{\nu}+k_{\nu})
 - i(p_{\mu}+k_\mu) \sigma_{\nu\lambda}\, p_{\nu} k_{\lambda}\,,\nonumber\\
 {T_8}_{\mu} &= -\gamma_{\mu}\sigma_{\nu\lambda}p_{\nu}k_{\lambda}
 - \pslash k_{\mu} + \kslash p_{\mu}\,,
 \end{align}
 where
 $\sigma_{\mu\nu} = \frac{1}{2}\left[ \gamma_{\mu},\gamma_{\nu}\right]$. 
Since in this paper we will only consider
the soft gluon kinematics where the gluon momentum $q=0$ and $k=p$, these transverse tensors do not contribute here.
The full vertex structure in general kinematics will be considered in
a future paper.
%
\subsection{Lattice simulation details}
\label{sec:simulation}
In this study we use the same action and formulation for the
quark propagator as in Ref.~\cquark, and refer to that paper and
references therein for further details.  We employ the Wilson action
for the gauge sector and the 
Sheikholeslami--Wohlert (clover) fermion action
\cite{Sheikholeslami:1985ij},
\begin{equation}
S_{SW}= S_W-i\frac{a}{4}g_0 c_{SW} \sum_x \sum_{\mu\nu} {\overline{\psi}}(x)\sigma_{\mu\nu}F_{\mu\nu}(x)\psi(x)\,,
\end{equation}
where $S_W$ is the ordinary (unimproved) Wilson fermion action.  The
improvement coefficient $c_{SW}$ has been determined
nonperturbatively.
In order for the quark
propagator and quark--gluon vertex to be fully $\order(a)$ improved,
it is also necessary to improve the quark propagator
\cite{Capitani:2000xi}, and for this purpose we use
the $\order(a)$ improved ``rotated'' quark propagator,
\begin{equation} 
S_R(x,y) \equiv \bra\psi'(x)\psibar'(y)\ket 
 = \bbra (1+b_q am)^2(1-c_qa\Dslashfwd(x))S_0(x,y;U)(1+c_q a\Dslashbak(y))  \kket \; ,
\label{eq:rotprop}
\end{equation}
where the double brackets $\bbra\cdot\kket$ denote averaging over gauge
fields only, while $S(x,y;U)$ is the quark propagator evaluated on a
single gauge configuration $U$.
We use the tree-level values for the coefficients $b_q$ and $c_q$:
$b_q=c_q=1/4$.  We note that this propagator
differs from the improved propagator used in Refs.~\cite{Skullerud:2000un,
  Skullerud:2001aw,
  Skullerud:2002ge,Skullerud:2003qu,Kizilersu:2006et}.

The quark--gluon vertex is then determined by
\begin{align}
\Lambda_\mu^{a,\rm lat}(p,q)
 &= S_R(p)^{-1} V^a_\nu(p,q)S_R(p+q)^{-1}D(q)^{-1}_{\nu\mu}
\, ,
\label{eq:vtx-amputate}
\intertext{where the unamputated vertex $V$ is given by}
 V^a_\mu(p,q) &= \bbra S_R(p;U)A^a_\mu(q)\kket\,.
\end{align}

In Landau gauge it is not possible to implement
\eqqref{eq:vtx-amputate} fully since the inverse gluon propagator,
$D_{\mu\nu}^{-1}$, does not exist for $q\neq0$. Instead we compute the
transverse projected vertex
 \begin{equation} 
   \tilde\Lambda_{\mu}(p,q) = -ig_0\,\tilde\Gamma_\mu
   =  P_{\mu\nu}^T(q)\,\Lambda_\nu (p,q)  =
 \left( \delta_{\mu\nu}-\frac{q_\mu
   q_\nu}{q^2}\right)\,\Lambda_\nu(p,q) \,.
  \label{eq:transverse_vertex}
 \end{equation} 

In the present work  we use the same gauge ensembles as in \cquark,
which are a subset of
the gauge ensembles generated by the Regensburg QCD (RQCD)
collaboration (see, e.g.,
\cite{Bali:2012qs,Bali:2014gha,Bali:2014nma}).  These ensembles have
$N_f=2$ dynamical quarks with pion masses in the range 280--420\,MeV.
Three values of the lattice spacing, $a\approx0.081\,\text{fm}$, 
$a\approx0.071\,\text{fm}$ and $a\approx0.060\,\text{fm}$, are used.
Most of the calculations
have been carried out on a lattice volume of $32^3\times64$, but  we
have also used a $64^4$ lattice to check finite volume effects for one
of the parameter choices ($m_\pi\approx290$ MeV).

In addition, we have produced a quenched ensemble with lattice spacing
matching that of ensembles L07 and H07, i.e.,
with $a=0.07\,$fm, but with a larger valence quark mass
$m_\pi\approx1000$MeV.

The parameters used are listed in Table~\ref{tab:params}.  In order to
more easily refer to the different ensembles when presenting the
results, we have labelled them such that the first letter refers to
the pion mass [heavy (H), 420\,MeV, or light (L), 290\,MeV] and the
following two digits refer to the lattice spacing.  The quenched
ensemble is labelled Q07.
\begin{table}[thb]
\begin{tabular}{lcc|cccccc}
Name & $\beta$ & $\kappa$ & $a$ [fm] & $V$ & $m_\pi$ [MeV] & 
 $m_q$ [MeV] &  $\ncfg$ & $N_{\text{src}}$ \\ \hline
L08  & 5.20 & 0.13596 & 0.081 & $32^3\times64$ & 280 & 6.2 & 900 & 4 \\
H07  & 5.29 & 0.13620 & 0.071 & $32^3\times64$ & 422 & 17.0 & 900 & 4 \\
L07  & 5.29 & 0.13632 & 0.071 & $32^3\times64$ & 295 & 8.0 & 908 & 4 \\
L07-64& 5.29 & 0.13632 & 0.071 & $64^3\times64$ & 290 & 8.0 & 750 & 2 \\
H06  & 5.40 & 0.13647 & 0.060 & $32^3\times64$ & 426 &18.4 & 900 & 4 \\\hline
Q07  & 6.16 & 0.1340  & 0.071 & $32^3\times64$ & 1000 & 130 & 998 & 4
\end{tabular}
\caption{Lattice parameters used in this study.  The lattice spacings
  $a$ and pion masses $m_\pi$ and critical hopping parameters used to obtain
  the subtracted bare quark mass $m_q=1/(2\kappa)-1/(2\kappa_c)$ are
  all taken from Ref.~\cite{Bali:2014gha}.  The quenched ensemble Q07 was
  produced specifically for this study.  $N_{\text{src}}$ denotes the
  number of different point sources per configuration used to produce
  the quark propagators.} 
\label{tab:params}
\end{table}

For the gauge fixing we used an over-relaxation algorithm which iteratively 
maximizes the Landau-gauge functional
\begin{equation}
 F_U[g] = \frac{1}{4V}\sum_{x\mu} \Re\tr U^{g}_{x\mu}\, ,
\end{equation}
with $U^{g}_{x\mu} = g_x U_{x\mu} g^{\dagger}_{x+\hat{\mu}}$ and $g_x\in SU(3)$.
As stopping criterion we used
\begin{equation}
 \max_{x}\; \Re\tr\left[(\nabla_\mu A_{x\mu})(\nabla_\mu 
  A_{x\mu})^{\dagger}\right] < 10^{-9}\, ,
\end{equation}
where $A_{x\mu} \equiv \frac{1}{2iag_0}(U^g_{x\mu}-U_{x\mu}^{g\dagger}) \vert_{\mathrm{traceless}}$ and 
$\nabla_\mu A_{x\mu} \equiv \sum_\mu (A_{x\mu} - A_{x-\hat{\mu},\mu})$, as usual.


\subsection{Form factor extraction}
\label{sec:extract-formfactors}
In the soft gluon kinematics ($q_\mu=0, k_{\mu} = p_{\mu} $), the continuum vertex is given by
the three  non-transverse form factors $\lambda_1, \lambda_2 and
\lambda_3$,\footnote{$\lambda_4=0$ in this kinematics because of
  charge conjugation symmetry, which dictates that
  $\lambda_4(p^2,q^2,k^2)=-\lambda_4(k^2,q^2,p^2)$.}
\begin{equation}
  (\tilde\Lambda_\mu^a)_{\alpha\beta}^{ij}
  = t_{ij}^a\big(\overline{\Lambda}_\mu\big)_{\alpha\beta}
 = -ig_0 t_{ij}^a\,\left    (\lambda_1\, \left[\gamma_\mu\right]
                                 + \lambda_2 \, \left[-4\not\!p p_\mu\right]
                                 +  \lambda_3 \, \left[-2 i p_\mu\right]\,  \right)_{\alpha\beta}.
 \label{eq:sqk_cont_vertex}
\end{equation}

We can extract these individual form factors by taking appropriate contractions and traces: 
\begin{align}
%
\trd [ I\, (\overline\Lambda_\mu^a)] &=  -ig_0\,\left    (  \lambda_3 \, \left[-2 i p_\mu\right]\  \right),\\
%
\trd [\gamma_\nu \,(\overline\Lambda_\mu^a)] 
 &=  -ig_0 \,\left    (    \lambda_1\,\delta_{\nu\mu} +\lambda_2\, \left[-4\,p_\nu p_\mu\right]\,   \right), 
 \label{eq:trace-ga} 
\end{align}
where $\trd \equiv\frac{\tr}{4}$ denotes the trace over spin indices.
We note that \eqqref{eq:trace-ga} implies that
$\lambda_1$ and $\lambda_2$ are coupled, but we can disentangle them
by contracting it with the tensors $\delta_{\mu\nu}$ and $p_{\mu}p_{\nu}$
and solving the coupled equations. This yields the following
expressions,
\begin{align}
\lambda_1 &=   \frac{1}{(-ig_0)}\, \left\{ \,\,\, \frac{1}{3}\,\,\Big[\trd(\gamma_\mu \overline\Lambda_\mu)
  - \frac{p_{\mu}p_{\nu}}{p^2}\trd(\gamma_\nu \overline\Lambda_\mu)\Big]  \right\}\,,
\label{eq:l1-cont-covariant} \\
\lambda_2 &=  \frac{1}{(-ig_0)}\, \left\{ \frac{1}{12p^2}\Big[ \trd(\gamma_\mu \overline\Lambda_\mu)
  - 4\frac{p_{\mu}p_{\nu}}{p^2}\trd(\gamma_\nu \overline\Lambda_\mu)\Big]  \right\}\,,
\label{eq:l2-cont-covariant} \\
 \lambda_3 &= \frac{1}{(-ig_0)}\, \left\{ \frac{i}{2}\, \frac{p_\mu}{p^2} \trd(I\, \overline\Lambda_{\mu}) \right\}\,.
\label{eq:l3-cont-covariant}
\end{align}
In practice, expressions such as Eqs \eqref{eq:l1-cont-covariant} and
\eqref{eq:l2-cont-covariant} contain multiple terms that can lead to
numerically poorly determined form factors.  It may therefore be
desirable to sacrifice covariance and 
restrict ourselves to specific momentum configurations in order to
obtain simple expressions involving only a single term.
For instance, with the choices of (a) $\nu=\mu$ and $p_\mu=0$ and
(b) $\nu \ne \mu$ in  \eqqref{eq:trace-ga}  we obtain the following
non-covariant, single term expressions for the continuum $\lambda_1$
and $\lambda_2$ form factors respectively, which we will make use of
in this analysis: 
\begin{align}
 \lambda_1  &=   \frac{1}{(-ig_0)}\, \left\{  \,\,\Big[\trd(\gamma_\alpha \overline\Lambda_\mu)\Big]\,    \Bigg|_{\substack{\alpha=\mu\\p_\mu=0}}   \right\} \,,
  \label{eq:l1-cont-noncov} \\
 \lambda_2 &=  \frac{1}{(-ig_0)}\, \left\{    - \frac{1}{4\,p^2}\,\frac{p_{\alpha}p_{\mu}}{p^2}\, \Big[ \trd(\gamma_\alpha \overline\Lambda_\mu) \, \Big|_{\alpha \neq\mu} \Big]  \right\} \,,
  \label{eq:l2-cont-noncov} 
 %
\end{align}

The quark--gluon vertex computed on the lattice using
\eqqref{eq:vtx-amputate} is not renormalised.  To renormalise the
vertex, we impose a momentum subtraction scheme whereby the leading
form factor $\lambda_1$ takes on its tree-level value at a given
renormalisation scale $\mu$,
\begin{equation}
  \lambda^R_1(\mu^2,0,\mu^2) = 1\,.
\end{equation}
This fixes the renormalisation constant $Z_1$ such that
$\Gamma^{\text{lat}}_\mu(p,q,k)= Z_1\Gamma^R_\mu(p,q,k)$, which in turn
determines the renormalisation of all the form factors.

\subsection{Tree-level correction}
\label{sec:treelevel}

In order to determine the lattice equivalents of the continuum
expressions \eqref{eq:l1-cont-noncov}, \eqref{eq:l2-cont-noncov} and
\eqref{eq:l3-cont-covariant} for the form factors $\lambda_1,
\lambda_2 and \lambda_3$, and to 
estimate and reduce their lattice artefacts, we need to compute the
structure of the lattice tree-level vertex.
Here we will outline the main results, while the details are given in
Appendix~\ref{app:tree-level}.

The tree-level quark--gluon vertex associated with the Sheikholeslami--Wohlert
action without color and group indices is given by  \cite{Heatlie:1991kg,Capitani:1995qn}
\begin{align}
\overline\Lambda_{0,\nu}^{(0)}(p,q,k) &= (-ig_0)\, \left\{  
                \bm{ \gamma}_\nu \cos\left( \frac{a(p_\nu+k_\nu)}{2} \right)
               -i \,  \sin\left( \frac{a(p_\nu+k_\nu)}{2} \right)  \,\bm{I} 
-i \frac{\csw}{2} \, \cos\left( \frac{a q_\nu}{2}\right) \,\sum_{\lambda} \bm{\sigma_{\nu\lambda}}  \,\sin\left( a q_\lambda \right)   \right\}\,,
\label{eq:vtx-tree1}
\end{align}
for the case of
the `unimproved' propagator $S_0(x,y)=\braket{\psi(x)\psibar(y)}$.
The improved vertex obtained using the rotated propagator
of \eqqref{eq:rotprop} is given at tree level by
\begin{align}
  \overline{\Lambda}^{(0)}_{R,\mu}(p,q,k)
  &= (1+b_q am)\,[S_R^{(0)}(p)]^{-1}\,S_0^{(0)}(p)\,
  \left[ \Lambda_{0,\mu}^{a(0)}(p,q,k)\right]\,
  S_0^{(0)}(k)\,[S_R^{(0)}(k)]^{-1}\,,
\label{eq:tree-imp} 
\end{align}
where $S_0^\z(p)$ is the tree-level unimproved dimensionless Wilson quark
propagator, while $S_R^\z(p)$ is the level expression for the
improved propagator of \eqqref{eq:rotprop}.
In the soft gluon kinematics with $q=0, p=k$, this reduces to
\begin{align}
  \overline\Lambda^{(0)}_{R,\mu}(p,0,p)
  =  (-ig_0)&\,\frac{(1+b_q am)}{(1+am/2)^2}\,\frac{1}{ \big(1+ c_q^2 a^2 K^2(p) \big)^4}  \nonumber\\
  \times &  \bigg\{  
         {\boldsymbol{ \gamma_\mu}}\, \Big[ \big(1+c_q^2 a^2K^2(p)\big)^2
           C_\mu(p) \Big]  \nonumber\\
         &  \,\, \bm{-4a^2K_\mu\Kslash(p)}\,
         \Big[ 2c_q^2 C_\mu(p)
           - c_q\big(1-c_q^2a^2K^2(p) \big) \Big]  \nonumber\\
         & \,\,  \bm{-2i aK_\mu}\, \Big[ -2c_q^2 a^2K^2(p)
   +\frac{1}{2}\big(1-c_q^2 a^2K^2(p) \big) -2c_q
   \big(1-c_q^2a^2K^2(p)\big)\, C_\mu(p) \Big]
 \bigg\}         \,,
  \label{eq:tree-imp6} 
\end{align}
where we have defined the lattice momentum variables
\begin{align}
  K_\mu(p) &= \frac{1}{a}\sin(p_{\mu}a)\,,
  & C_\mu(p) &= \cos(p_{\mu}a)\,. \label{def:Kmu-Cmu}
\end{align}
We note in particular that there are two separate tensor structures
appearing in the tree-level vertex \eqref{eq:tree-imp6} which in the
continuum become equal to $L_{2,\mu}$, and likewise for $L_{3,\mu}$.
In Eqs~\eqref{eq:lam2-corr} and \eqref{eq:lam3-corr} below, these are
associated with separate form factors $\lambda_i^\z$ and
$\overline{\lambda}_i^\z$ respectively.
It should be noted that all of these
are proportional to the lattice spacing $a$ and hence vanish in the
naive continuum limit, and likewise that the deviation between
$\lambda_1^\z$ and the
continuum, tree-level value of 1 vanishes in the same limit.  However, their magnitude can be large at finite lattice spacing.  

Due to asymptotic freedom, we expect the nonperturbative vertex to
approach its tree-level form at large momentum, and hence be
dominated by lattice artefacts.  We attempt to reduce these by
dividing the lattice $\lambda_1$ by its tree-level form, while
subtracting off the tree-level expression from the raw data for
$\lambda_2$ and $\lambda_3$.

Making a comparison between \eqqref{eq:tree-imp6} and the continuum vertex
\eqqref{eq:sqk_cont_vertex}, we
end up with the tree-level corrected, lattice equivalents of the expressions
\eqref{eq:l1-cont-noncov}, \eqref{eq:l2-cont-noncov},
\eqref{eq:l3-cont-covariant},
\begin{align}
\lambda_1(p^2,0,p^2)  
& =  \frac{{\bm{\mbox{Im}}}}{g_0}\,
\left\{  \,\,\Big[\trd(\gamma_\alpha \overline\Lambda_\mu)\Big]\,
\bigg|_{\substack{\alpha=\mu\\p_\mu=0}}   \right\} \bigg/ \lambda_1^{(0)}
 \label{eq:l1-lat-noncov-corr} \\[2mm]
 \lambda_2 (p^2,0,p^2)
  &=  \frac{{\bm{\mbox{Im}}}}{g_0}\, \left\{    - \frac{1}{4\,K(p)^2}\,\frac{K_{\alpha}(p)K_{\mu}(p)}{K(p)^2}\, \Big[ \trd(\gamma_\alpha \overline\Lambda_\mu) \, \Big|_{\alpha \neq\mu} \Big]  \right\} 
  -\left(\lambda_2^\z +\overline\lambda_{2(\mu)}^{(0)}\right)  
  \label{eq:l2-lat-noncov-corr} \,, \\[2mm]
 \lambda_3(p^2,0,p^2) 
  &= \frac{{\bm{\mbox{Re}}}}{(-g_0)}\, \left\{ \frac{1}{2}\, \frac{K_\mu(p)}{K^2(p)} \trd(I\, \overline\Lambda_{\mu}) \right\}
  -\left(\lambda_3^\z +\overline\lambda_{3(\mu)}^{(0)}\right) 
\label{eq:l3-lat-noncov-corr} \,,
\end{align}
where the lattice tree-level form factors are given by
\begin{align}
  \lambda_1^{(0)}
  & =  F(p)\big(1+ c_q^2 a^2 K^2(p) \big)^2\,,
 \label{eq:lam1-corr} \\
\lambda_2^{(0)}+ \overline\lambda_{2(\mu)}^{(0)}
& = a^2F(p)
\,\bigg[ - c_q\,\big(1-c_q^2a^2K^2(p) \big)
  + 2c_q^2C_\mu(p)\bigg]\,,
 \label{eq:lam2-corr} \\
 \lambda_3^{(0)} + \overline\lambda_{3,(\mu)}^{(0)}
 & = \frac{a}{2}F(p)
 \bigg[\big(1-c_q^2 a^2K^2(p)\big)^2- 4c_q^2 a^2K^2(p)
  - 4c_q\big(1-c_q^2a^2K^2(p)\big)C_{\mu}(p) \,\bigg] \,,
\label{eq:lam3-corr}   
\intertext{and the common prefactor $F(p)$ is given by}
F(p) &=
\frac{(1+b_q am)}{(1+am/2)^2}\,\frac{1}{ \big(1+c_q^2 a^2 K^2(p)\big)^4}\,.
\end{align}
These ($\lambda_1,\lambda_2,\lambda_3)$  are the complete form factors required to determine the quark-gluon vertex in the soft gluon limit, and 
Eqs (\ref{eq:l1-lat-noncov-corr},\ref{eq:l2-lat-noncov-corr},\ref{eq:l3-lat-noncov-corr}) define the exact procedures to calculate them on the lattice.

\subsection{One-loop continuum form factors}
\begin{figure}[H]
    \begin{center}
     \begin{subfigure}[t]{0.3\textwidth}
        \centering
       \includegraphics*[width=\textwidth]{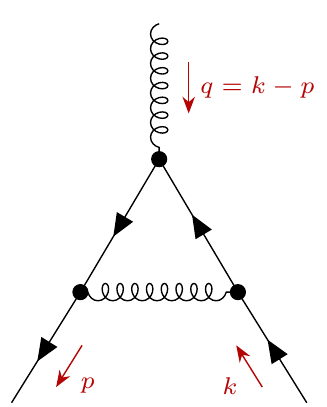}
        \caption{ Abelian contribution}
        \label{fig:oneloop-abelian}      
    \end{subfigure}
\begin{subfigure}[t]{0.3\textwidth}
        \centering      
       \includegraphics*[width=\textwidth]{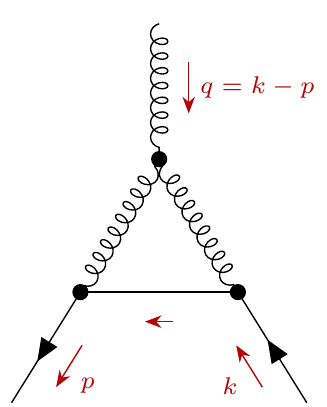}
        \caption{Non-abelian contribution }
        \label{fig:oneloop-nonabelian}      
    \end{subfigure}
      \caption{Complete one-loop diagrams of the quark-gluon vertex.}
\label{fig:one-loop-ver}
  \end{center}
\end{figure}
In interpreting the lattice data for the various quark-gluon vertex form factors, knowledge of the corresponding
perturbative expressions will be useful. Therefore, for completeness we report on the results for the
Landau gauge one-loop Euclidean-space form factors $\lambda_{1,2,3}$ in the
soft gluon kinematics in the momentum subtraction (MOM) scheme, where
$\mu$ is the renormalisation scale, see Fig.~\ref{fig:one-loop-ver}:
\begin{align}
\lambda_1^{pert}(p^2,0,p^2;\mu^2)&= 1-\frac{\alpha_\mu}{4\pi}\, \frac{C_A}{4}\, \left( 
- \frac{m^2}{p^2} +  \frac{m^2}{\mu^2} - 3 \ln\left( \frac{p^2+m^2}{\mu^2+m^2} \right) 
+ \frac{m^4}{p^4}\ln \left(1 + \frac{p^2}{m^2} \right) -  \frac{m^4}{\mu^4}\ln \left(1 + \frac{\mu^2}{m^2} \right)\right)+ {\cal{O}}(\alpha^2)\,,\label{eq:l1-pert} \\
\lambda_2^{pert}(p^2,0,p^2) &= \frac{\alpha_\mu}{4\pi}\frac{C_A}{8p^2}\left(
   1- 2\frac{m^2}{p^2}
 + 2\frac{m^4}{p^4}\ln \left(1 + \frac{p^2}{m^2} \right)  \right) + {\cal{O}}(\alpha^2)\,,
 \label{eq:l2-pert}\\
 \lambda_3^{pert}(p^2,0,p^2) &=  \frac{\alpha_\mu}{4\pi}\, 3C_F\,\frac{m}{p^2}\,
\left(1-  \frac{m^2}{p^2} 
  \ln\left(1 + \frac{p^2}{m^2} \right) \right) + {\cal{O}}(\alpha^2)\,.  \label{eq:l3-pert}
\end{align}
$C_A$ and $C_F$ are the eigenvalues of the quadratic Casimir operator
in the fundamental representation and adjoint representation,
respectively: $C_A=N$, $C_F=N^2-1/2N$, with $N=3$ for SU(3).
Note that $\lambda_3$ is proportional to the quark mass and a non-vanishing $\lambda_3$, as is observed in lattice simulations,
is an indication of chiral symmetry breaking.
\begin{figure}[thb]
    \begin{center}
     \begin{subfigure}[t]{0.32\textwidth}
        \centering
       \includegraphics*[width=\textwidth]{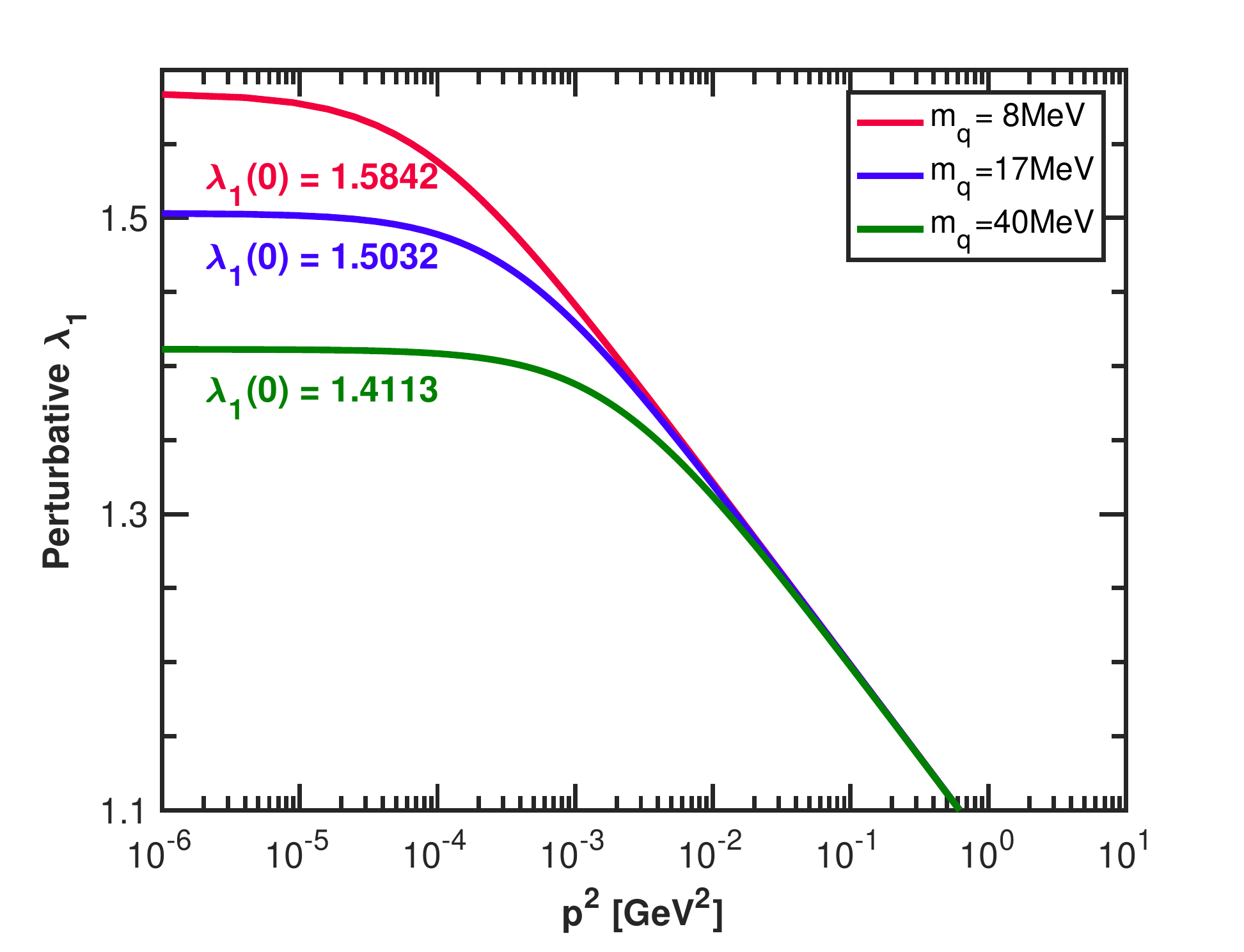}
        \caption{$\lambda_1$.}
        \label{fig:pertlam1}      
    \end{subfigure}
\begin{subfigure}[t]{0.32\textwidth}
        \centering      
       \includegraphics*[width=\textwidth]{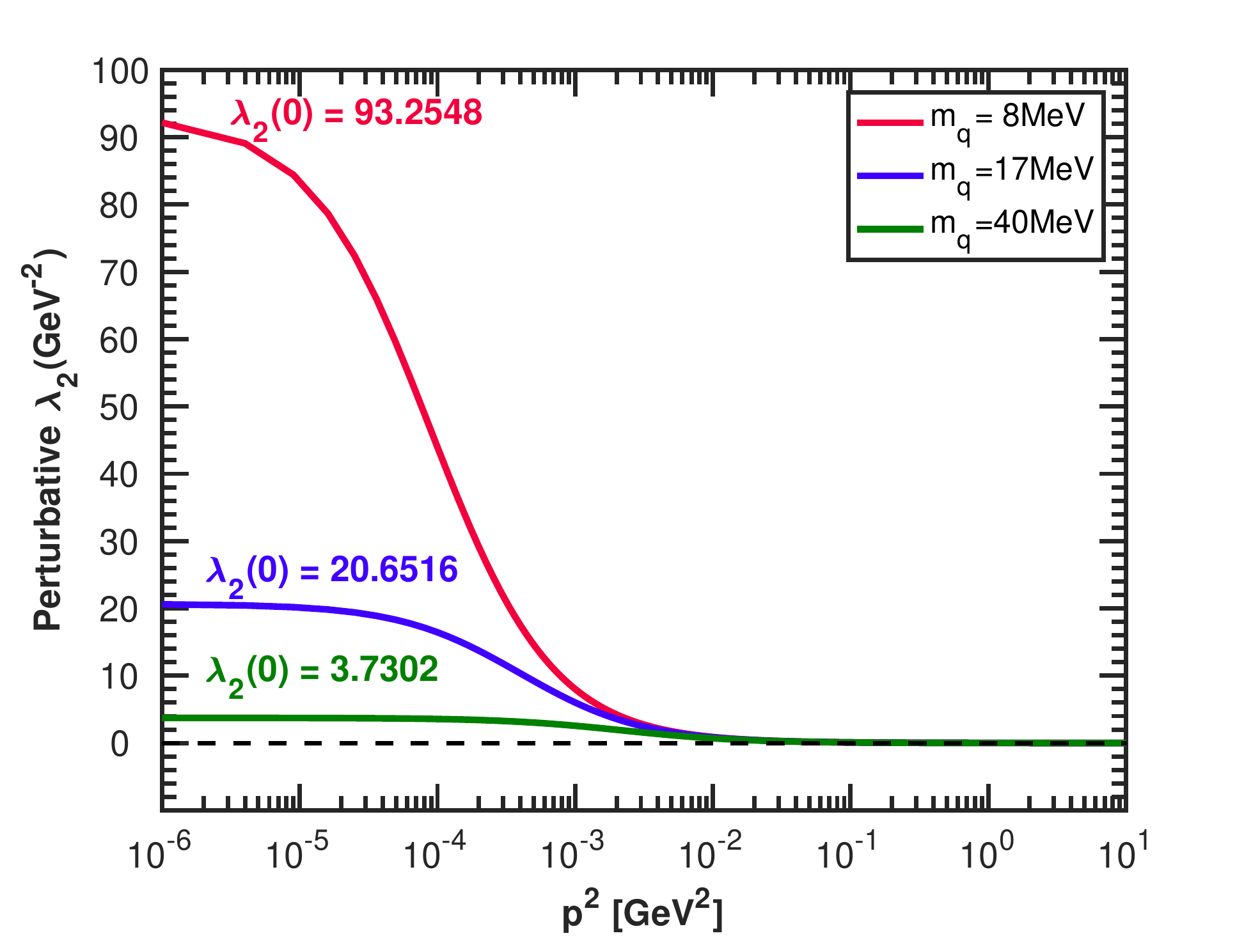}
        \caption{$\lambda_2$.}
        \label{fig:pertlam2}      
    \end{subfigure}
    \begin{subfigure}[t]{0.32\textwidth}
        \centering      
       \includegraphics*[width=\textwidth]{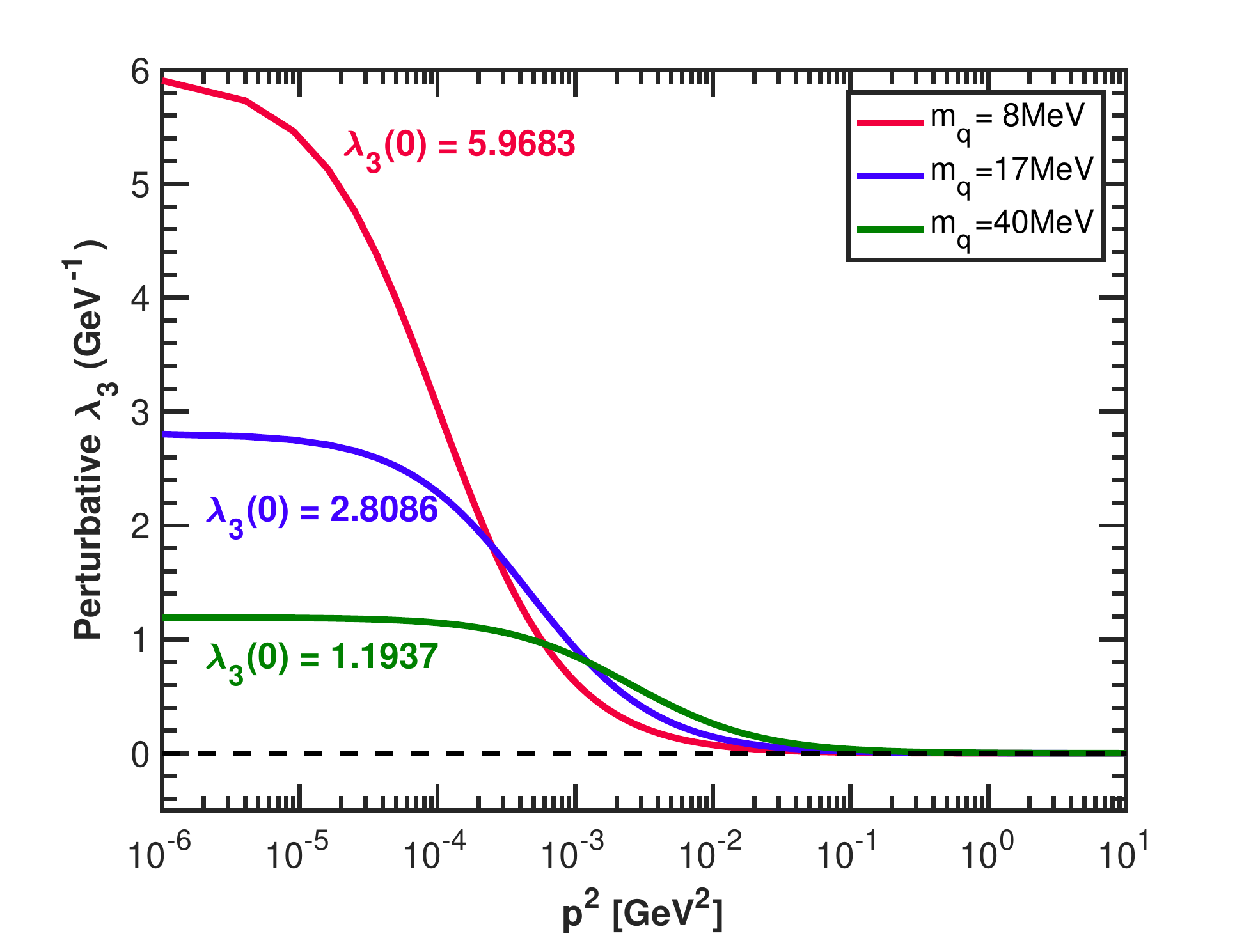}
        \caption{$\lambda_3$.}
        \label{fig:pertlam3}      
    \end{subfigure}
      \caption{One-loop perturbative form factors  of the quark-gluon vertex.}
\label{fig:one-loop-lambda}
  \end{center}
\end{figure}

The one-loop expressions are plotted in Figure~\ref{fig:one-loop-lambda},
for three different values of the quark mass.  The value of the form
factors in the limit $p\to0$ is also shown in the graph.  We note that
all three form factors are enhanced in the infrared, and that this
enhancement increases with decreasing quark mass for all three.  This
has also been observed in other kinematics \cite{Bermudez:2017bpx}. In
the case of $\lambda_3$, however, we note that this mass ordering only occurs
in the very far infrared, $p\lesssim15\,$MeV, and that the mass ordering is
the opposite for $p\gtrsim50\,$MeV, which includes all momenta
available for the lattice volumes considered in this study.

\section{Results}
\label{sec:results}
In the following, we will present our results for the quark-gluon
vertex in such a way as to carefully trace
the effects of tree-level  correction (Sec.~\ref{TL-corrected}),
unquenching (Sec.~\ref{sec:QuenchedvsDyn}), mass
(Sec.~\ref{sec:massdep}), volume (Sec.~\ref{sec:volumedep})
and lattice spacing (Sec.~\ref{sec:latspacing}) on each of the three
form factors.

\subsection{Tree-level-corrected  versus uncorrected}
\label{TL-corrected}

\begin{figure}[thb]
    \centering
 \begin{subfigure}[t]{0.495\textwidth}
        \centering
       \includegraphics*[width=\textwidth]{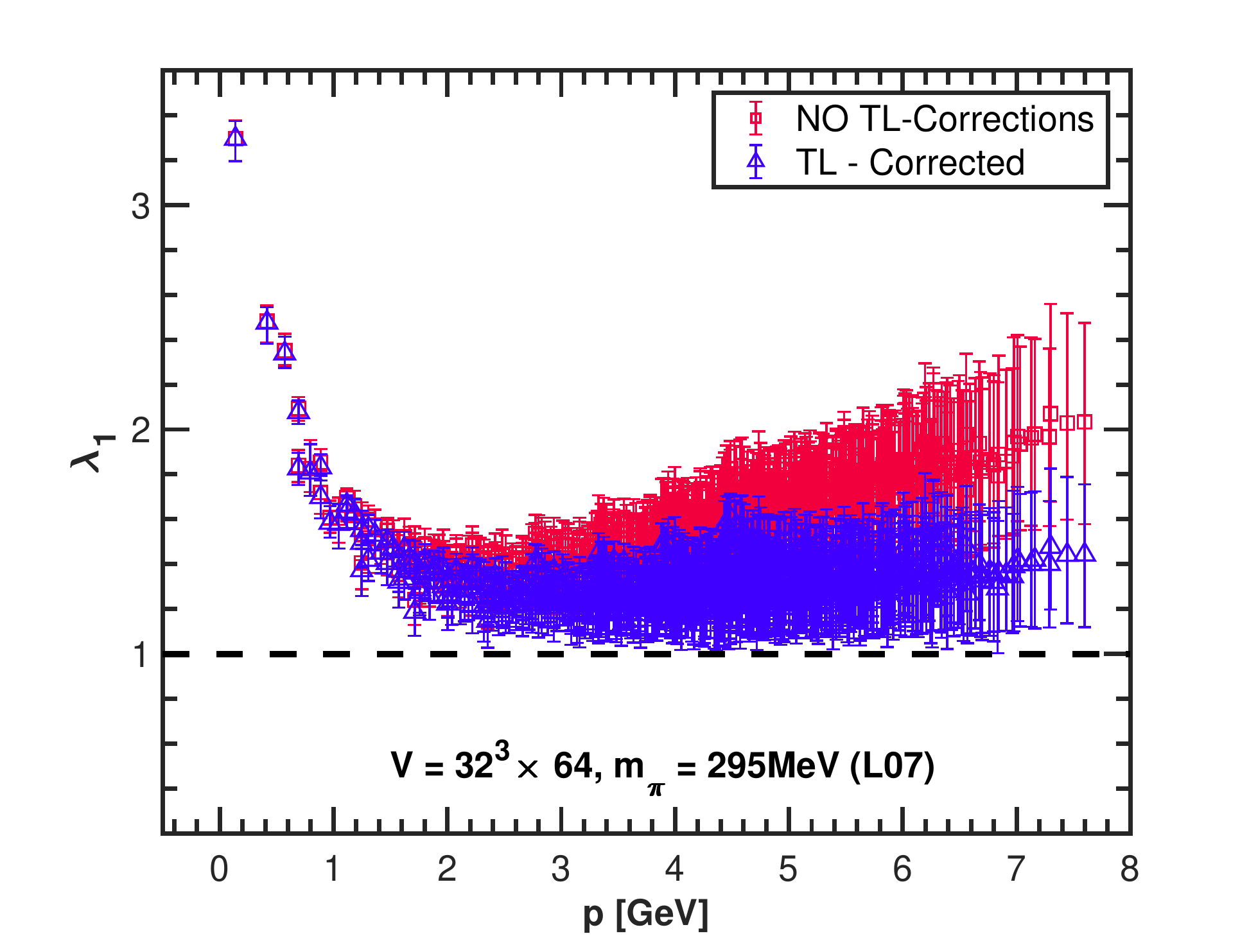}
        \caption{ }
        \label{fig:lambda1_vs_p_L07tl}      
    \end{subfigure}
\begin{subfigure}[t]{0.495\textwidth}
        \centering
       \includegraphics*[width=\textwidth]{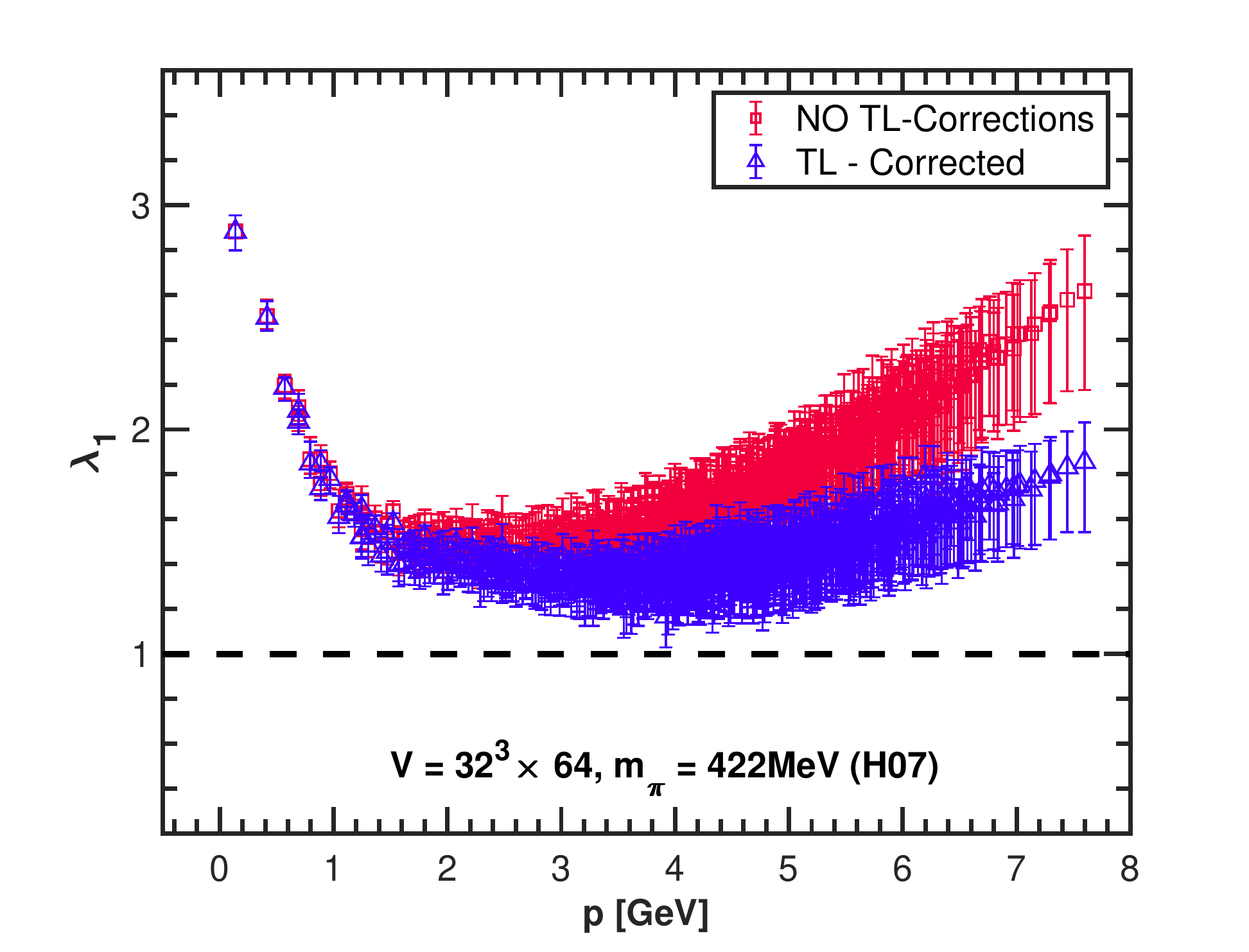}
         \caption{ }
        \label{fig:lambda1_vs_p_H07tl}      
    \end{subfigure}
    \begin{subfigure}[t]{0.495\textwidth}
        \centering
       \includegraphics*[width=\textwidth]{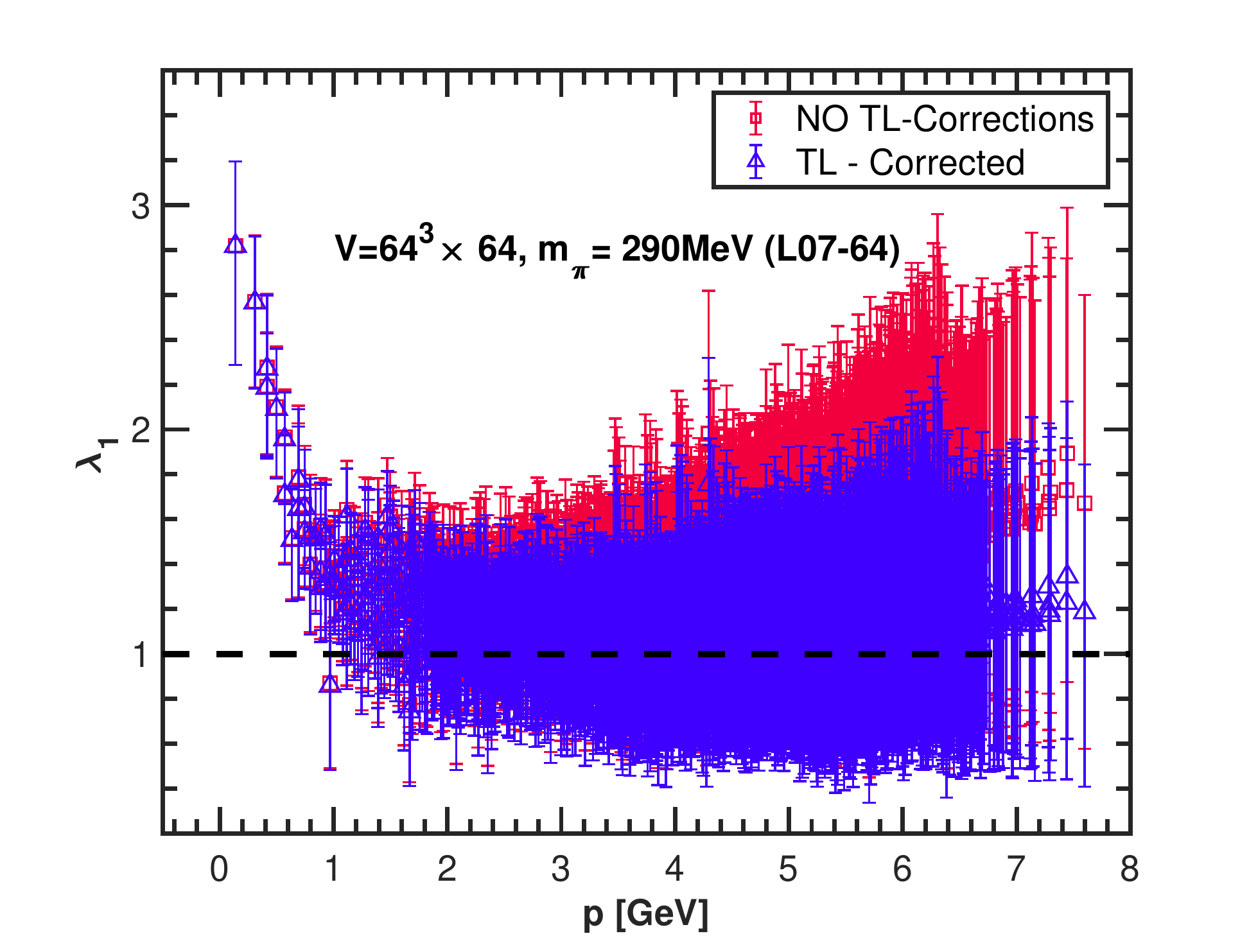}
 \caption{ }
 \label{fig:lambda1_vs_p_L07-64tl}      
           \end{subfigure}
      \begin{subfigure}[t]{0.495\textwidth}
        \centering
       \includegraphics*[width=\textwidth]{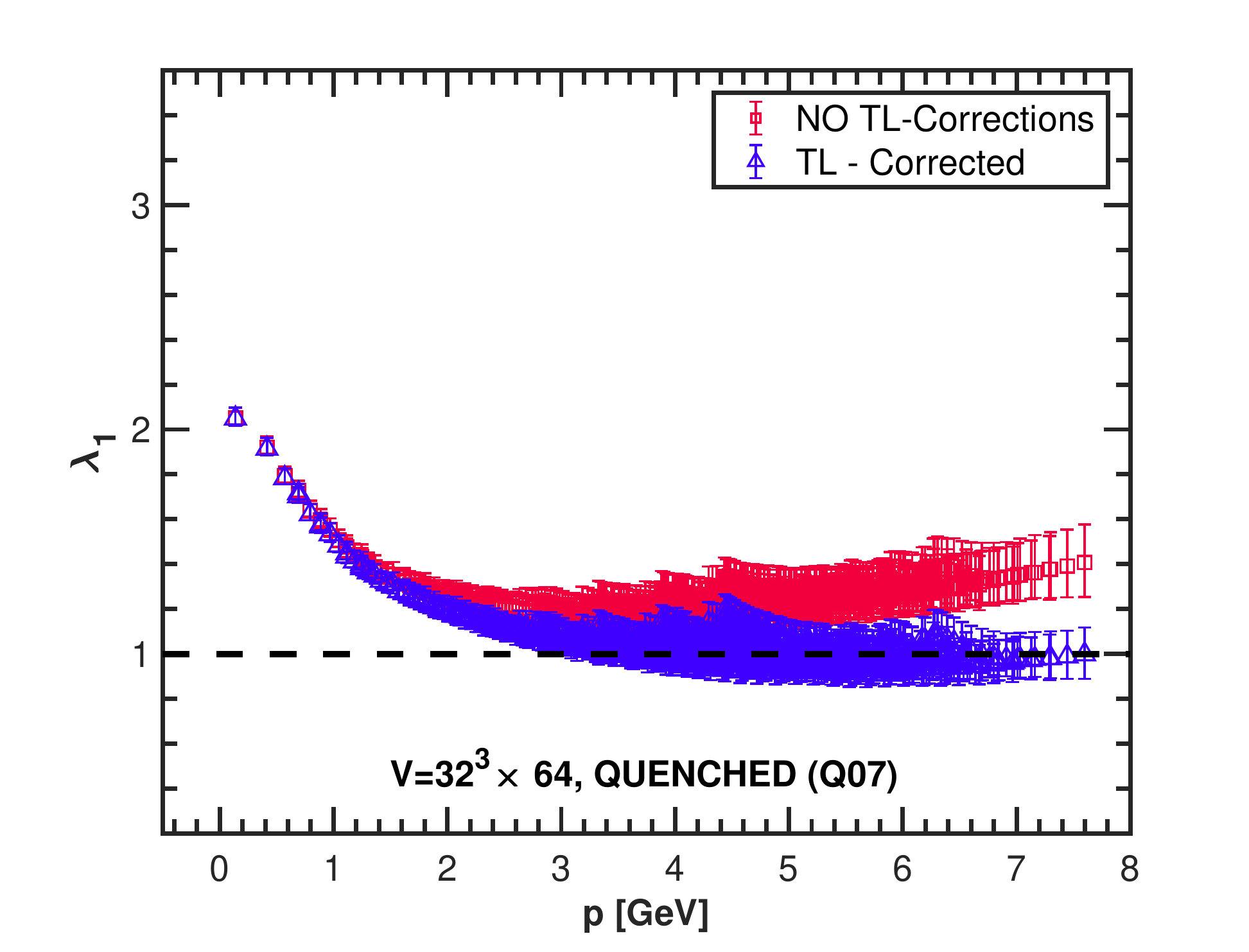}
 \caption{ }
        \label{fig:lambda1_vs_p_quenchedtl}
    \end{subfigure}         
      \caption{The leading, unrenormalised form factor
        $\lambda_1(p^2,0,p^2)$ with and without tree-level (TL) corrections
        as a function of momentum $p$, for different lattice
        ensembles with $a=0.07\,$fm: (\subref{fig:lambda1_vs_p_L07tl}) L07;
        (\subref{fig:lambda1_vs_p_H07tl}) H07;
        (\subref{fig:lambda1_vs_p_L07-64tl}) L07-64;
        (\subref{fig:lambda1_vs_p_quenchedtl}) Q07.
      }
\label{fig:lambda1_TL}
\end{figure}

We first consider the effect of the tree-level corrections on the vertex, starting with the leading vertex component $\lambda_1$.  Representative results for
$\lambda_1$ as a function of momentum $p$ with and without tree-level correction are shown in Fig.~\ref{fig:lambda1_TL}.   
Fig.~\ref{fig:lambda1_vs_p_quenchedtl} shows quenched results, $N_f=0$, while Figs.~\ref{fig:lambda1_vs_p_L07tl}  and ~\ref{fig:lambda1_vs_p_L07-64tl}  show results 
for $a=0.07\,\mathrm{fm}, m_{\pi} \simeq 295\,$MeV on the $32^3 \times 64$ and $64^3 \times 64$ lattices respectively. For comparison, Fig.~\ref{fig:lambda1_vs_p_H07tl}  shows the results for the heavier pion mass, $m_{\pi} = 422\,$MeV.  

Due to asymptotic freedom, we expect that $\lambda_1(p)\to1$ (plus
logarithmic corrections) at large $p$, and we can see that the
tree-level corrections have the effect of bringing the lattice data
closer to this continuum-like form for all cases and reducing the
violations of rotational symmetry compared to the uncorrected data.
In some cases, (notably for the largest quark mass $m_\pi=422\,$MeV),
we see indications that significant lattice artefacts remain even
after tree-level correction and, therefore, careful study of the
lattice spacing dependence will be necessary to determine the
large-momentum behaviour of the vertex.  For the larger volume
($64^3\times 64$), the fluctuations of $\lambda_1$ are quite large,
Fig.~\ref{fig:lambda1_vs_p_L07-64tl}.  In all cases the
tree-level correction is small below 1\,GeV but becomes pronounced for
$p\gtrsim1\,$GeV ($pa\gtrsim0.4$).


\begin{figure}[thb]
    \centering
 \begin{subfigure}[t]{0.495\textwidth}
        \centering
       \includegraphics*[width=\textwidth]{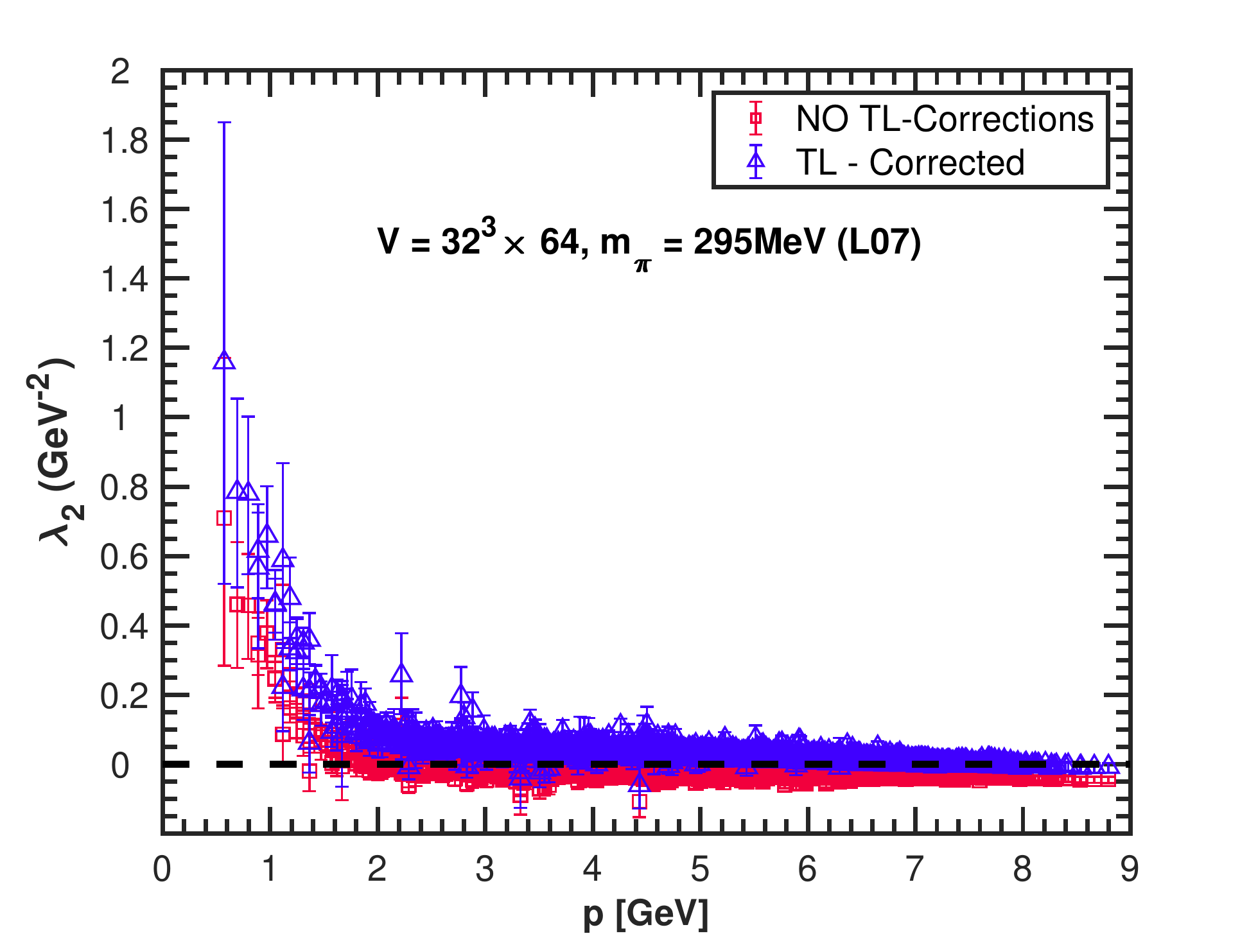}
       \caption{ }  
        \label{fig:lambda2_vs_ap_lin_b529_a007_m295}      
    \end{subfigure}
\begin{subfigure}[t]{0.495\textwidth}
        \centering
       \includegraphics*[width=\textwidth]{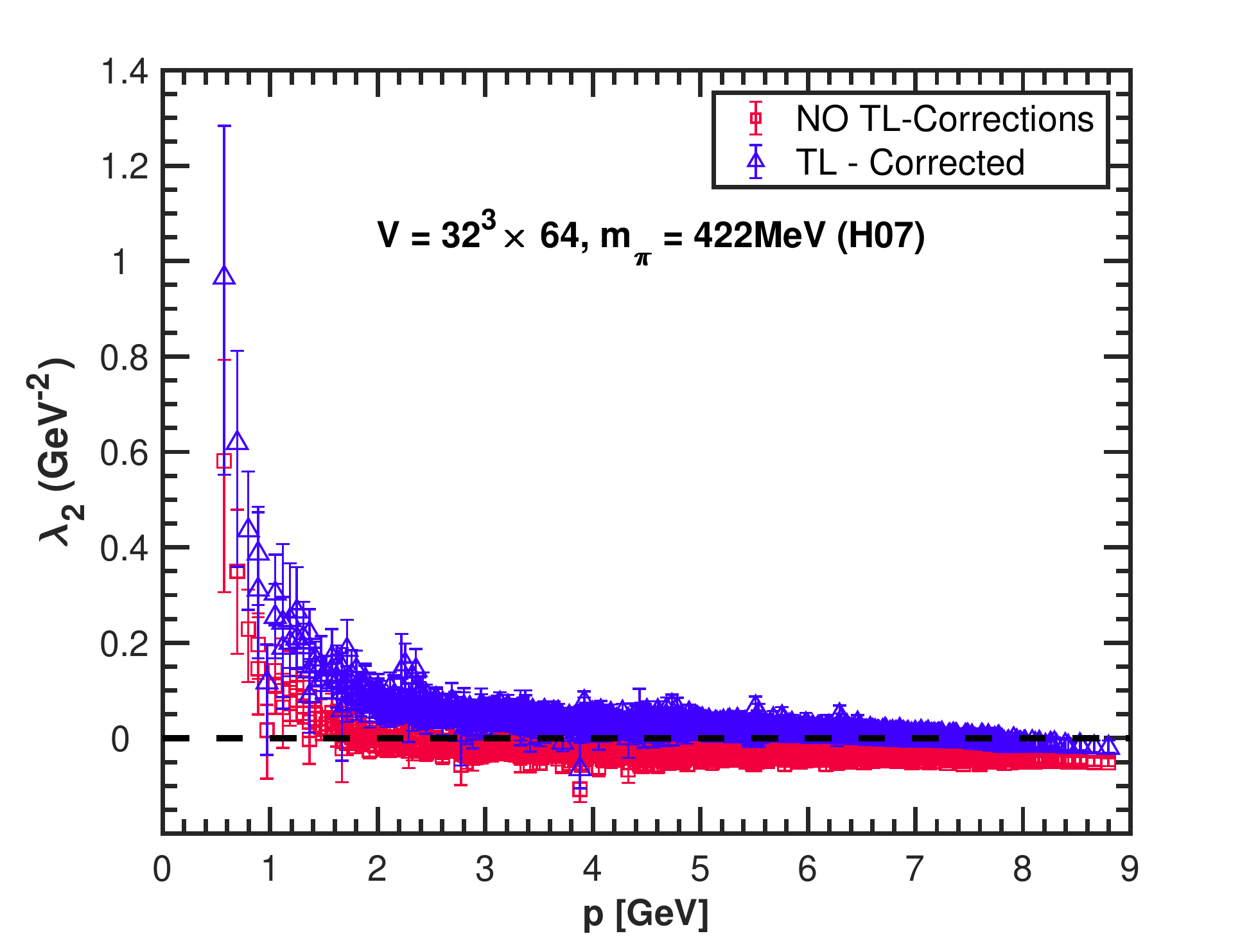}
      \caption{ }  
        \label{fig:lambda2_vs_ap_lin_b529_a007_m422}      
    \end{subfigure}
    \begin{subfigure}[t]{0.495\textwidth}
        \centering
       \includegraphics*[width=\textwidth]{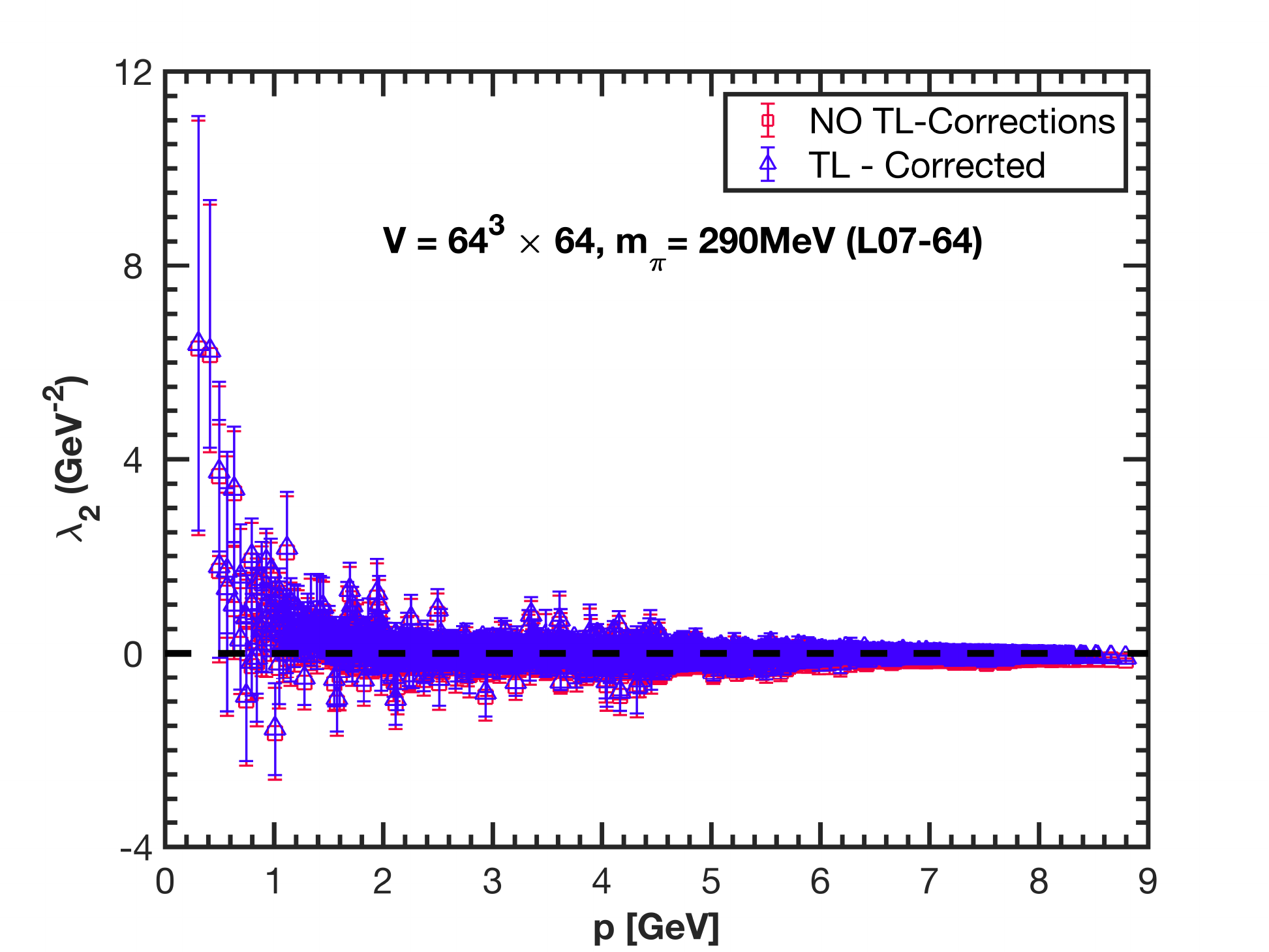}
 \caption{ }  
        \label{fig:lambda2_vs_ap_lin_b616_a007_mxxx}      
    \end{subfigure}
      \begin{subfigure}[t]{0.495\textwidth}
        \centering
       \includegraphics*[width=\textwidth]{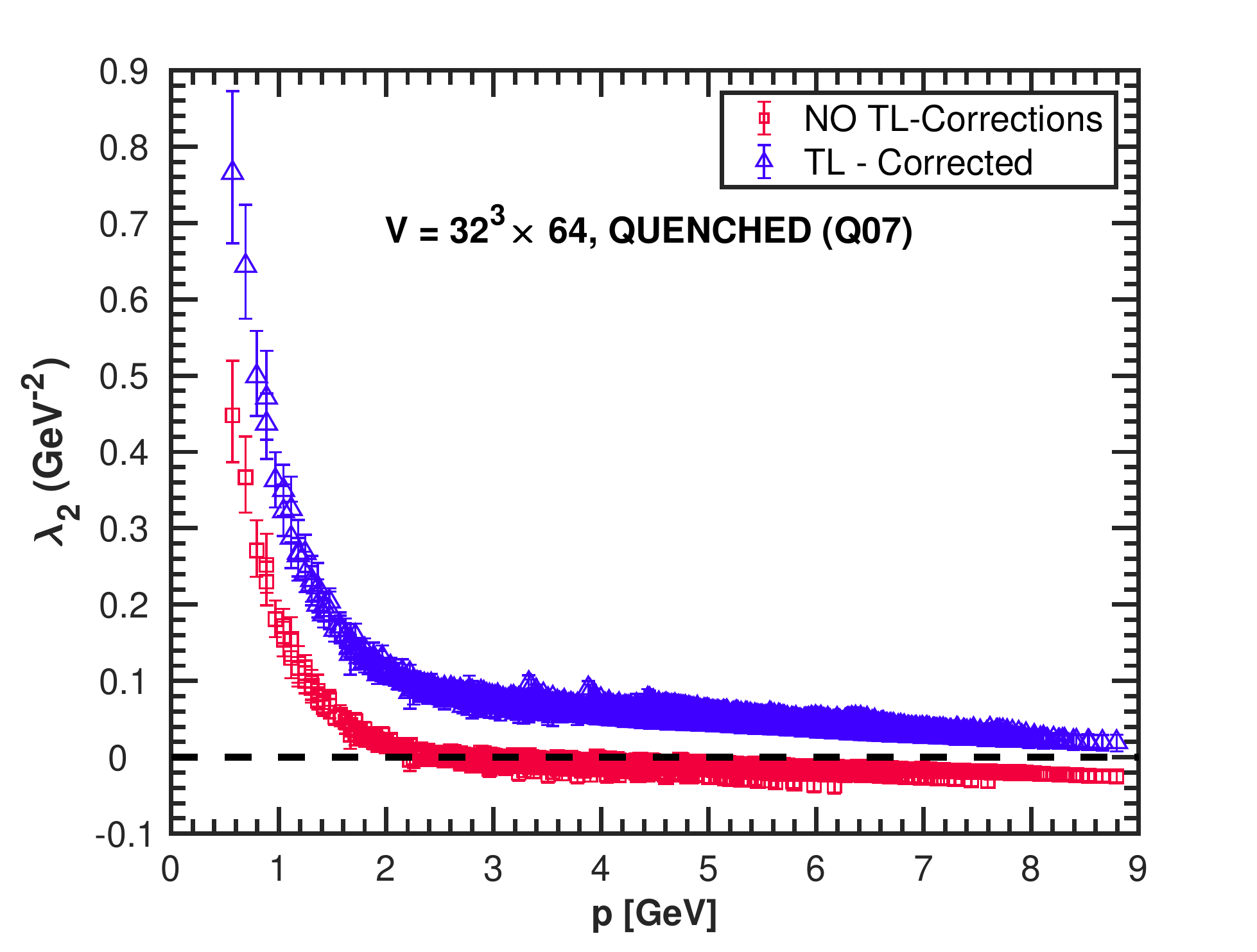}
 \caption{ }  
        \label{fig:llambda2_ren_vs_ap_lin_mixed}
    \end{subfigure}         
\caption{As Fig.~\ref{fig:lambda1_TL}, for the unrenormalised form
  factor $\lambda_2$.}
\label{fig:lambda2_TL}
\end{figure}

Using the same lattice ensembles as in Fig.~\ref{fig:lambda1_TL},
the effect of the tree-level corrections on the $\lambda_2$, are shown
in  Fig.~\ref{fig:lambda2_TL}. We see that
$\lambda_2$ tends to zero as $p$ increases, in contrast to
$\lambda_1$, which approaches a value close to 1 --- in line with
expectations from asymptotic freedom in both cases.
  On the larger volume $64^3\times 64$ lattice, the tree-level
  correction does not appear to have any effect on the $\lambda_2$
  lattice artifacts within the uncertainties of the data.  However,
  the different vertical scales complicate the comparison with the
  other figures. The response of the $\lambda_2$ to tree-level corrections  is noticeable  for all momenta.
 The data in Fig.~\ref{fig:lambda2_TL} also show that $\lambda_2$ is strongly enhanced at low momenta, although the volume and mass dependence
 of $\lambda_2$ are difficult to disentangle. The data also suggest that $\lambda_2$ is enhanced for dynamical simulations relative to the quenched case.

\begin{figure}[thb]
    \centering
 \begin{subfigure}[t]{0.495\textwidth}
        \centering
       \includegraphics*[width=\textwidth]{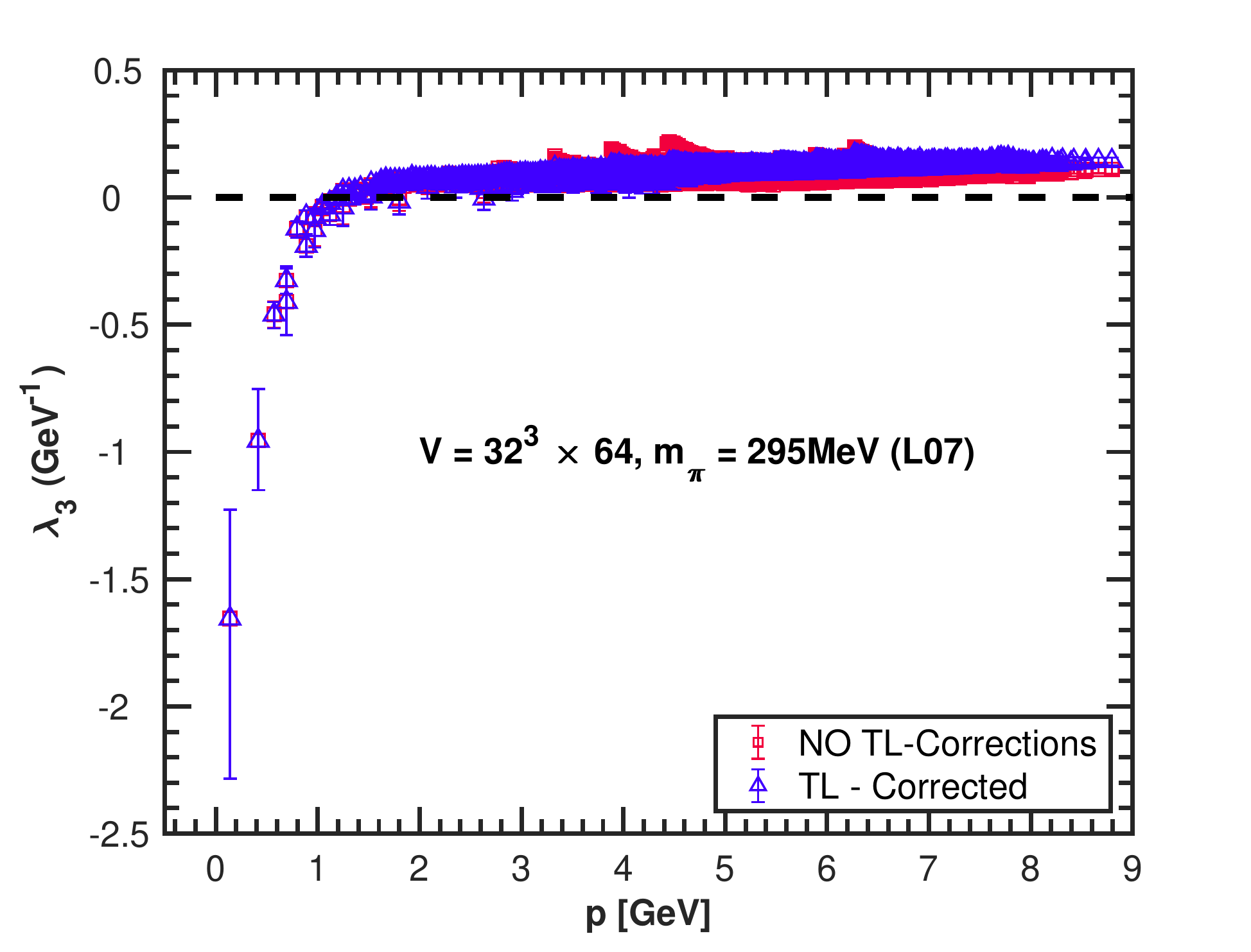}
 \caption{ }  
        \label{fig:lambda3_vs_ap_lin_b529_a007_m295}      
    \end{subfigure}
\begin{subfigure}[t]{0.495\textwidth}
        \centering
       \includegraphics*[width=\textwidth]{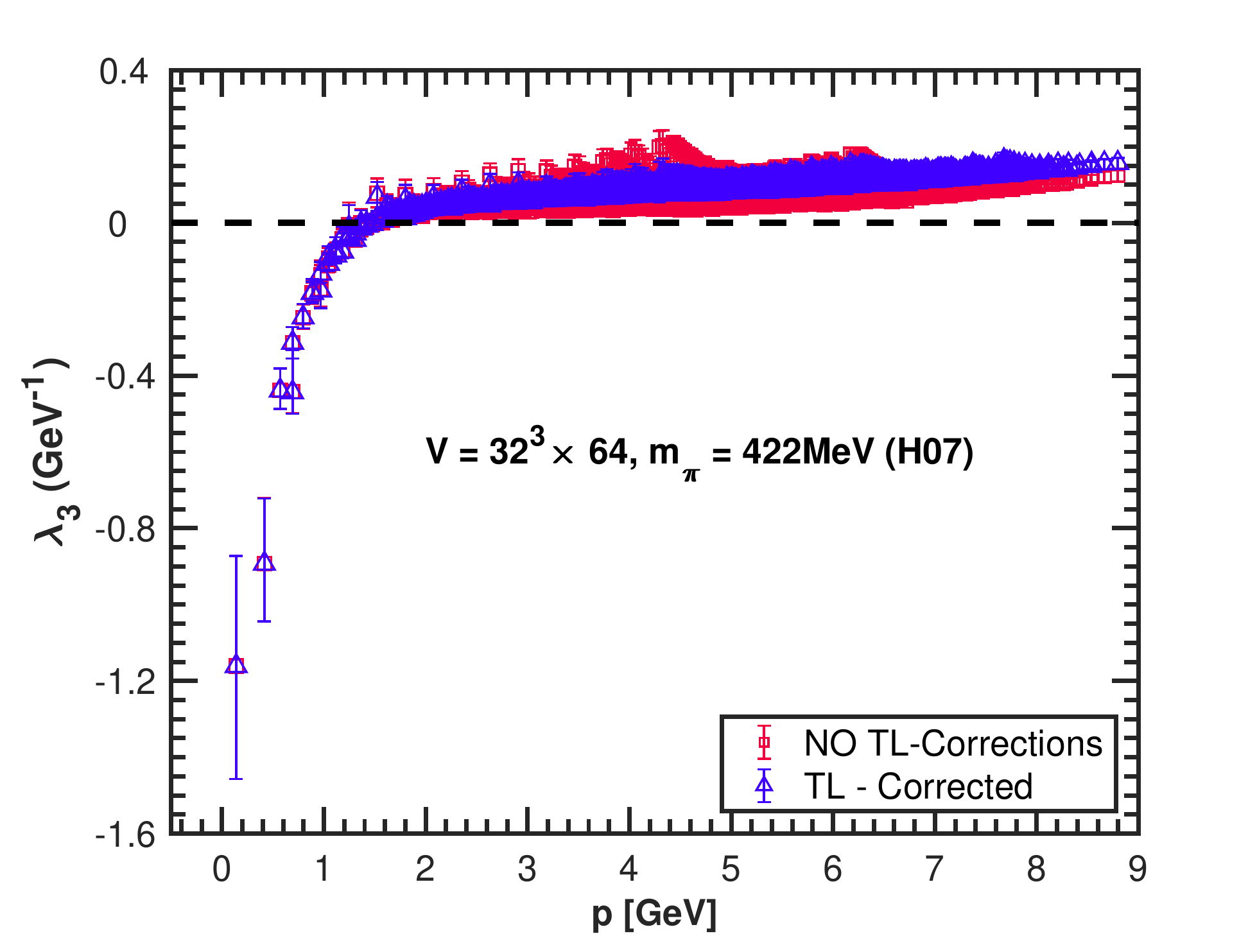}
 \caption{ }  
        \label{fig:lambda3_vs_ap_lin_b529_a007_m422}      
    \end{subfigure}
\begin{subfigure}[t]{0.495\textwidth}
        \centering
       \includegraphics*[width=\textwidth]{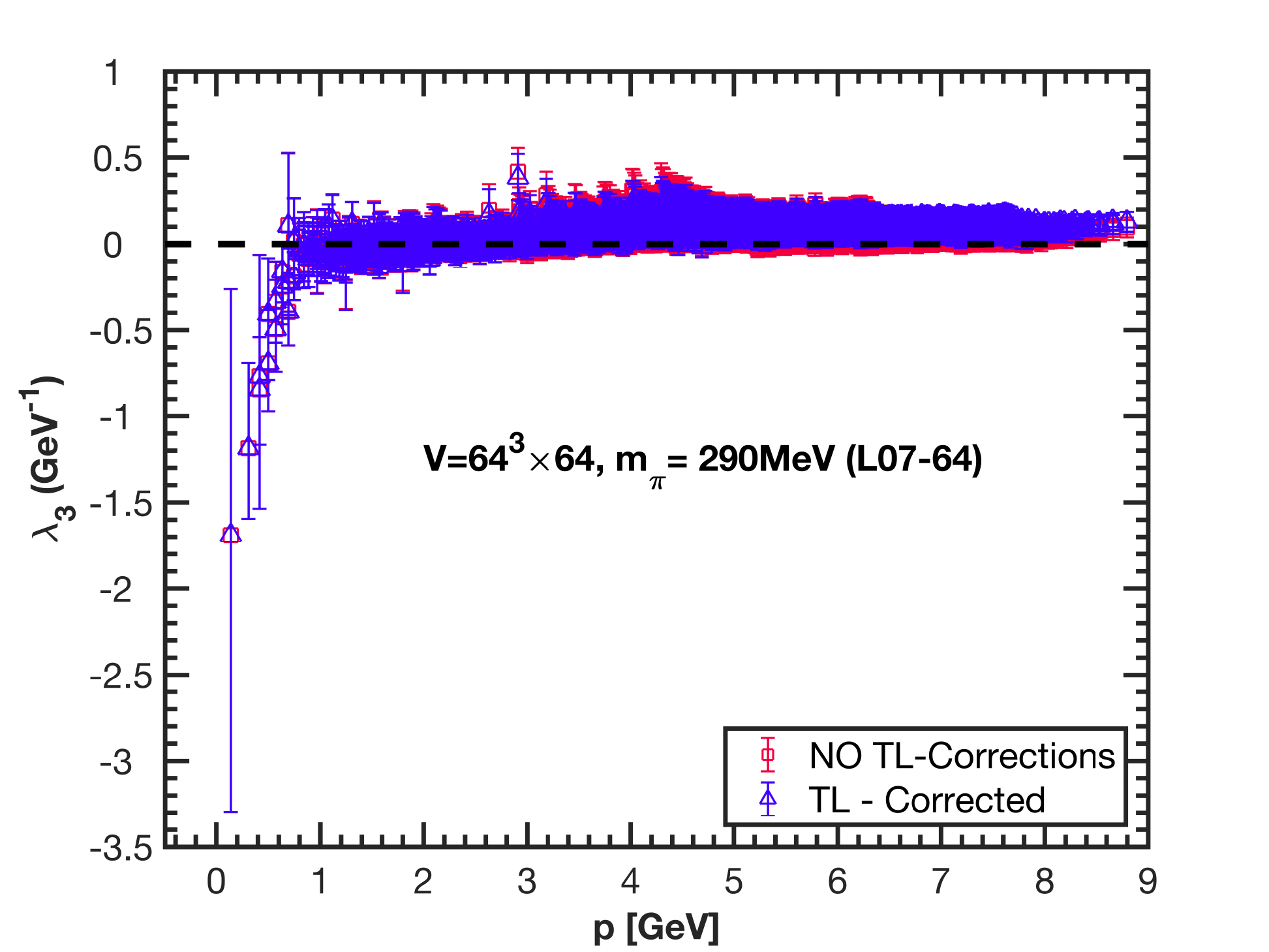}
 \caption{ }  
        \label{fig:lambda3_vs_ap_lin_b529_a007_largeV}      
    \end{subfigure}   
      \begin{subfigure}[t]{0.495\textwidth}
        \centering
       \includegraphics*[width=\textwidth]{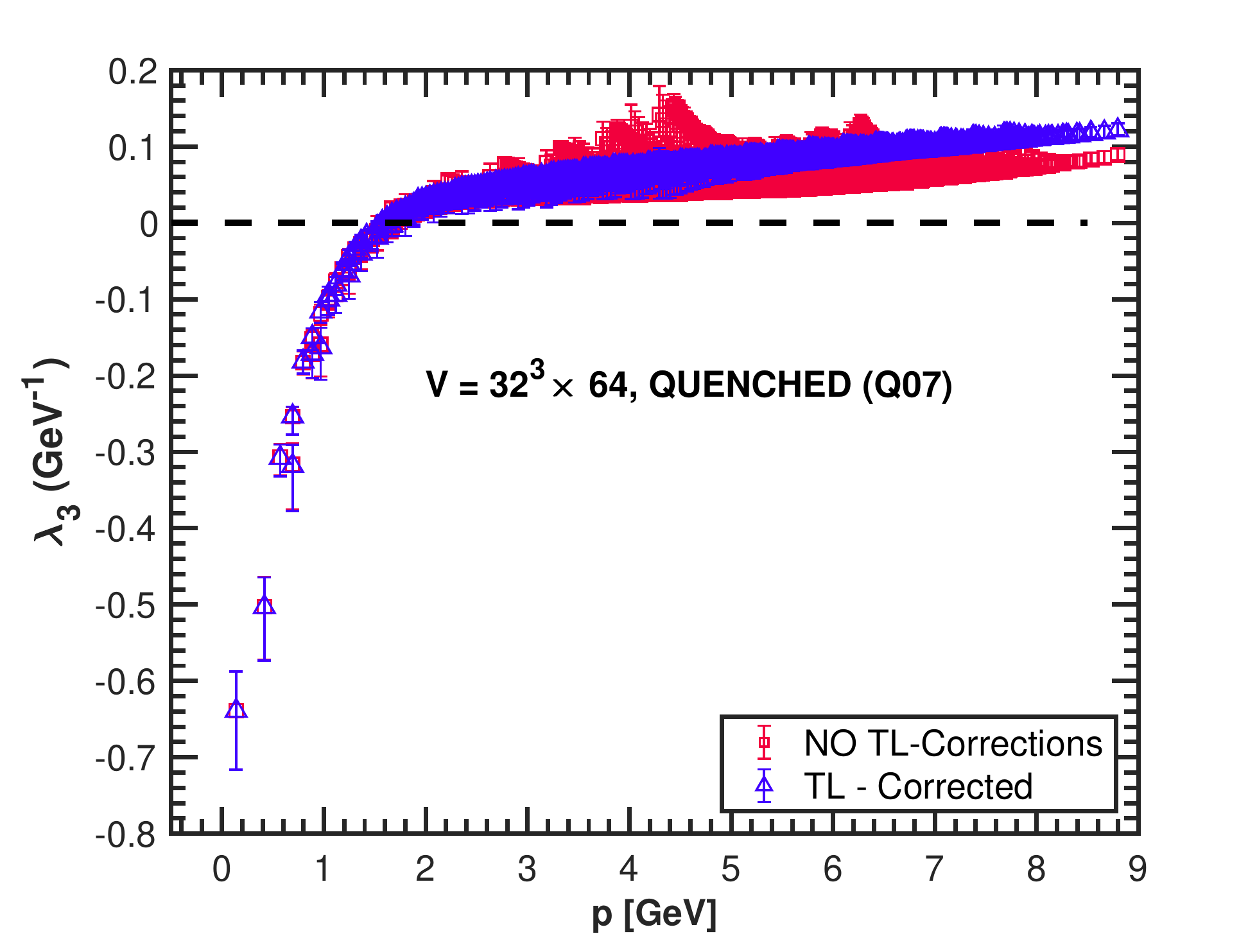}
 \caption{ }  
        \label{fig:lambda3_vs_ap_lin_a007_quenched}      
    \end{subfigure}    
\caption{As Fig.~\ref{fig:lambda1_TL}, for the unrenormalised form
  factor $\lambda_3$.}
\label{fig:lambda3_TL}
\end{figure}

In Fig.~\ref{fig:lambda3_TL} the effect of the
tree-level corrections on $\lambda_3$ are shown.  They are pronounced
at  large momentum of around 2 GeV onwards, and have almost no effect
for momenta  less than 2\,GeV.
The corrections pushes $\lambda_3$ upwards in the ultraviolet, and
$\lambda_3$ does not vanish at higher momenta as one would expect.
This appears to be the case
for all ensembles.  We conclude that in the case of $\lambda_3$, the
tree-level correction does not work satisfactorily for $pa\gtrsim1.5$,
corresponding to momenta greater
than $\sim4\,$GeV, and we will discard our results for
$pa\gtrsim1.5$ ($p\gtrsim4\,$GeV) as unreliable in the absence of a careful,
controlled continuum extrapolation.

%
%

\begin{figure}[thb]
    \centering
      \begin{subfigure}[t]{0.495\textwidth}
        \centering
       \includegraphics*[width=\textwidth]{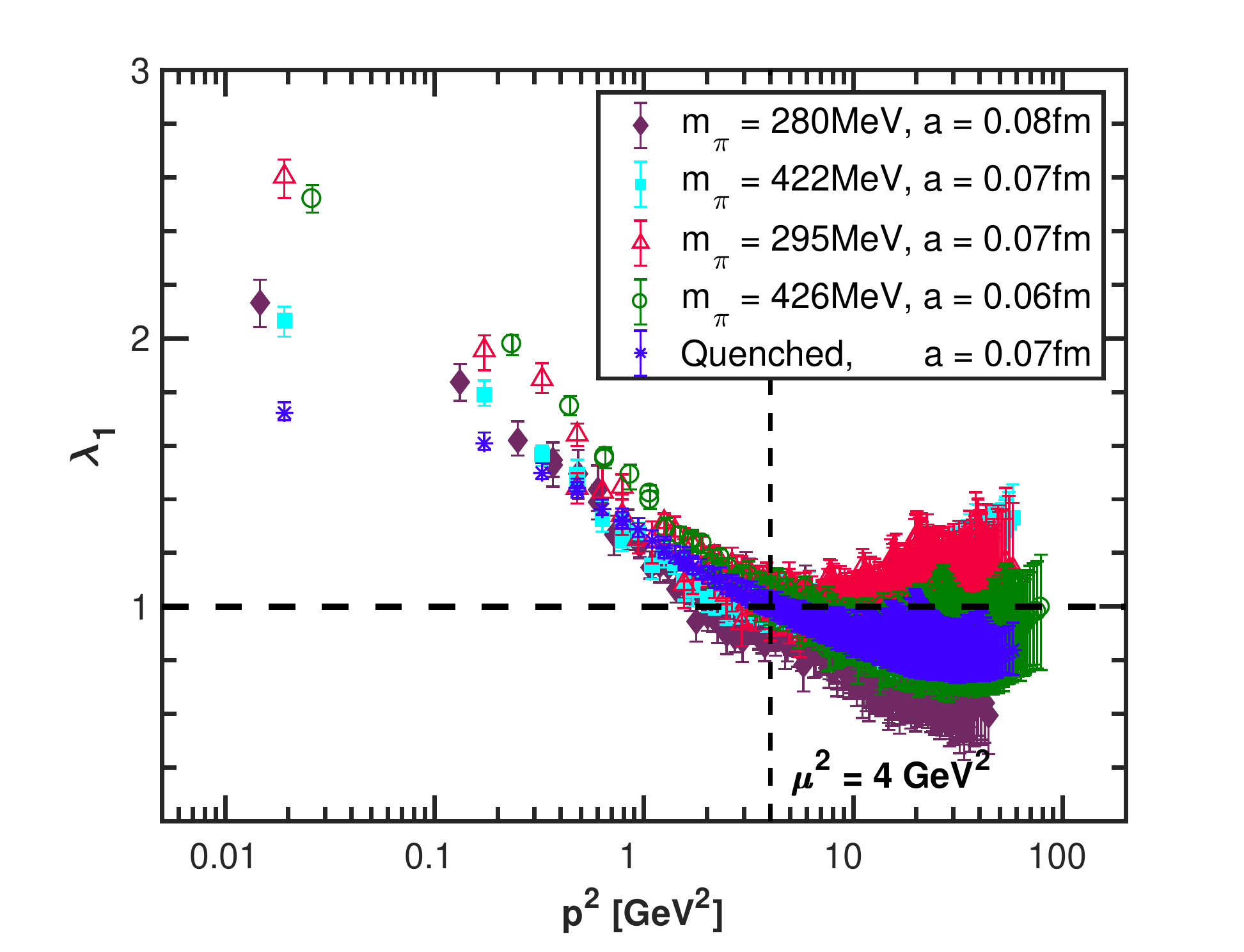}
        \caption{$\lambda_1$}  
        \label{fig:lambda1_ren_vs_ap_lin_mixed}
    \end{subfigure}         
  \begin{subfigure}[t]{0.495\textwidth}
        \centering
       \includegraphics*[width=\textwidth]{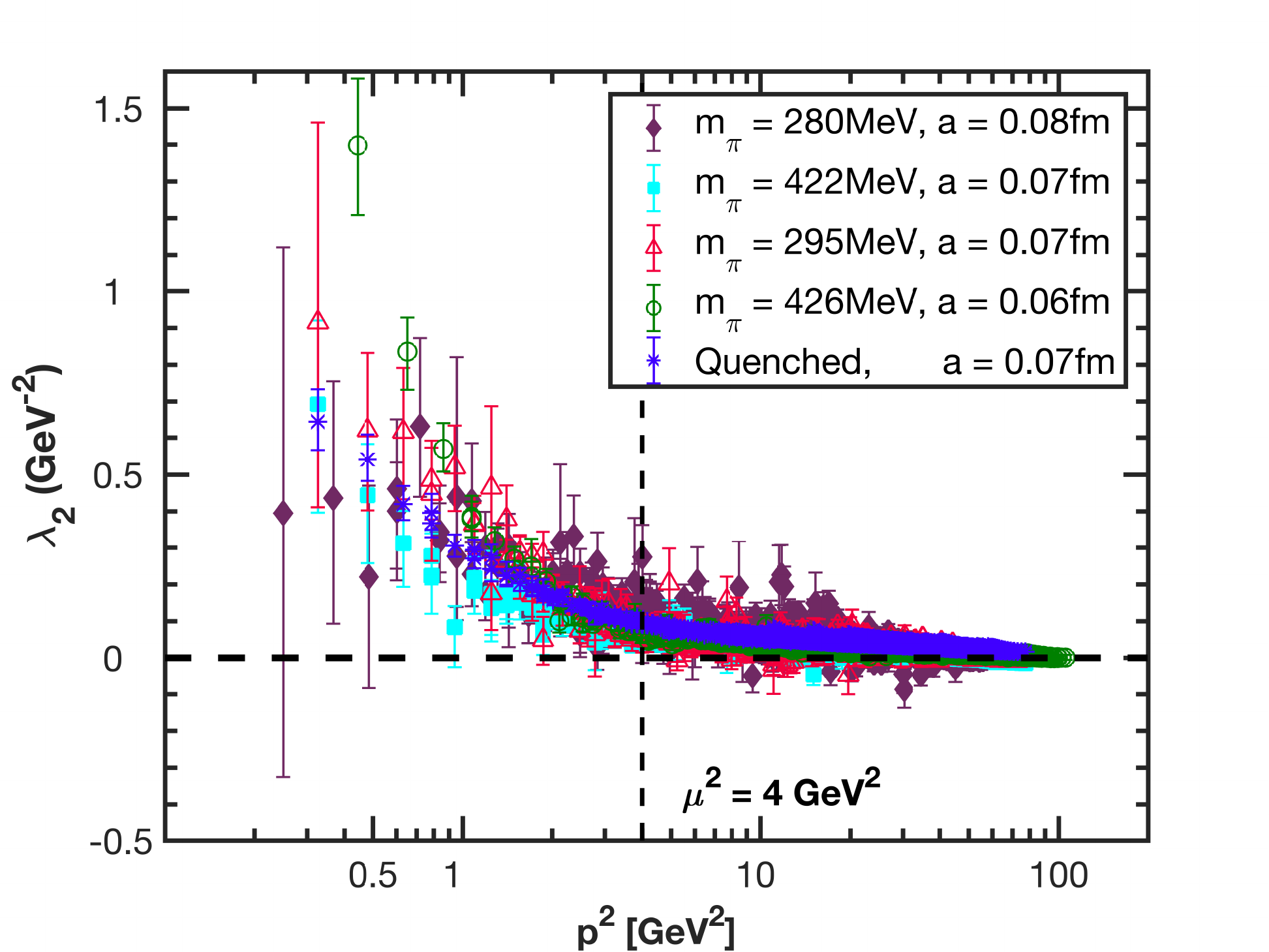}
        \caption{$\lambda_2$. }
        \label{fig:lambda2_ren_vs_ap_lin_mixed}
    \end{subfigure}         
  \begin{subfigure}[t]{0.495\textwidth}
        \centering
       \includegraphics*[width=\textwidth]{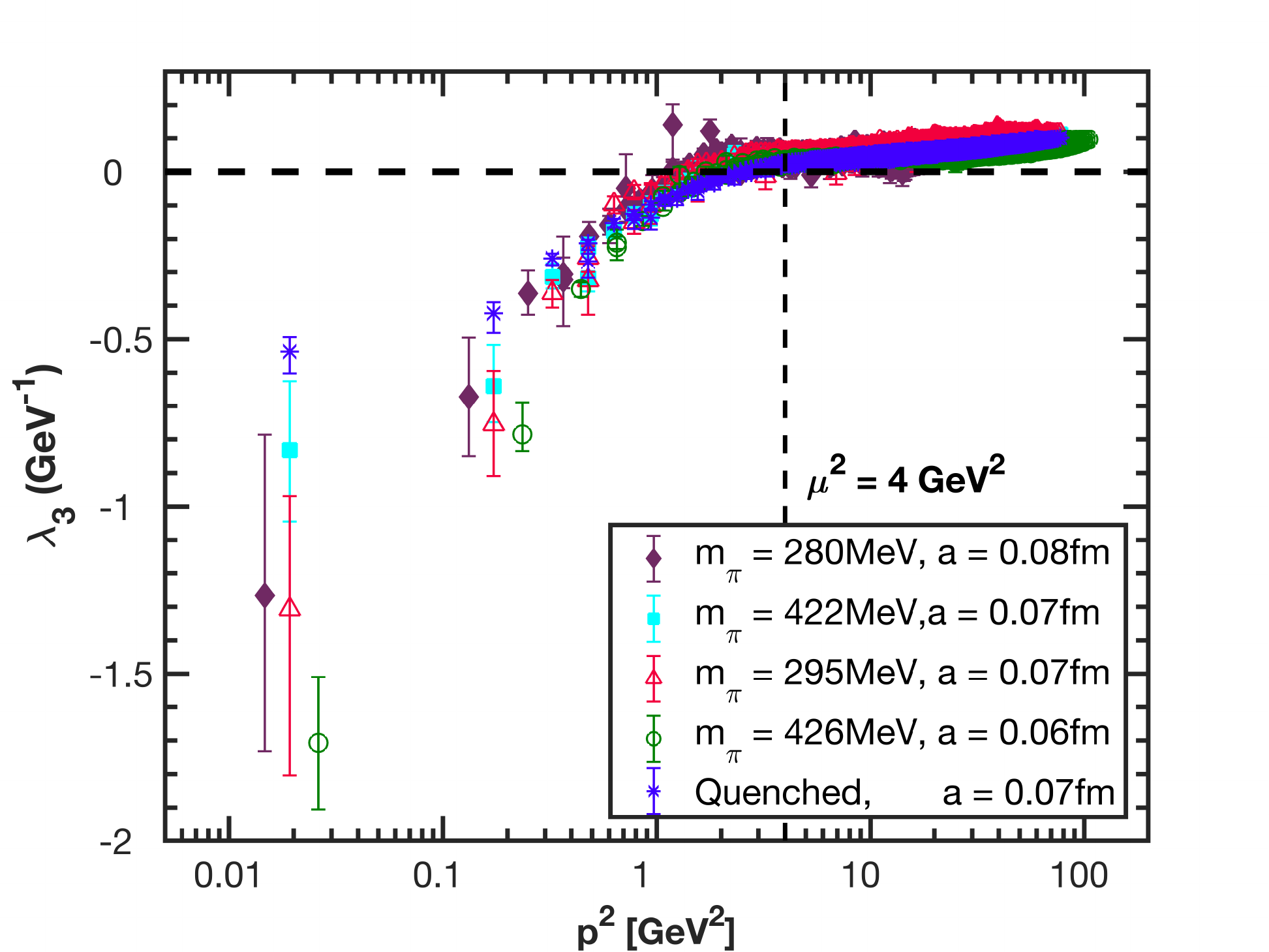}
        \caption{$\lambda_3$. }
        \label{fig:lambda3_ren_vs_ap_lin_mixed}
    \end{subfigure}         
      \caption{The form factors $\lambda_1, \lambda_2, \lambda_3$ with
        tree-level corrections versus momentum $p^2$.  All form
        factors have been renormalised at 2\,GeV.}
\label{fig:lambda_all_TL}
\end{figure}

The results for all our ensembles are collated in Figs.~\ref{fig:lambda_all_TL}. This figure shows only minor variations with the quark mass and lattice spacing,
with indications that decreasing the quark mass leads to a slightly
stronger enhancement in the infrared, while decreasing the lattice
spacing may have a small effect in the same direction.  Indeed, for
all form factors we see that their magnitude is largest for the H06 ensemble
($m_\pi=426\,\mathrm{MeV}, a=0.06\,$fm).  We will discuss the quark mass
and lattice spacing dependence in more detail in sections
\ref{sec:massdep} and \ref{sec:latspacing}.

The most striking feature of $\lambda_2$ in Fig.~\ref{fig:lambda2_ren_vs_ap_lin_mixed} is that this
form factor appears to diverge more strongly in the infrared as we
approach the continuum limit.  
Our results for $\lambda_3$ are summarised in
Fig.~\ref{fig:lambda3_ren_vs_ap_lin_mixed}.  We find that the magnitude of $\lambda_3$
in the infrared increases with increasing quark mass, which is
expected since according to perturbation theory this form factor
breaks chiral symmetry and is 
proportional to the quark mass,  \eqqref{eq:l3-pert}.  Our result
also suggest that its infrared magnitude increases as we approach the
continuum limit.
%

From now on we will only show tree-level corrected data.
All data shown from here on have
been renormalised at 2\,GeV, and to make the
presentation clearer we have  averaged data with nearby values of
momentum, with a momentum bin size $a\Delta p = 0.05$.

\subsection{Quenched  versus dynamical}
\label{sec:QuenchedvsDyn}

\begin{figure}[htb]
\centering
   \begin{subfigure}[t]{0.495\textwidth}
        \centering
       \includegraphics*[width=\textwidth]{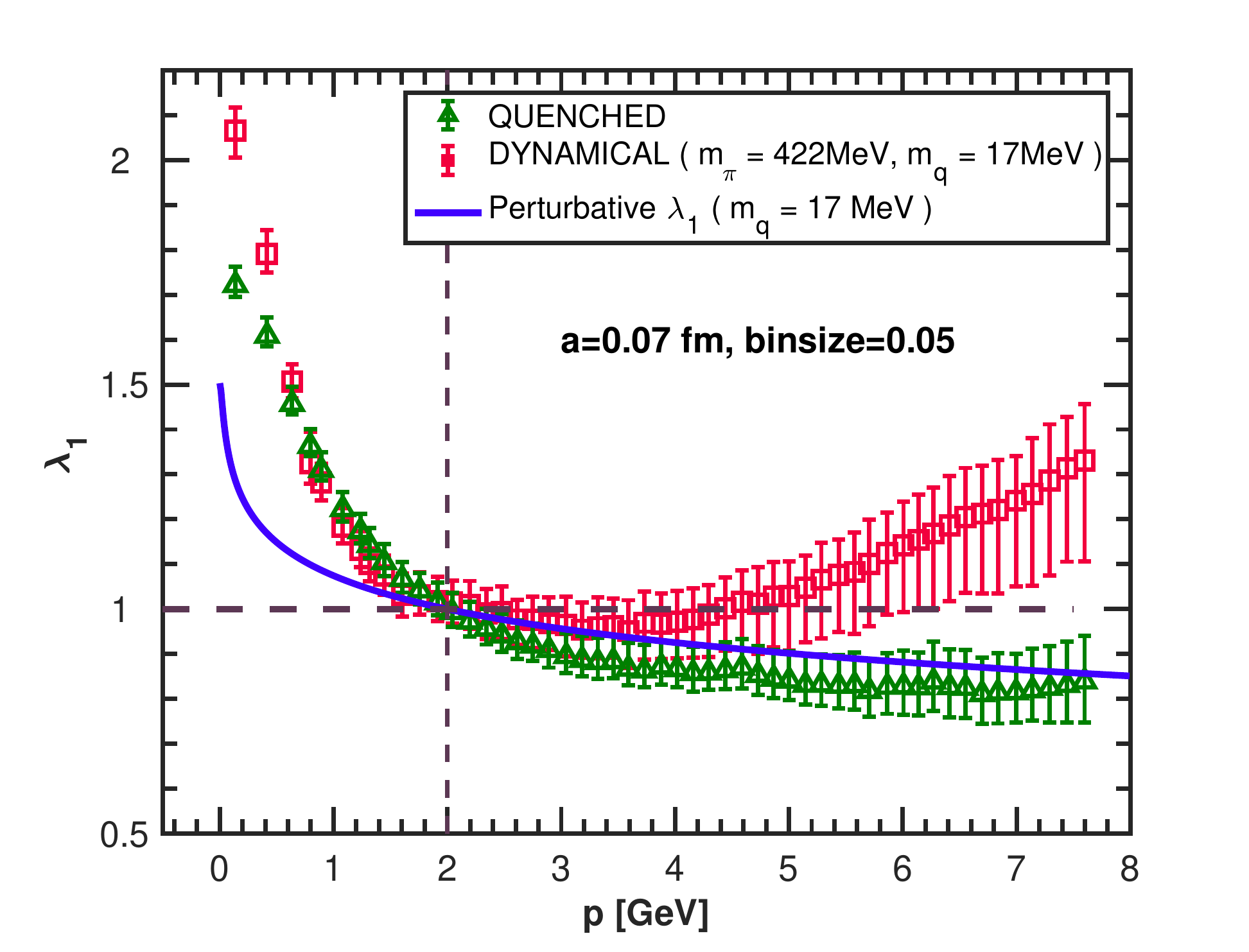}
        \caption{ }  
        \label{fig:lambda1_vs_ap_lin_b529_a007_m422_q_vs_unq}      
    \end{subfigure}
  \begin{subfigure}[t]{0.495\textwidth}
        \centering
       \includegraphics*[width=\textwidth]{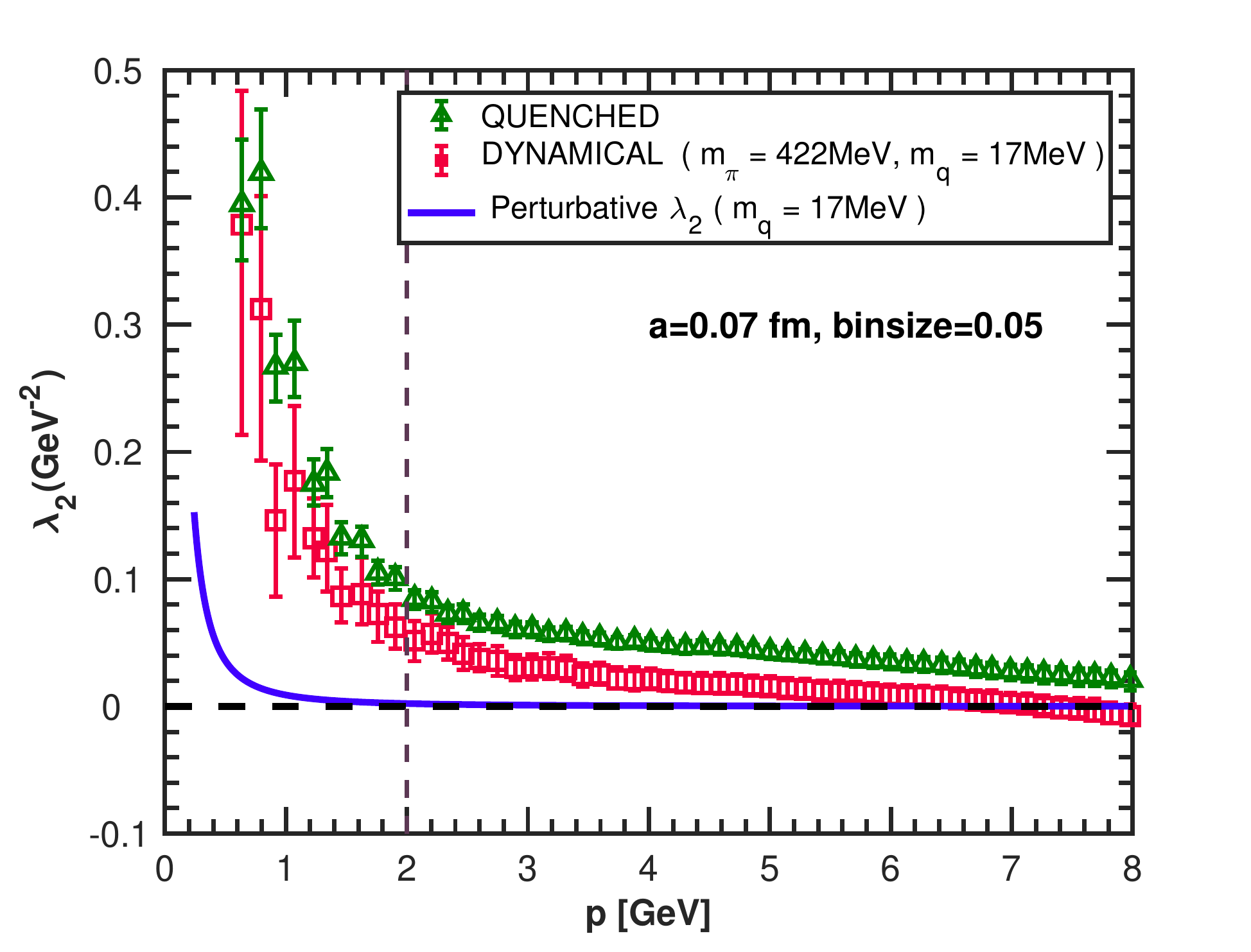}
        \caption{}  
        \label{fig:lambda2_vs_ap_lin_b529_a007_m422_q_vs_unq}      
    \end{subfigure}
 \begin{subfigure}[t]{0.495\textwidth}
        \centering
       \includegraphics*[width=\textwidth]{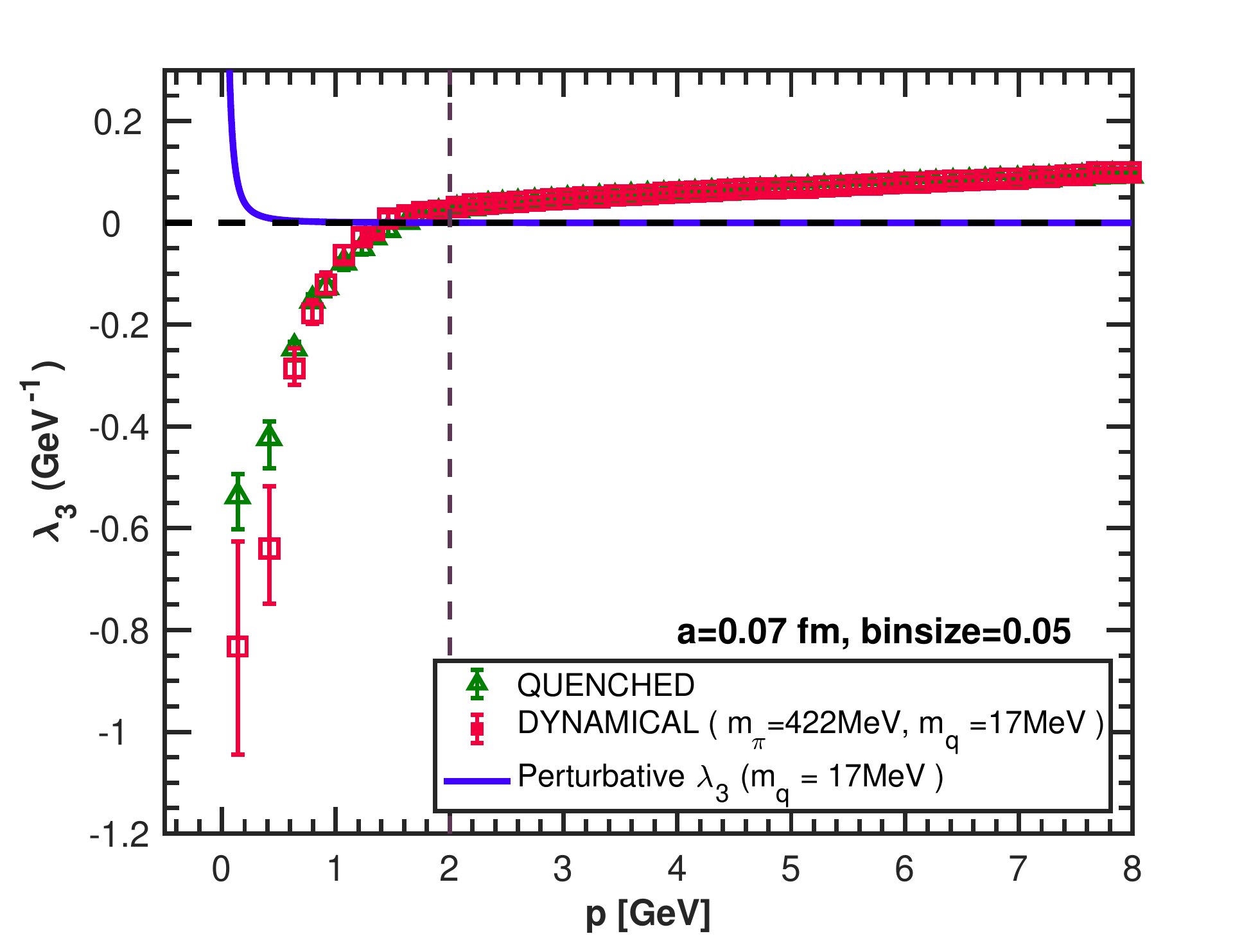}
        \caption{}  
        \label{ig:lambda3_vs_ap_lin_b529_a007_m422_q_vs_unq}     
         \end{subfigure}
\caption{Quenched and dynamical form factors versus momentum $p$.}
\label{fig:quench_vs_unq}
\end{figure}

We now turn to a comparison of results from our quenched $(N_f=0)$ and dynamical
($N_f=2$) ensembles at $a=0.07\,$fm.  Since the
valence quark mass in the quenched case is quite large, we use the H07
($m_\pi=422\,$MeV) ensemble in this comparison.  The results are shown
in Fig.~\ref{fig:quench_vs_unq}.
The one-loop perturbative expression \eqref{eq:l1-pert}, evaluated at
$\alpha=0.3\approx\alpha_{\overline{\text{MS}}}(2\,\mathrm{GeV})$~\cite{Bethke:2009jm,Bethke:2015etp}, $m_q= 17$\,MeV and
with $\mu=2\,$GeV,  is also plotted.  We have used the subtracted bare
quark mass rather than the renormalised mass in this calculation, but
this makes little difference as varying $m_q$ even by a factor 2 
has a negligible effect for the range of momenta we are considering
here.  We will now discuss each form factor in turn.
 
 \begin{description}
\item[$\lambda_1$]  Although  we find that the qualitative behaviour is the same in both quenched and unquenched
cases,   we see that with dynamical fermions the vertex is more
strongly infrared enhanced than in the quenched case, as seen in
Fig.~\ref{fig:lambda1_vs_ap_lin_b529_a007_m422_q_vs_unq}.  However,
a more detailed comparison, with different valence quark masses, will
be required to disentangle the effects of dynamical fermions and
valence quark masses.
Even at one-loop order the perturbative contribution to $\lambda_1$
shows IR enhancement, and at one loop level in Landau gauge this is
purely due to the non-abelian contribution. 
\item[$\lambda_2$]
This form factor has a  large uncertainty in the IR and as the momentum increases its value approaches zero and the uncertainties become small (see Fig.~\ref{fig:lambda2_vs_ap_lin_b529_a007_m422_q_vs_unq}).
Here we clearly see that the dynamical quarks only have a small effect on this form factor, leading
at most to a slight enhancement. 
We also note that our lattice results $\lambda_2$ are orders of magnitude larger than the
one-loop perturbative expression, which only exhibits a very small increase in the deep infrared. Like  in $\lambda_1$,  the entire contribution in Landau gauge to $\lambda_2$ at one-loop level is purely non-abelian. 
\item[$\lambda_3$]
As was the case for $\lambda_1$, we find that inclusion of dynamical fermions
leads to a significant enhancement of this form factor in the
infrared,  however  above 1\,GeV the difference between dynamical and quenched fermions diminishes.  At lower momentum this form
factor is negative, and as the momentum increases it approaches zero.
We find a zero crossing between 1 and 2\,GeV.  We
note that at this point,
the uncertainties associated with the tree-level correction are not
yet significant, so it appears unlikely that this zero crossing is entirely
due to lattice artefacts.  The one-loop expression has contributions from both abelian and 
non-abelian diagrams, and the combined effect is positive in the
IR region, in contrast to what is found from the lattice data.  Since
we expect the perturbative behaviour to be reproduced at large
momentum, this suggests that $\lambda_3$ should approach zero from
above in the ultraviolet, and hence have a zero crossing in the
intermediate-momentum region.
\end{description}
%

%
%

\begin{figure}[ht]
\centering
   \begin{subfigure}[t]{0.495\textwidth}
        \centering
       \includegraphics*[width=\textwidth]{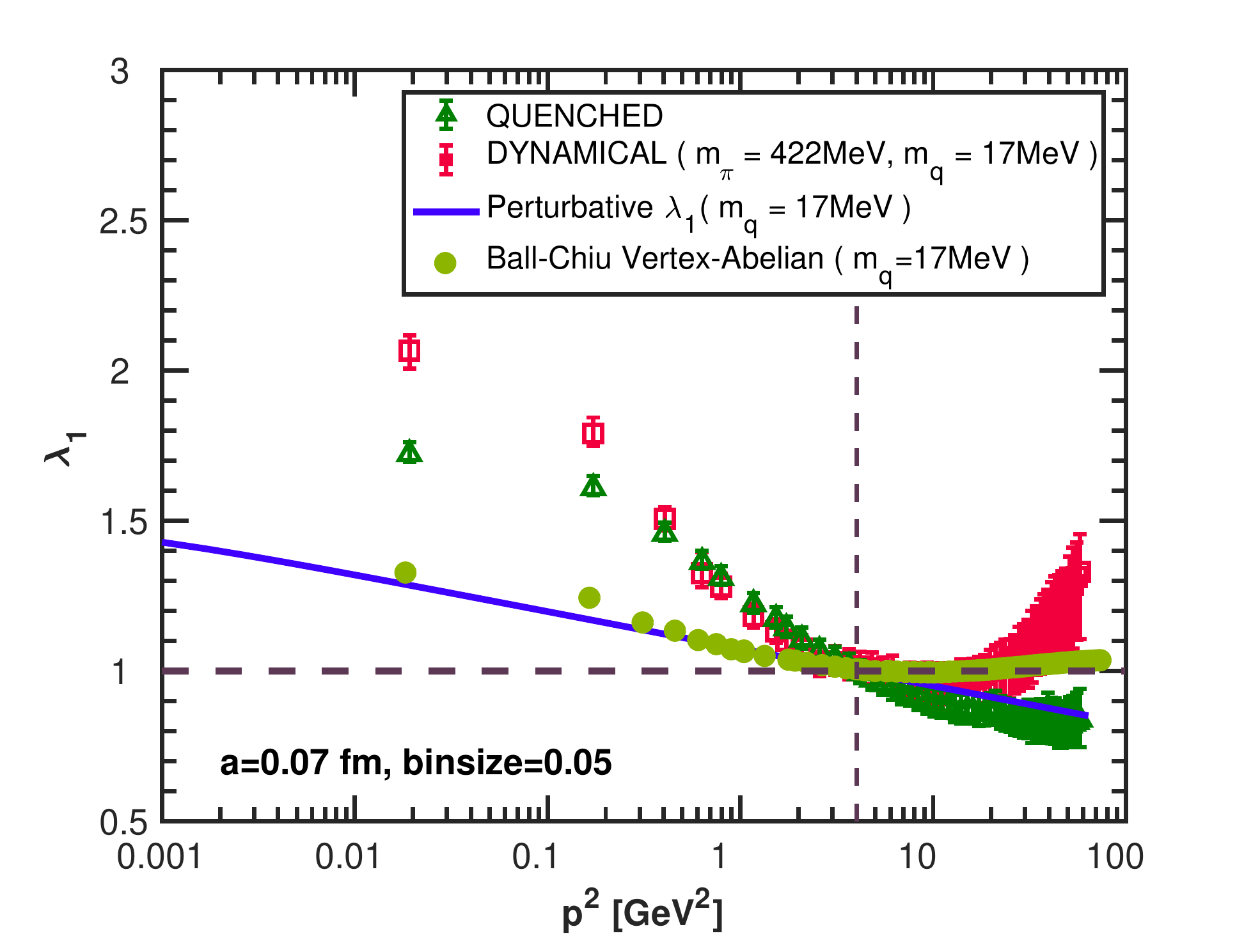}
        \caption{}  
        \label{fig:lambda1_vs_ap_lin_b529_a007_m422_pert_BC}      
    \end{subfigure}
  \begin{subfigure}[t]{0.495\textwidth}
        \centering
       \includegraphics*[width=\textwidth]{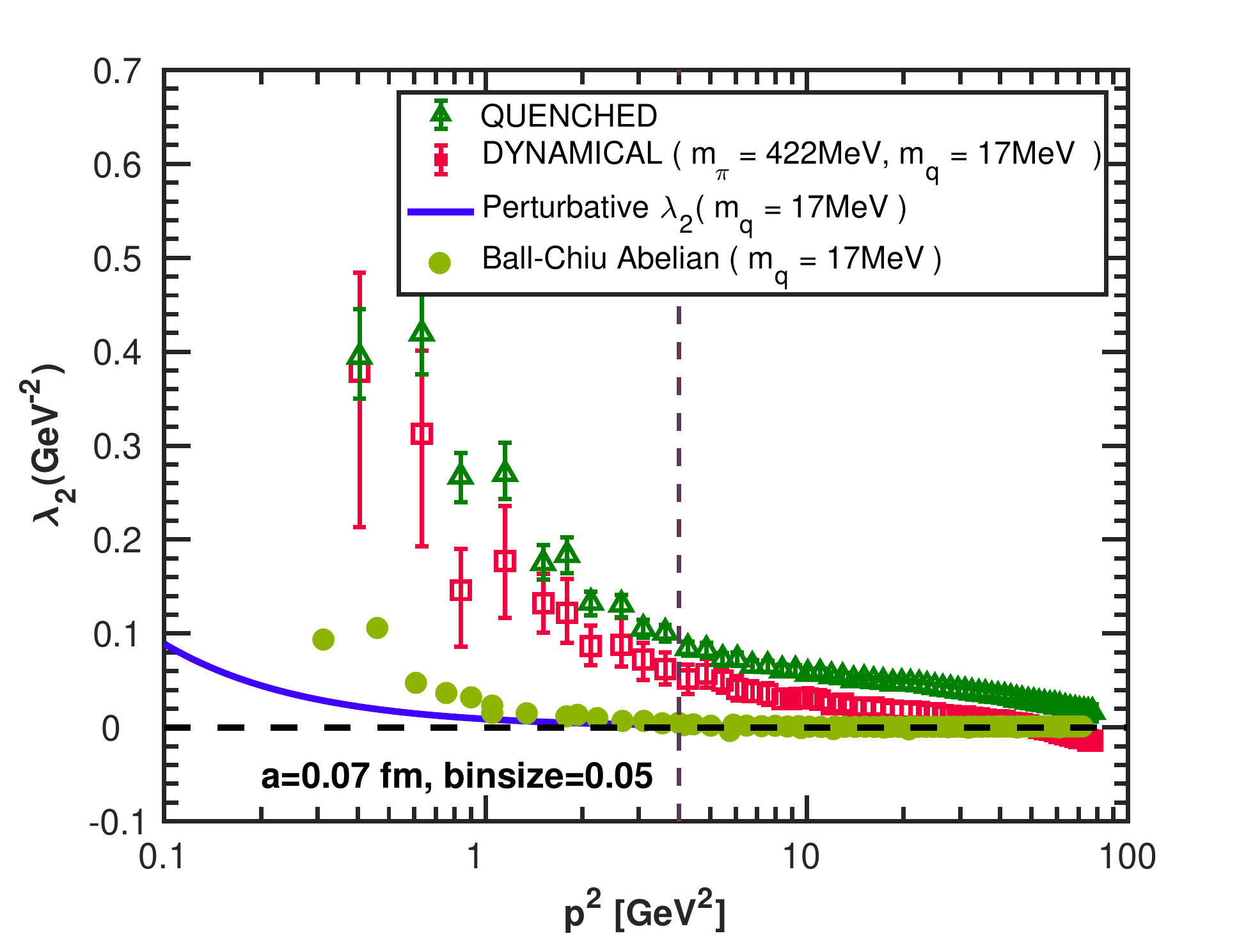}
        \caption{}  
        \label{fig:lambda2_vs_ap_lin_b529_a007_m422_pert_BC}      
    \end{subfigure}
 \begin{subfigure}[t]{0.495\textwidth}
        \centering
       \includegraphics*[width=\textwidth]{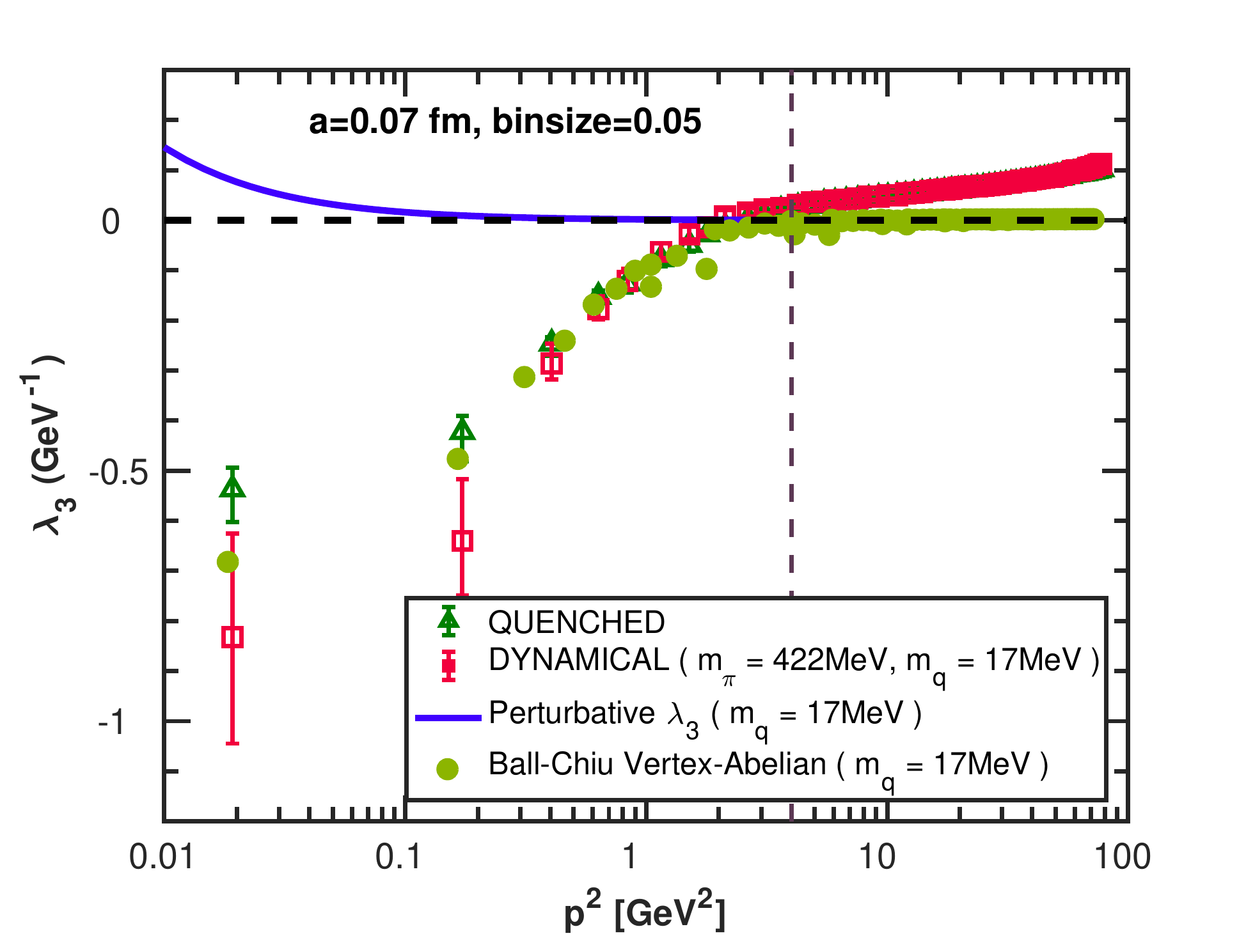}
        \caption{}  
        \label{lambda3_vs_ap_lin_b529_a007_m422_pert_BC}
         \end{subfigure}
\caption{The form factors $\lambda_1, \lambda_2$ and $\lambda_3$ for
  quenched and dynamical fermions, renormalised at 2\,GeV, the abelian
  Ball--Chiu vertex and the one-loop order form factors as functions
  of momentum squared.}
\label{fig:quench_vs_unq_pert_BC}
\end{figure}

Fig.~\ref{fig:quench_vs_unq_pert_BC} shows results for all three form
factors for the H07 (dynamical) and Q07 (quenched) ensembles, along
with the abelian Ball--Chiu vertex~ \cite{Ball:1980ay} and the
one-loop perturbative contributions to the quark-gluon vertex.   The
Ball--Chiu vertex is calculated from the quark propagator
\begin{equation}
  S(p)= \frac{1}{i\pslash A(p^2)+B(p^2)} = \frac{Z(p^2}{i\pslash+M(p^2)}\,,
\end{equation}
using  the expressions
\begin{align}
  \lambda_1^{BC} &= A(p^2)\,,\\
  \lambda_2^{BC} &= -\frac{1}{2}\,d A(p^2)/dp^2\,,\\
  \lambda_3^{BC} &= dB(p^2)/dp^2\,.
\end{align}
The quark wave function $A(p^2)=1/Z(p^2)$ and the mass function
$M(p^2)=B(p^2)/A(p^2)$ are calculated using the $N_f=2$ lattice
results from Ref.~\cquark.

We observe a hierarchy in the infrared enhancement of $\lambda_1$
where the largest enhancement is for the dynamical lattice ensembles,
followed by the quenched lattice data, the Ball--Chiu vertex and the
one-loop perturbative expression in that order.  The same hierarchy is
found for $\lambda_2$, while for $\lambda_3$ the quenched and
dynamical lattice results and the Ball--Chiu are roughly equal.  More
work will be required to disentangle the effects of explicit and
spontaneous chiral symmetry breaking and quark dynamics in this form
factor. We will study the quark mass dependence of $\lambda_3$ further in
section~\ref{sec:massdep}.

The IR hierarchy casts light on the reliability of the various approximations  when they are  used to calculate non-perturbative quantities.  
These results combined with the Slavnov--Taylor identity \eqref{Eq:STI} show that the abelian Ball--Chiu vertex is not able to saturate the identity and, 
therefore, the quark--ghost kernel $H$ in  \eqqref{Eq:STI} has sizeable
contributions in the infrared region.

\subsection{Mass dependence}
\label{sec:massdep}

\begin{figure}[thb]
\centering
   \begin{subfigure}[t]{0.495\textwidth}
        \centering
       \includegraphics*[width=\textwidth]{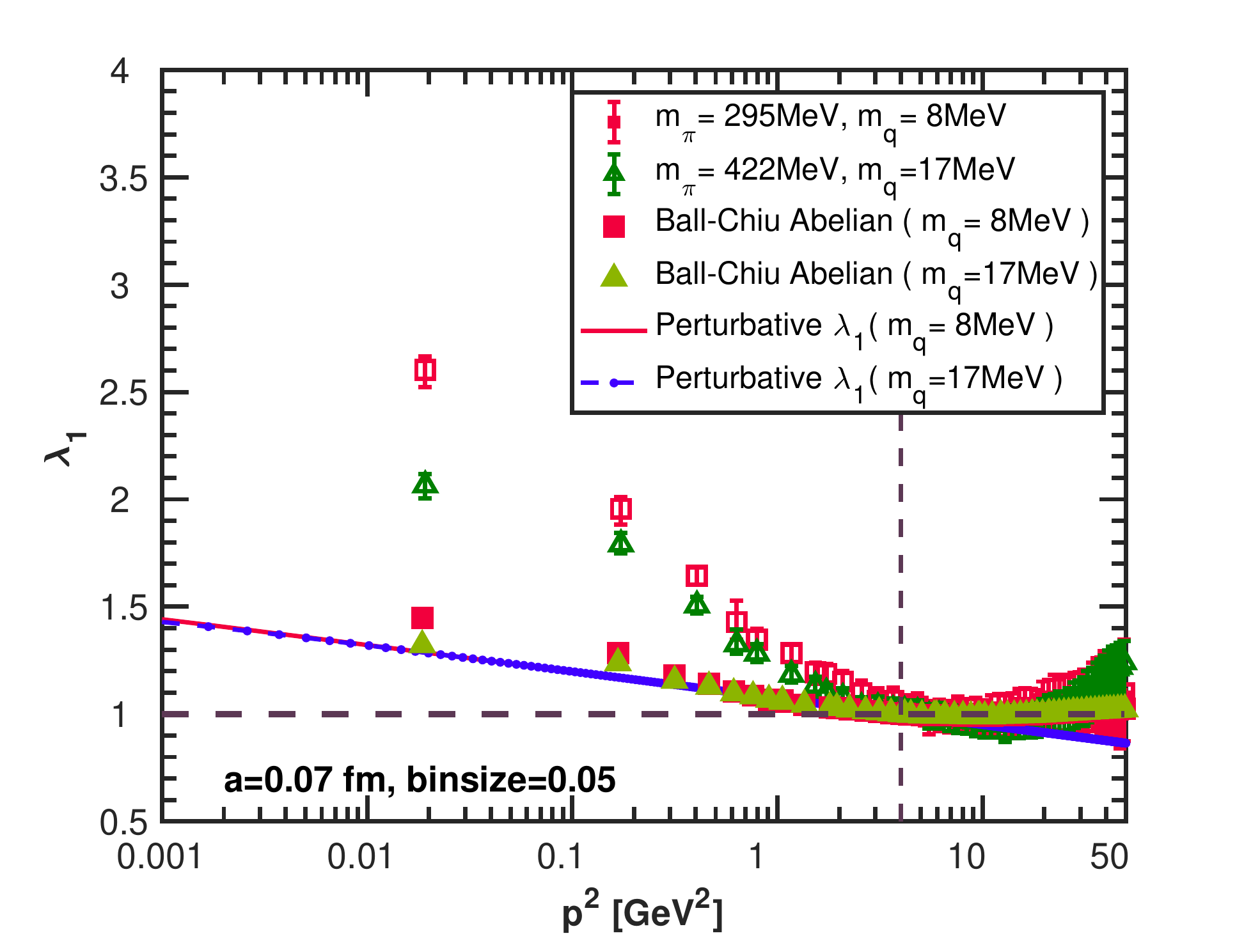}
        \caption{ $\lambda_1$ }  
        \label{fig:lambda1_vs_ap_lin_b529_a007_m295_mass}      
    \end{subfigure}
  \begin{subfigure}[t]{0.495\textwidth}
        \centering
       \includegraphics*[width=\textwidth]{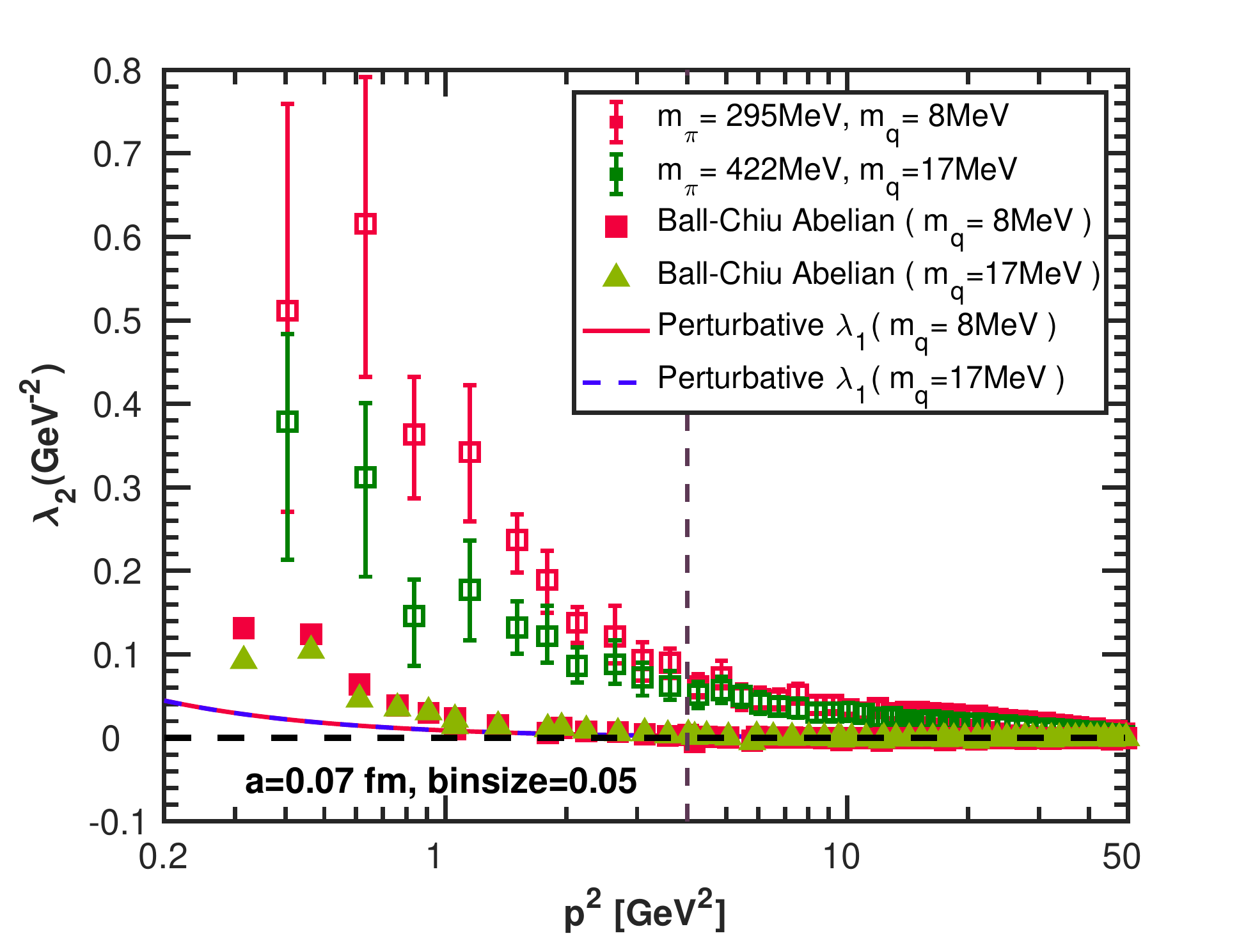}
        \caption{$\lambda_2$ }  
        \label{fig:lambda2_vs_ap_lin_b529_a007_m295_mass}      
    \end{subfigure}
 \begin{subfigure}[t]{0.495\textwidth}
        \centering
       \includegraphics*[width=\textwidth]{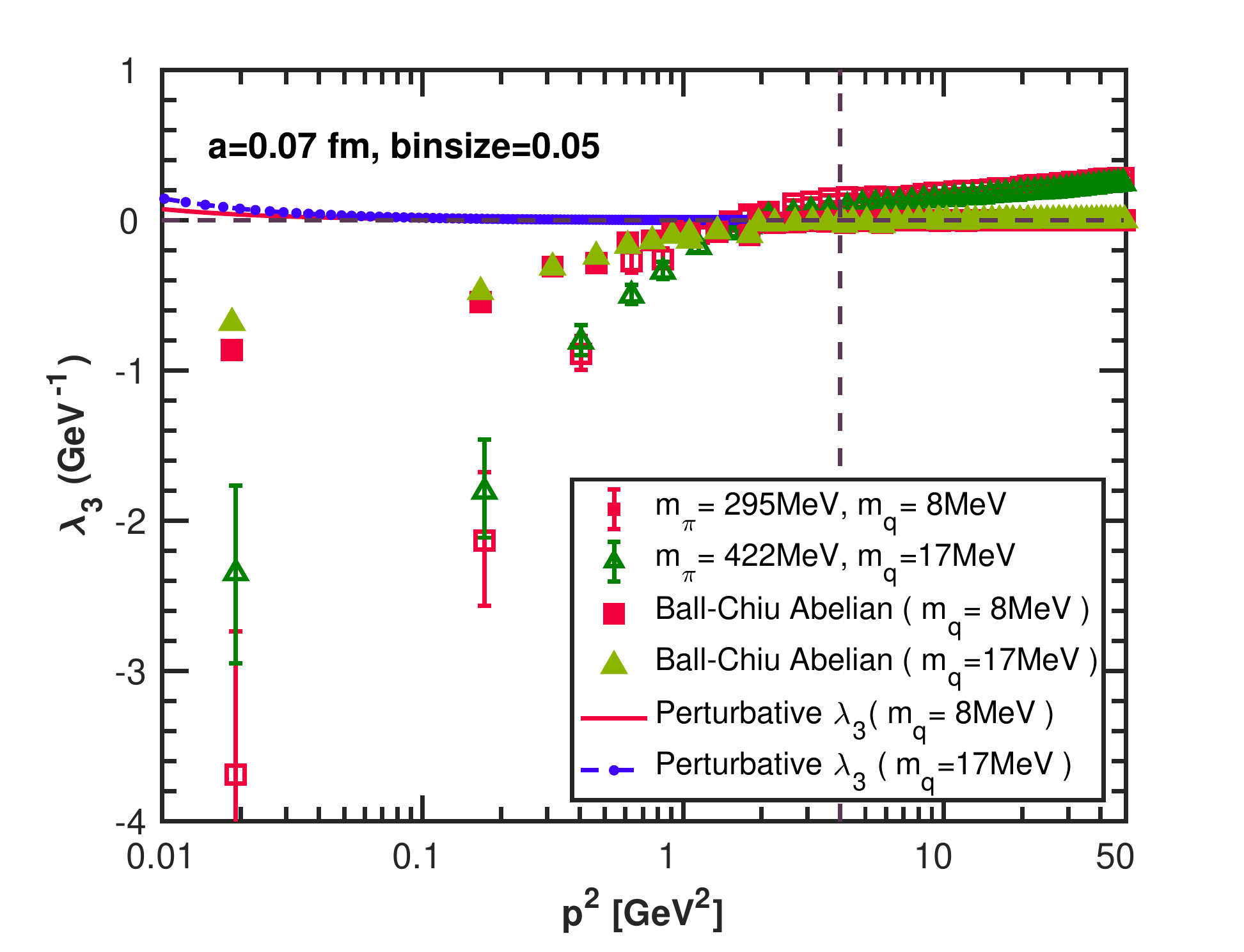}
        \caption{$\lambda_3$ }  
        \label{ig:lambda3_vs_ap_lin_b529_a007_m295_mass}      
         \end{subfigure}
\caption{The form factors $\lambda_1, \lambda_2$ and $\lambda_3$ for the
  L07 ($m_{\pi} = 295\,$MeV) and H07 ($m_{\pi} = 422\,$MeV) ensembles,
  together with the abelian Ball--Chiu vertex and the one-loop order
  form factors, as functions of momentum squared.}
\label{fig:mass_dependence}
\end{figure}

In Fig.~\ref{fig:mass_dependence} we show results from the L07
($m_{\pi} = 295\,$MeV) and H07 ($m_{\pi} = 422\,$MeV) ensembles, which
differ only by the quark mass.  We also show the corresponding
Ball--Chiu vertices and one-loop order perturbative results.

For all form factors we see that the lighter quark mass gives rise to
a larger infrared enhancement.  This is particularly interesting in
the case of $\lambda_3$, where the one-loop expression \eqref{eq:l3-pert}
increases with increasing quark mass in this momentum region.
However, it should be noted that the one-loop expression also
shows a stronger enhancement with decreasing quark mass (albeit with
the opposite sign) in the deep
infrared as shown in Fig.~\ref{fig:pertlam3}.  We also note
that the abelian Ball--Chiu vertex exhibits the same effect, and that
the same quark mass dependence was observed for $\lambda_3$ in the
quenched case \cite{Skullerud:2003qu}.  We therefore consider this
mass dependence to be robust.

In the case of $\lambda_1$ and $\lambda_2$, although the one-loop
perturbative contribution for the lighter and heavier masses look the
same,  on  closer inspection one notes that in the deep IR they differ
from each other following the same trend, namely that they are more
enhanced the smaller the quark mass (see Fig.~\ref{fig:one-loop-lambda}).

\subsection{Volume dependence}
\label{sec:volumedep}
\begin{figure}[thb]
  \centering
    \begin{subfigure}[t]{0.495\textwidth}
        \centering
        \includegraphics*[width=\textwidth]{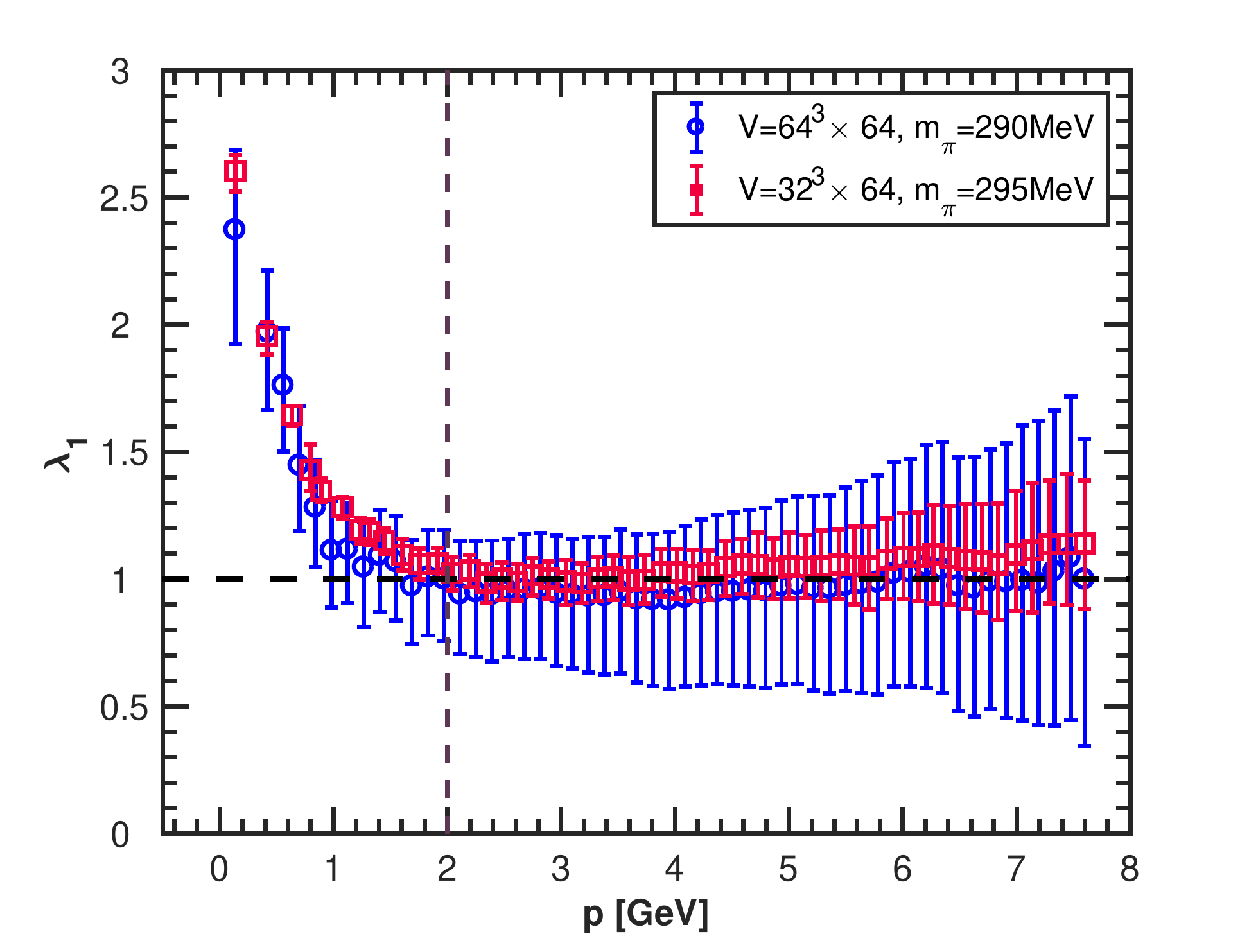}
         \caption{$\lambda_1$ }
        \label{fig:lambda1_volume_vs_ap_lin_b529_a007_m290_ren}
   \end{subfigure}
       \begin{subfigure}[t]{0.495\textwidth}
        \centering
        \includegraphics*[width=\textwidth]{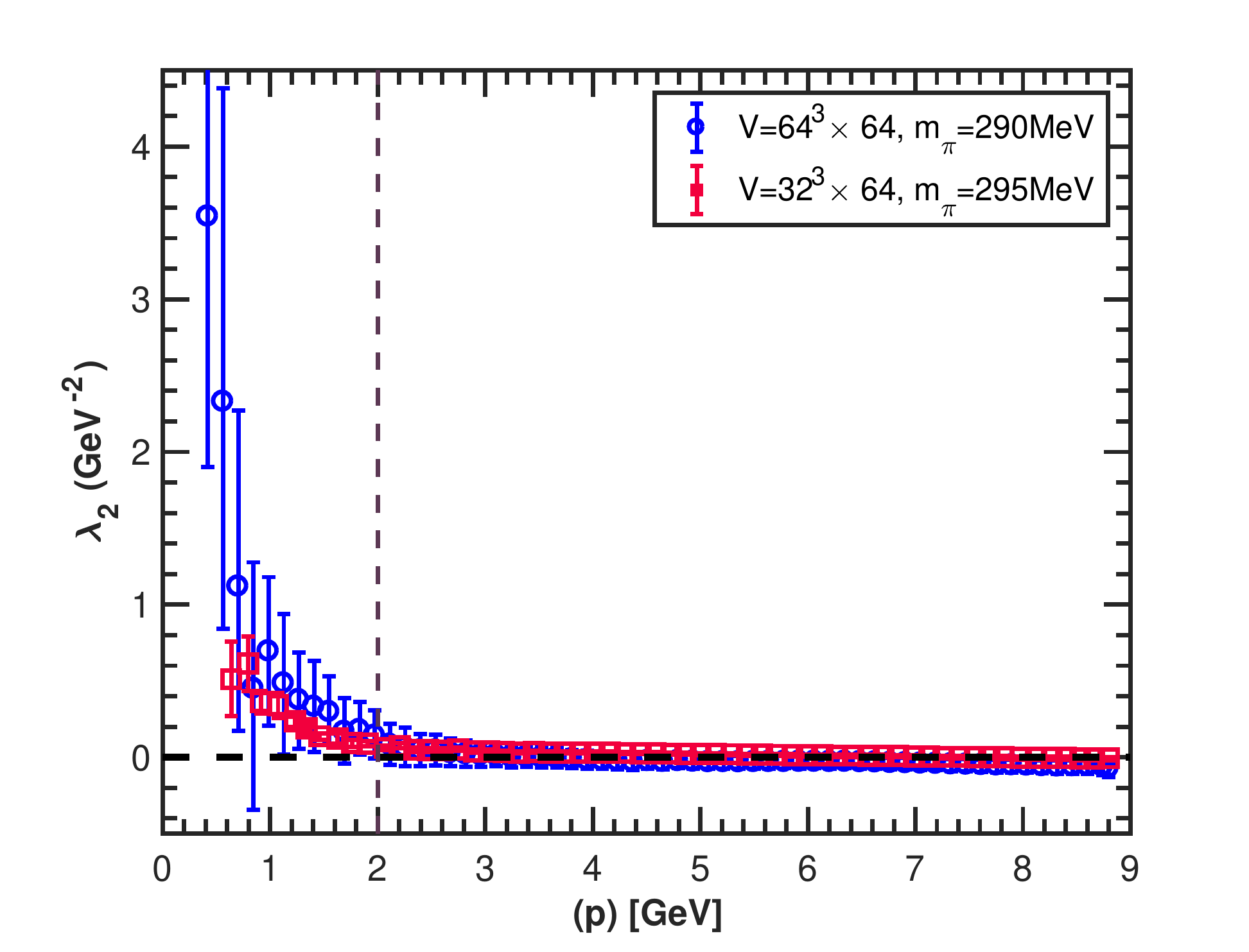}
         \caption{$\lambda_2$ }
        \label{fig:lambda2_GeV_volume_vs_ap_lin_b529_a007_m295_ren}
    \end{subfigure}
   \begin{subfigure}[t]{0.495\textwidth}
        \centering
        \includegraphics*[width=\textwidth]{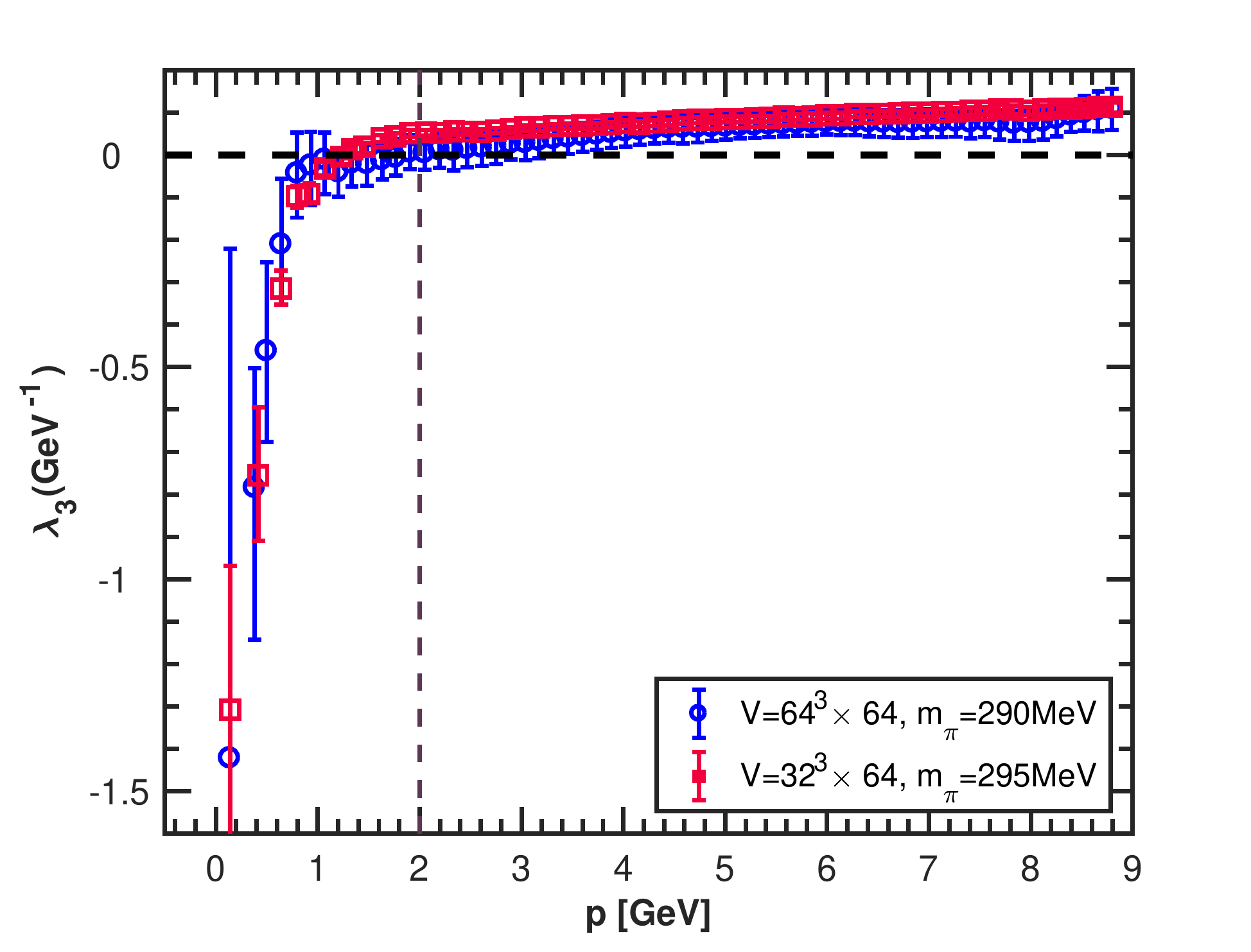}
         \caption{$\lambda_3$ }
        \label{fig:lambda3_volume_vs_ap_lin_b529_a007_m295_ren}
    \end{subfigure}  
\caption{Volume dependence of the form factors $\lambda_1, \lambda_2$
  and $\lambda_3$, from the L07 and L07-64 ensembles, with
  $a=0.07\,$fm and $m_\pi\approx290\,$MeV.}
 \label{fig:volume_dependence}
\end{figure}

In Fig.~\ref{fig:volume_dependence} we compare the results for two different volumes with the same quark mass and lattice spacing, $a=0.07$fm, $m_\pi = 295\,$MeV on the $64^4$ and $32^3\times 64$ lattices respectively. 

In all cases, the uncertainties for the larger volume are so large
that we are not able to draw any definitive conclusions; however, we
do not see any evidence of a significant finite volume effect for
$\lambda_1$ or $\lambda_3$.  For $\lambda_2$ our results suggest
that this form factor is more strongly enhanced in the infrared as the
volume increases; however, the results for the two volumes remain
consistent with each other within the uncertainties.

\subsection{Lattice spacing}
\label{sec:latspacing}
\begin{figure}[thb]
  \centering
    \begin{subfigure}[t]{0.495\textwidth}
        \centering
        \includegraphics*[width=\textwidth]{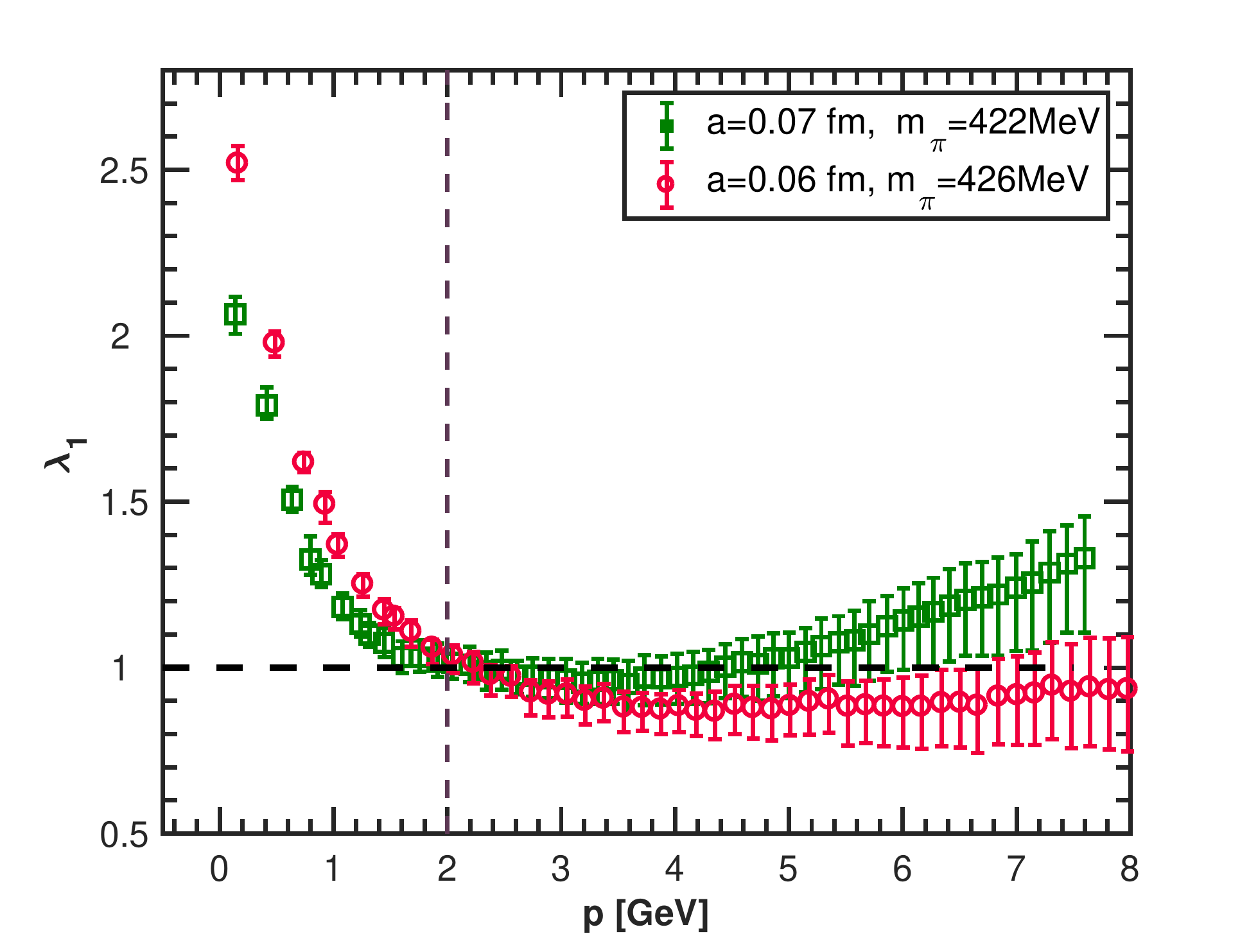}
         \caption{}
        \label{fig:llambda1_latticeSpacing_vs_ap_log_b529_a007_m422_ren}
    \end{subfigure}
     \begin{subfigure}[t]{0.495\textwidth}
        \centering
        \includegraphics*[width=\textwidth]{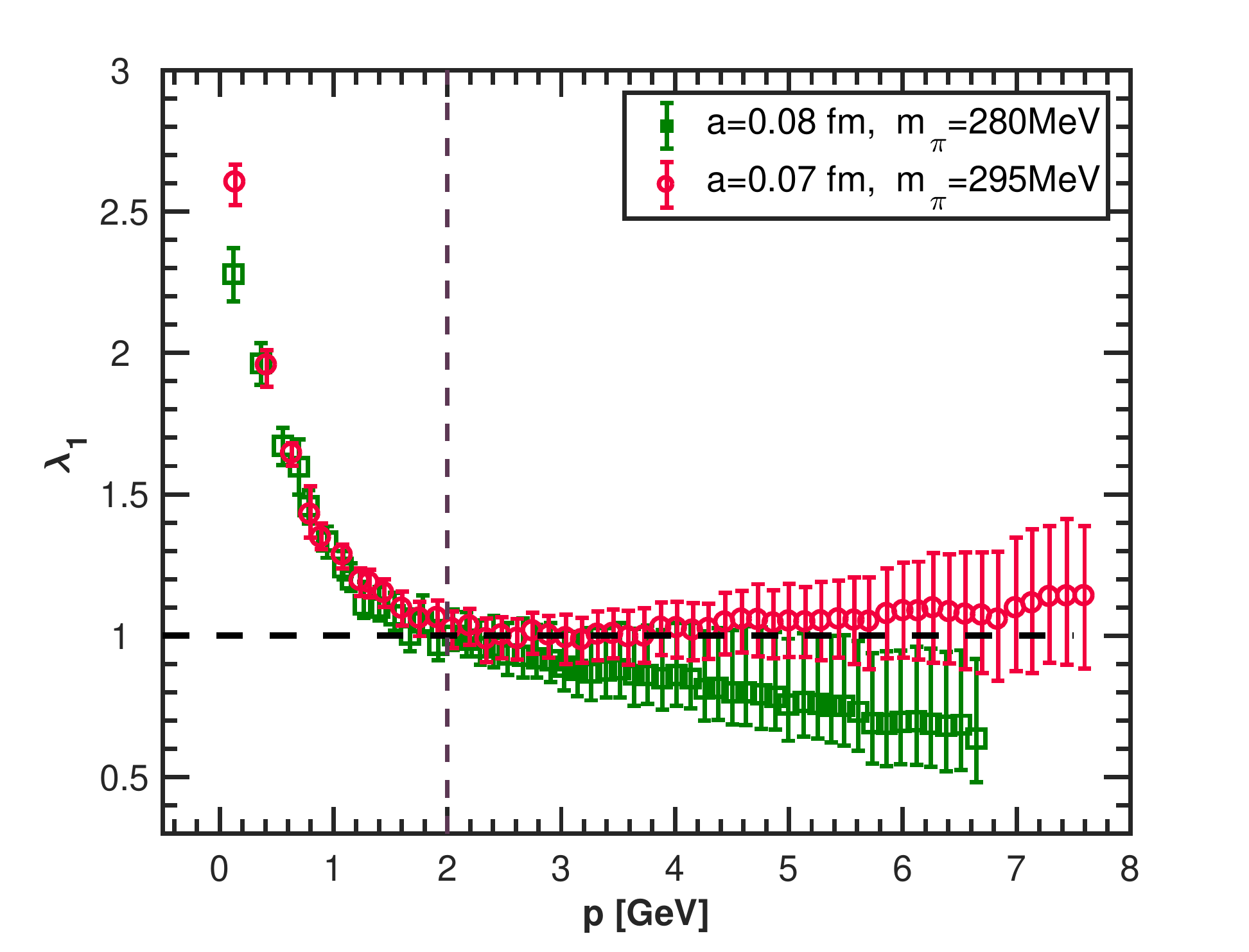}
         \caption{ }
        \label{fig:llambda1_latticeSpacing_vs_ap_log_b529_a007_m295_ren}
    \end{subfigure}
    \begin{subfigure}[t]{0.495\textwidth}
        \centering
        \includegraphics*[width=\textwidth]{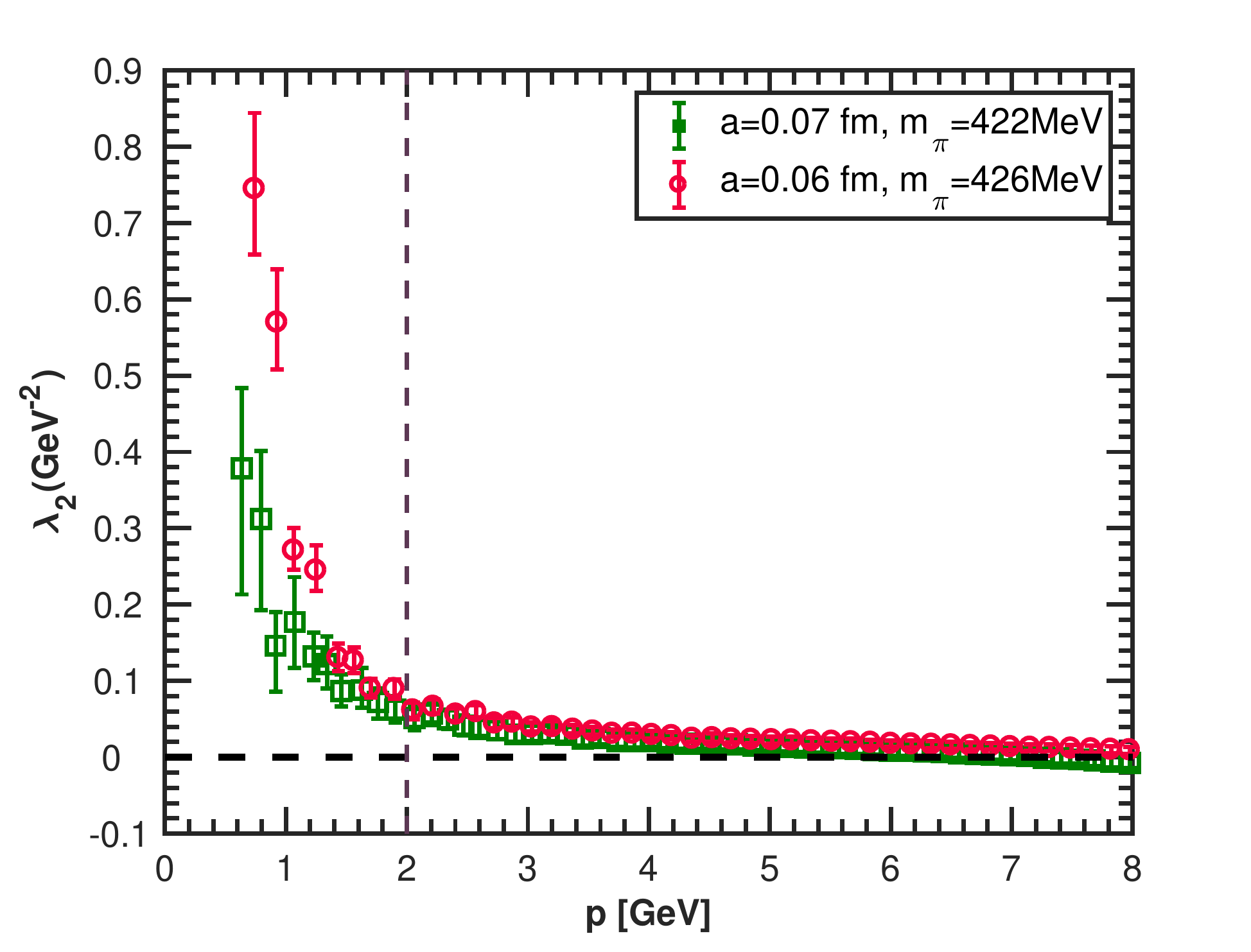}
         \caption{}
        \label{fig:llambda2_latticeSpacing_vs_ap_log_b529_a007_m422_ren}
    \end{subfigure}
     \begin{subfigure}[t]{0.495\textwidth}
        \centering
        \includegraphics*[width=\textwidth]{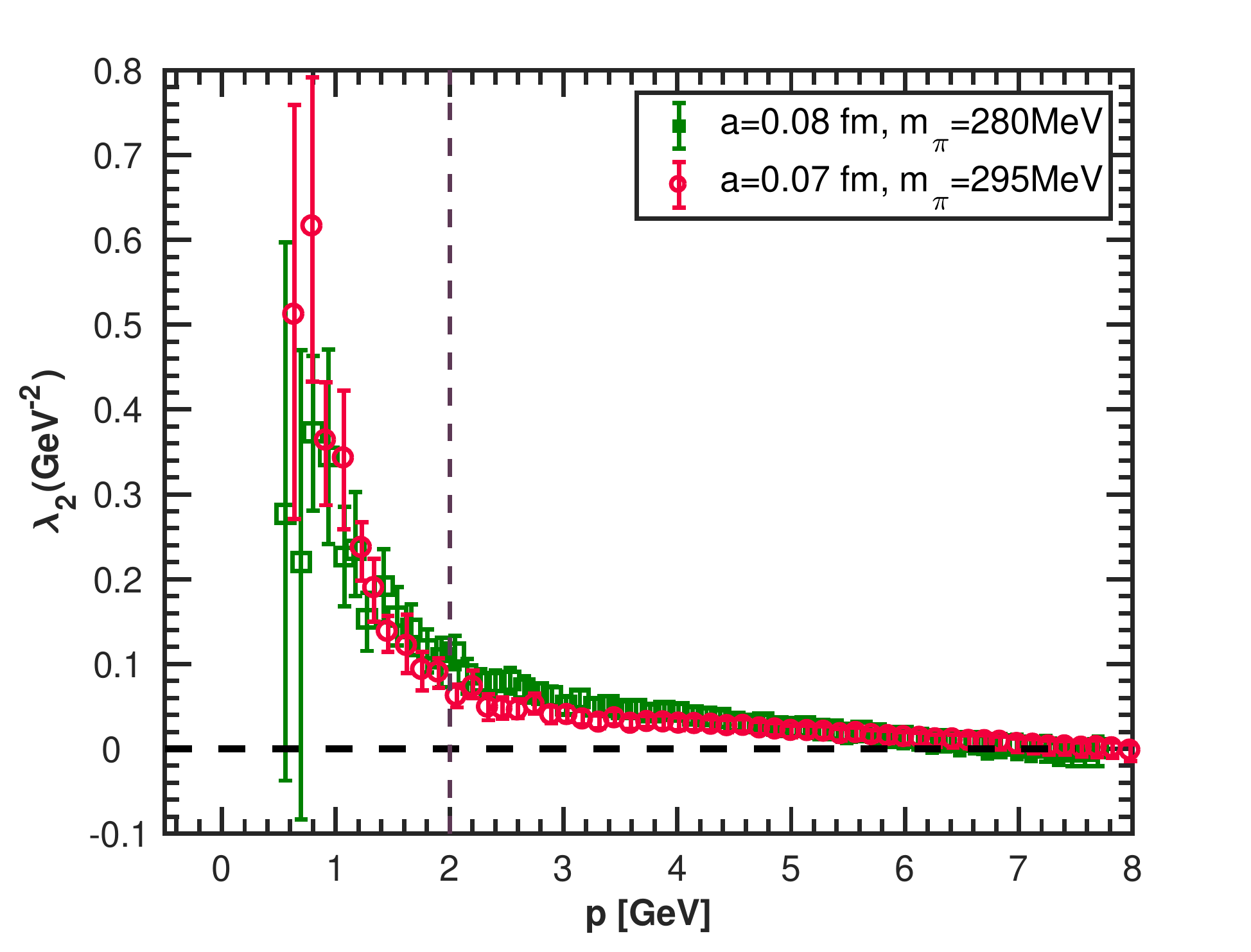}
         \caption{}
        \label{fig:llambda2_latticeSpacing_vs_ap_log_b529_a007_m295_ren}
    \end{subfigure}
    \begin{subfigure}[t]{0.495\textwidth}
        \centering
        \includegraphics*[width=\textwidth]{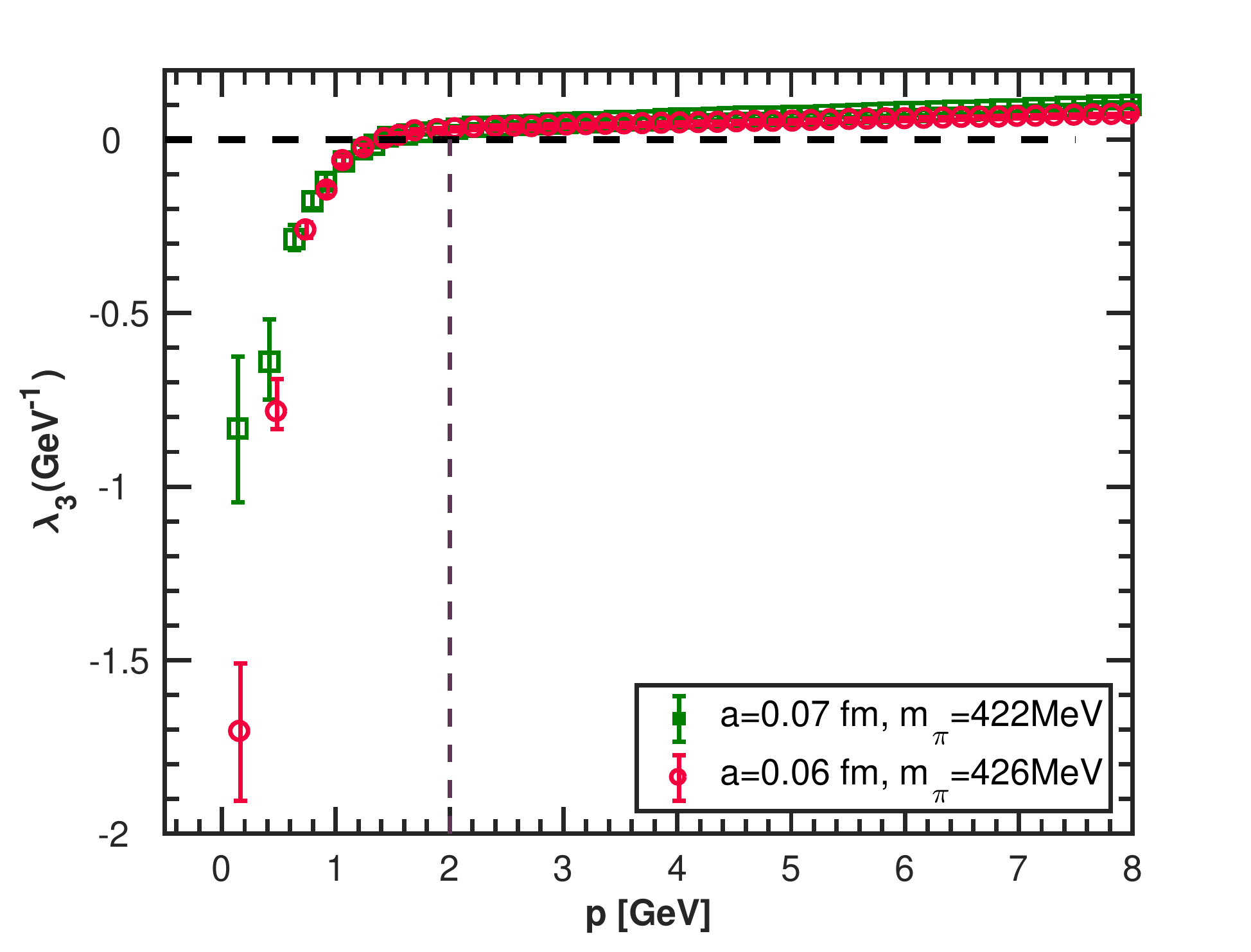}
         \caption{}
    \end{subfigure}
     \begin{subfigure}[t]{0.495\textwidth}
        \centering
        \includegraphics*[width=\textwidth]{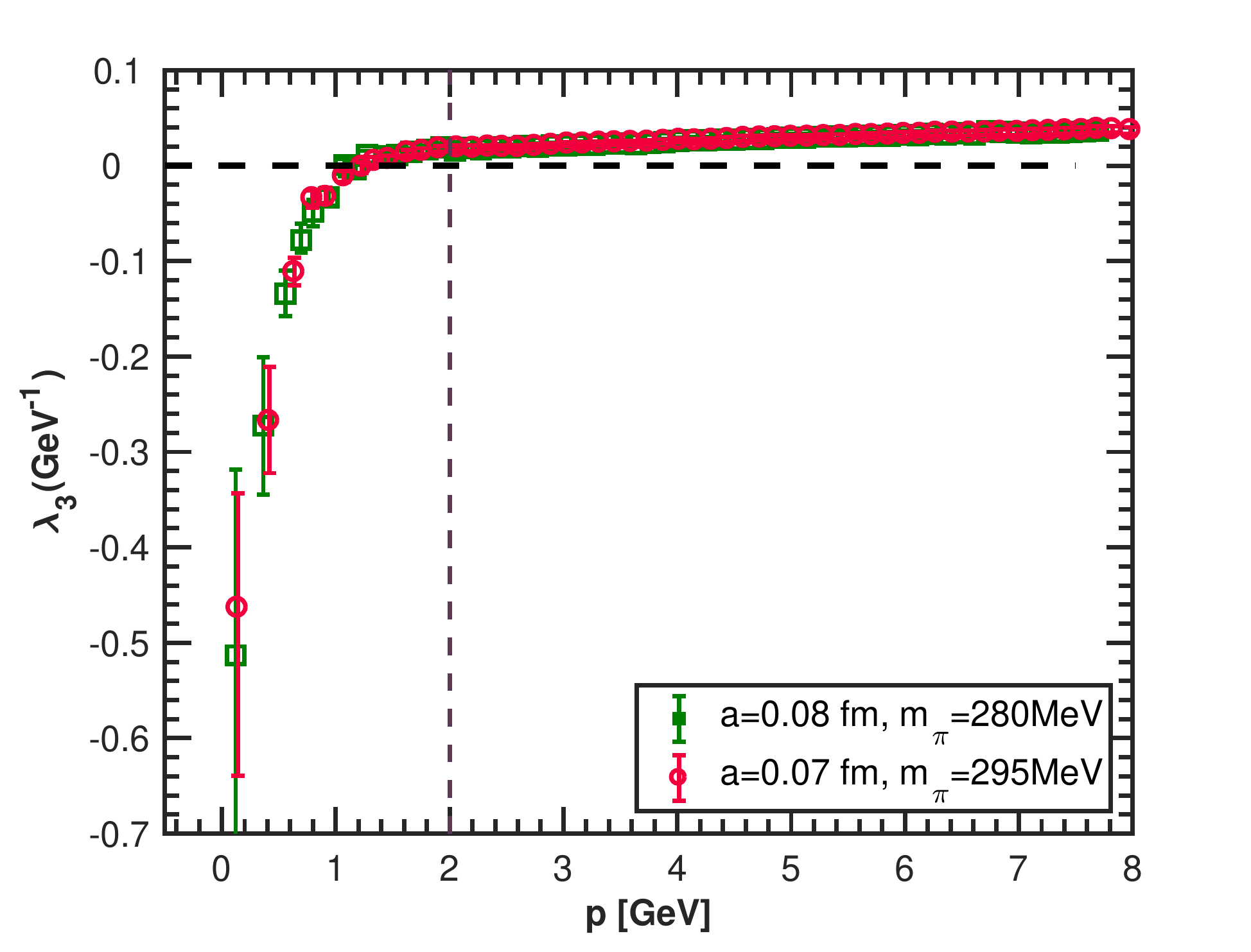}
         \caption{}
        \label{fig:llambda3_latticeSpacing_vs_ap_log_b529_a007_m295_ren}
    \end{subfigure}
\caption{Lattice  spacing dependence of the form factors, for the H06
  and H07 (heavier quark) ensembles (left) and the L07 and L08
  (lighter quark) ensembles (right).}
 \label{fig:lattice_spacing}
\end{figure}

In Fig.~\ref{fig:lattice_spacing} we compare results for different
lattice spacings, keeping the quark mass constant.
For comparison,  in Fig.~\ref{fig:lattice_spacing} the plots on the
left show results for the H06 and H07 ensembles with
$m_\pi\approx420\,$MeV, while the plots on the right are for the L07
and L08 ensembles, with $m_\pi\approx290\,$MeV.

For both quark masses we observe that all form factors have a larger
infrared enhancement for the smaller lattice spacing. The effect
is slightly larger for the heavier masses than for the lighter mass.

At large momentum we see that all form factors move closer to their
continuum tree-level value (1 for $\lambda_1$ and 0 for $\lambda_2$
and $\lambda_3$ as the lattice spacing is reduced, suggesting that the
expected perturbative behaviour will be reproduced in the continuum
limit.
\subsection{All form factors}
\label{sec:allformfactors}
\begin{figure}[thb]
  \centering
  \begin{subfigure}[t]{0.495\textwidth}
        \centering
       \includegraphics*[width=\textwidth]{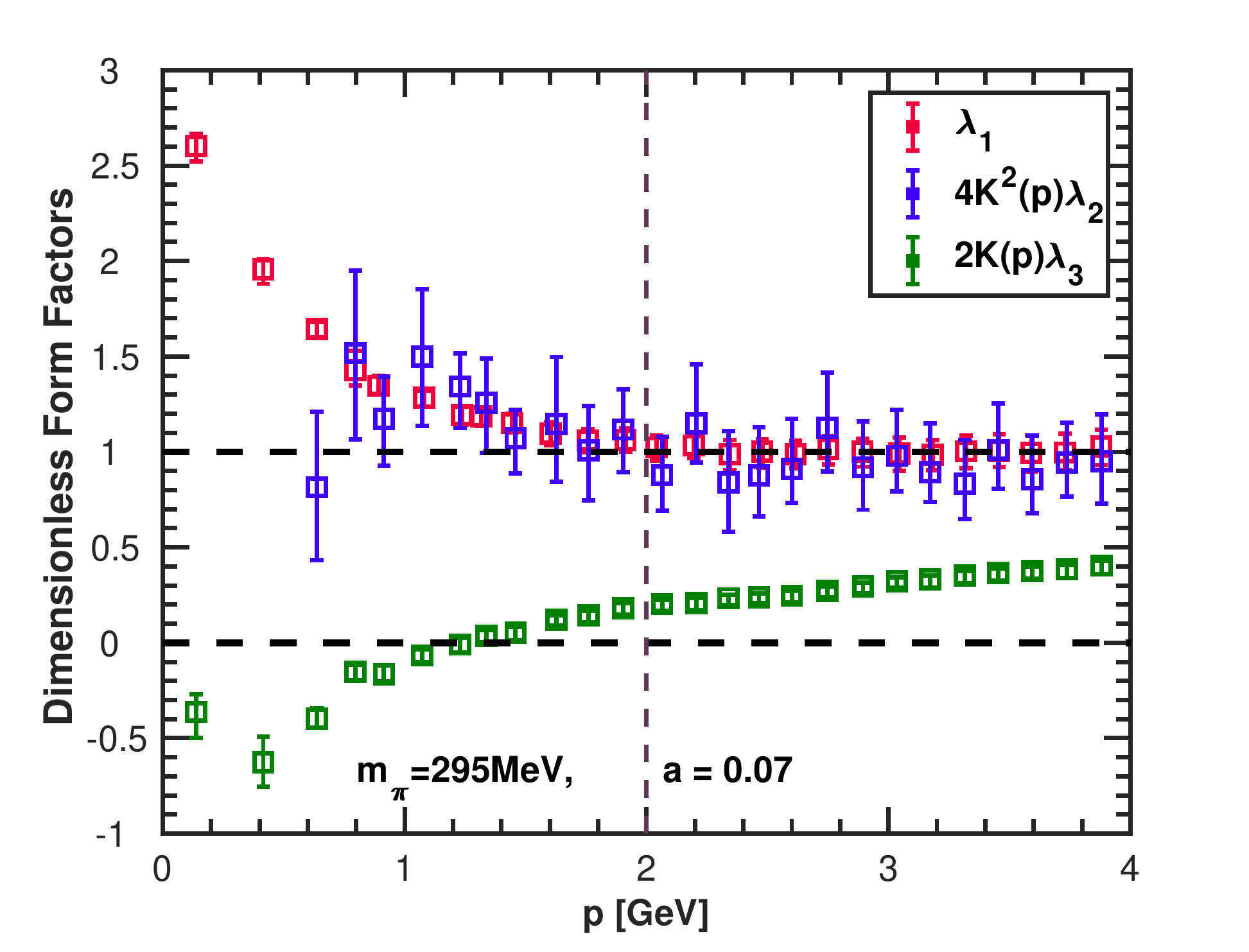}
\caption{}
        \label{fig:allLambdas_GeV_vs_ap_lin_m295_ren}      
    \end{subfigure}
    \begin{subfigure}[t]{0.495\textwidth}
        \centering
       \includegraphics*[width=\textwidth]{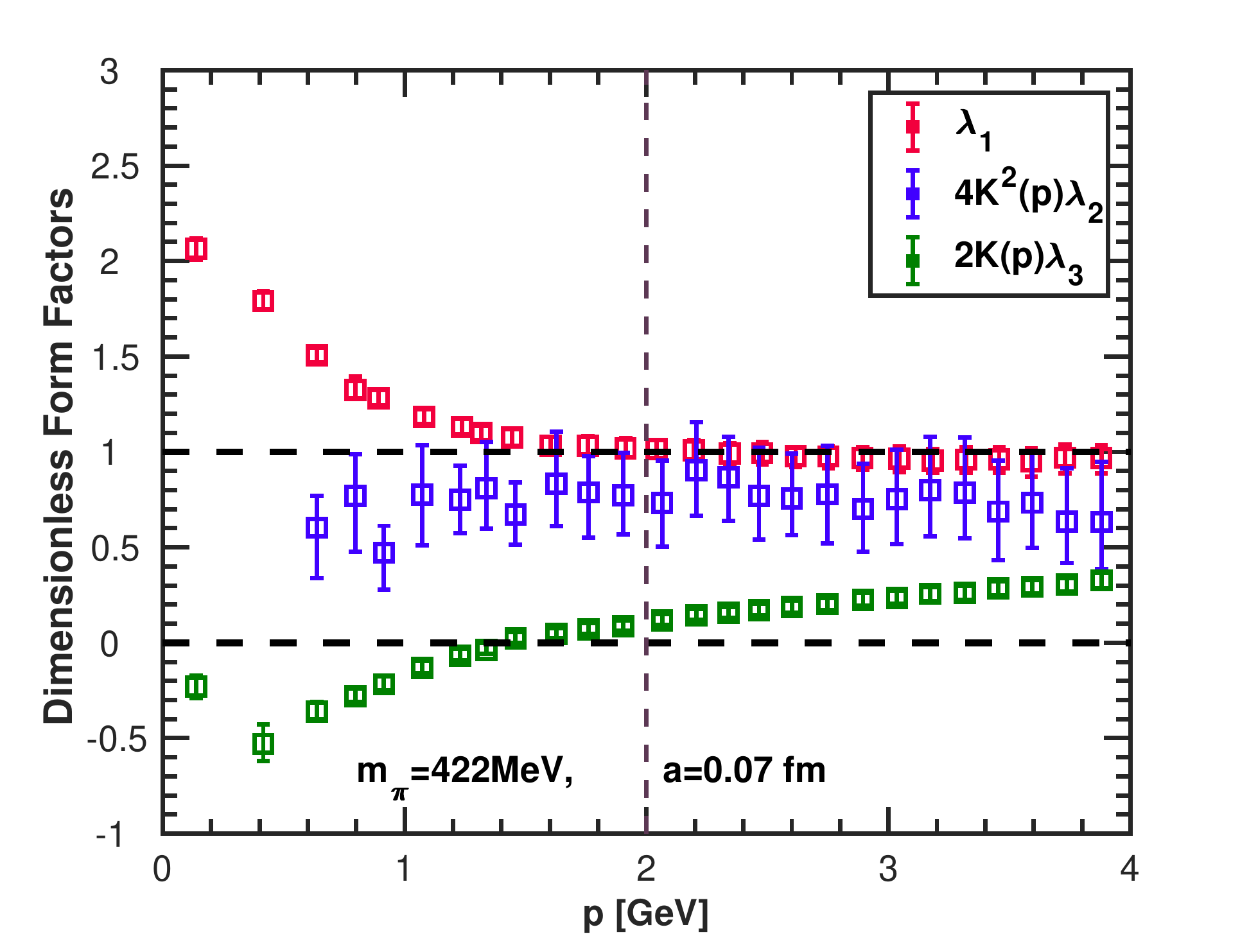}
        \caption{}
        \label{fig:allLambdas_GeV_vs_ap_lin_m422_ren}      
    \end{subfigure}
    \begin{subfigure}[t]{0.495\textwidth}
        \centering
       \includegraphics*[width=\textwidth]{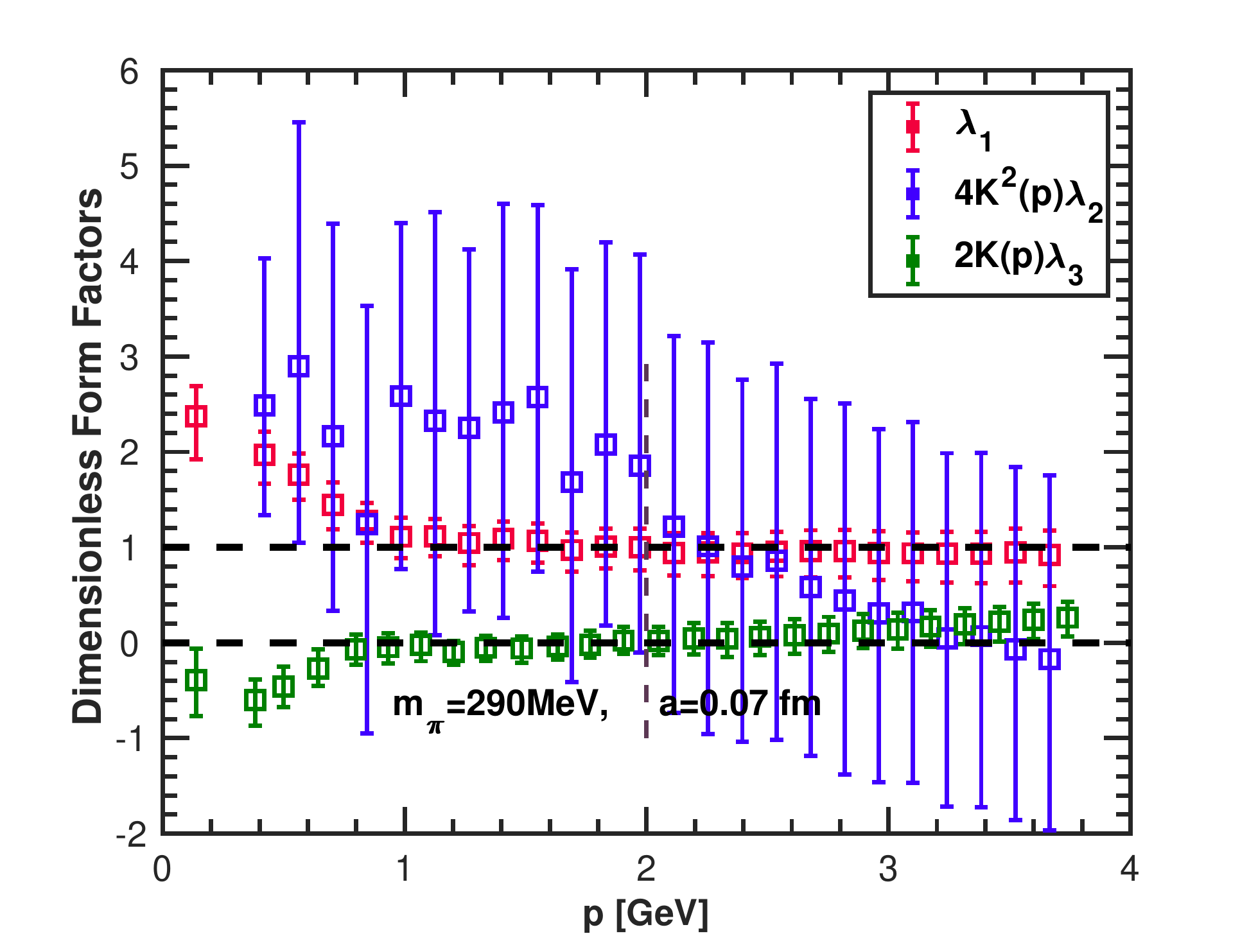}
\caption{}
        \label{fig:allLambdas_GeV_vs_ap_lin_m290_ren}      
    \end{subfigure} 
  \begin{subfigure}[t]{0.495\textwidth}
        \centering
       \includegraphics*[width=\textwidth]{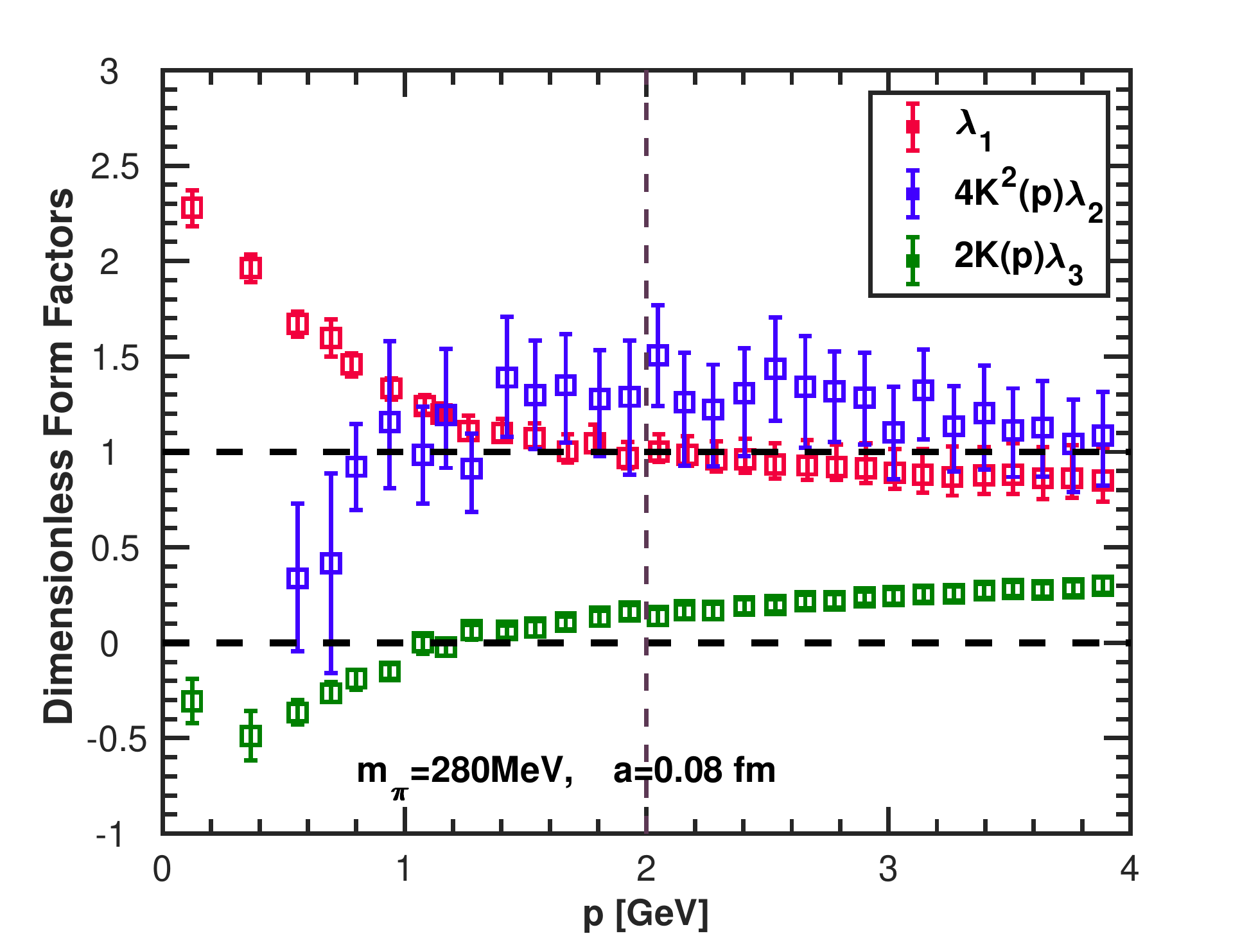}
\caption{}
        \label{fig:allLambdas_GeV_vs_ap_lin_m280_ren}    
    \end{subfigure} 
  \begin{subfigure}[t]{0.495\textwidth}
        \centering
       \includegraphics*[width=\textwidth]{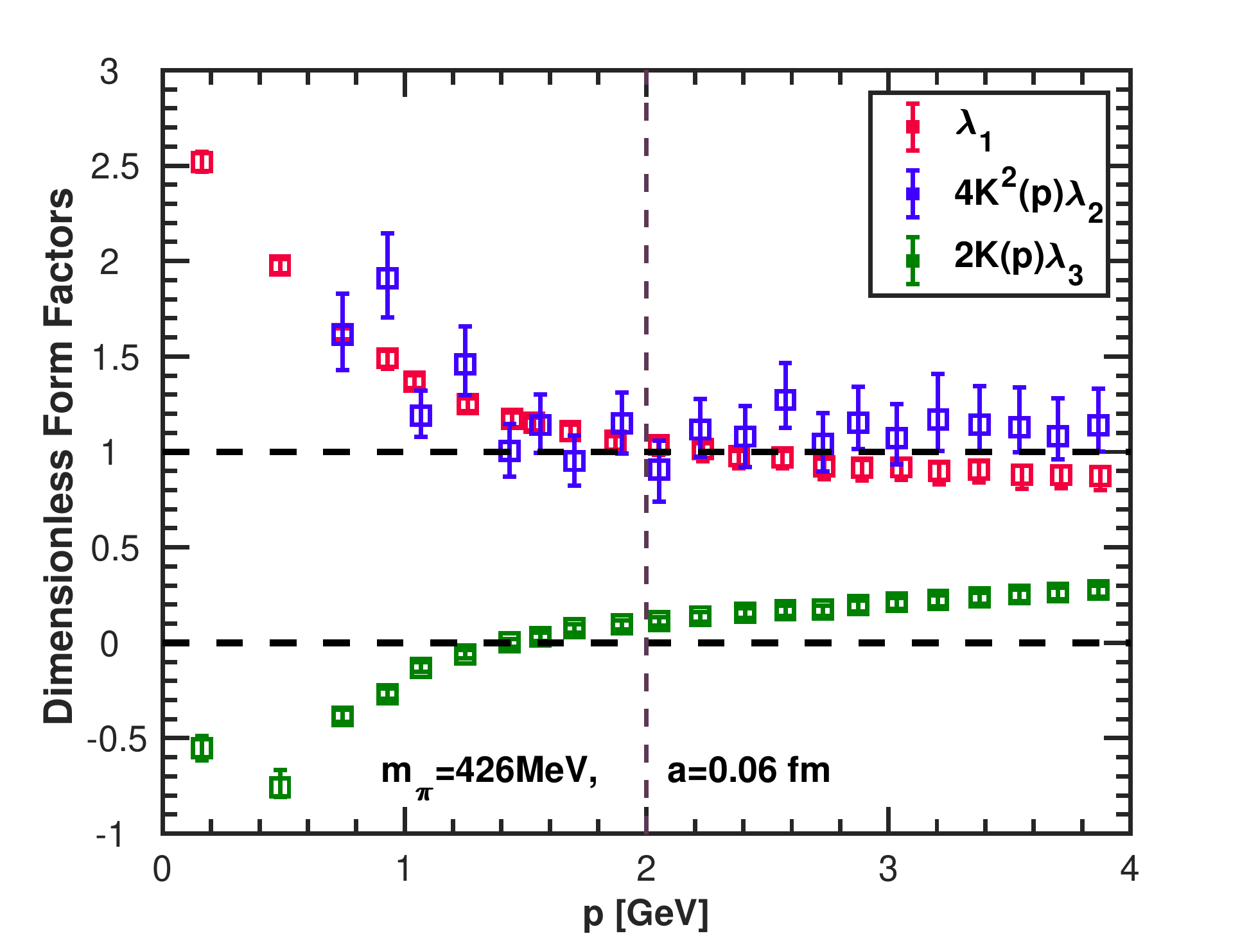}
\caption{}
        \label{fig:allLambdas_GeV_vs_ap_lin_m426_ren}      
    \end{subfigure} 
  \begin{subfigure}[t]{0.495\textwidth}
        \centering
       \includegraphics*[width=\textwidth]{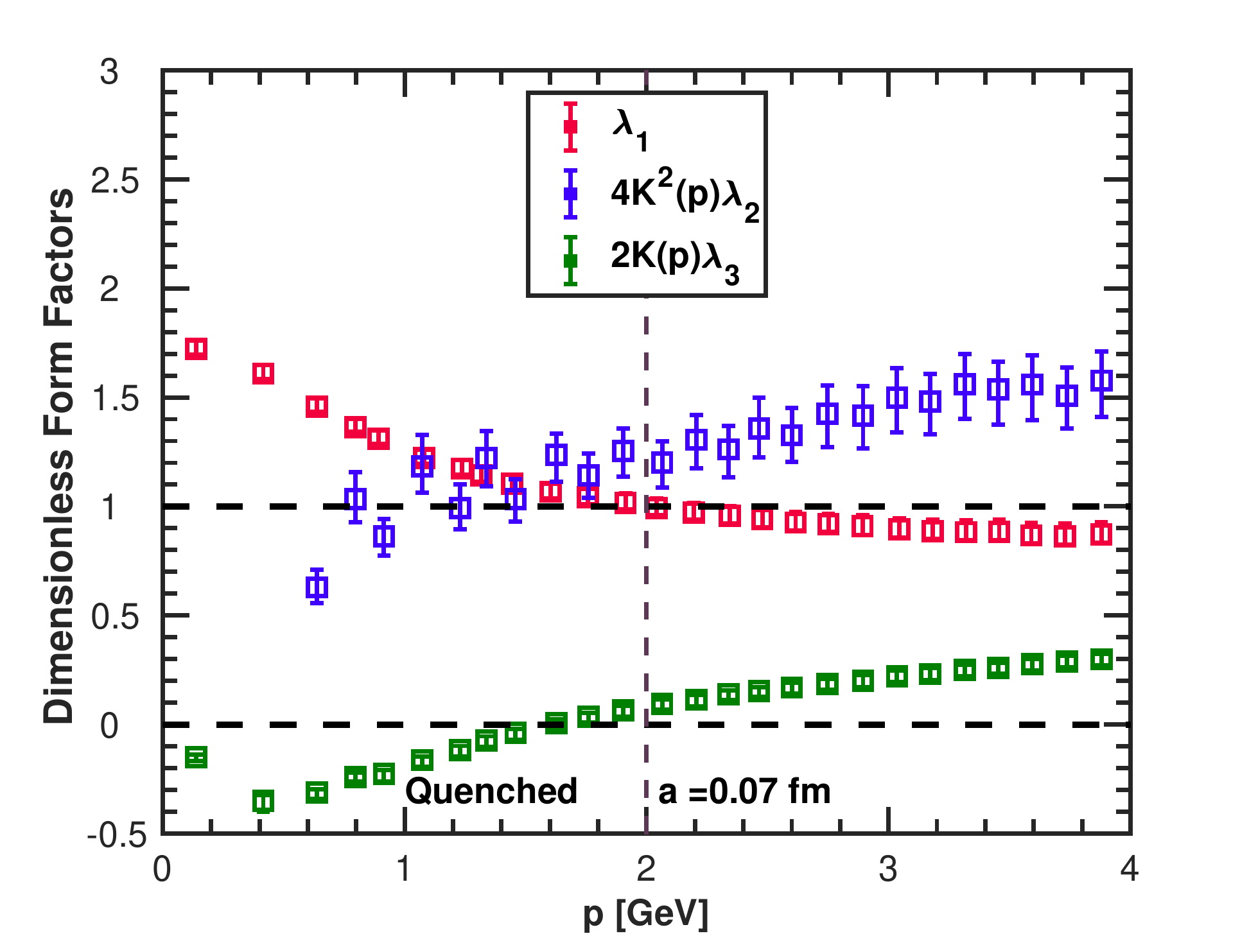}
\caption{}
        \label{fig:allLambdas_GeV_vs_ap_lin_mxxx_ren}      
    \end{subfigure} 
\caption{Dimensionless form factors $\lambda_1, 4K^2(p)\lambda_2$ and
  $2K(p)\lambda_3$, corresponding in the continuum to $\lambda_1,
  4p^2\lambda_2$ and 
  $2p\lambda_3$, for each ensemble versus momentum:
  (\subref{fig:allLambdas_GeV_vs_ap_lin_m295_ren}) ensemble L07;
  (\subref{fig:allLambdas_GeV_vs_ap_lin_m422_ren}) ensemble H07;
  (\subref{fig:allLambdas_GeV_vs_ap_lin_m290_ren}) ensemble L07-64;
  (\subref{fig:allLambdas_GeV_vs_ap_lin_m280_ren}) ensemble L08;
  (\subref{fig:allLambdas_GeV_vs_ap_lin_m426_ren}) ensemble H06;
  (\subref{fig:allLambdas_GeV_vs_ap_lin_mxxx_ren}) ensemble Q07.
  These measure the contribution of each form factor to the strength
  of the vertex.  Note that $K(p)$ and the naive momentum $p$ are
  equal in the continuum limit.}
\label{fig:allLambdas}      
\end{figure}
Finally, in Fig.~\ref{fig:allLambdas}  we show all form factors as function of momentum up to 4\,GeV for all lattice ensembles generated for this study. 
$\lambda_1$ is dimensionless but $\lambda_2$ and $\lambda_3$ have
dimensions of $p^{-2}$ and $p^{-1}$, respectively. In order to measure
the relative strength of $\lambda_2$ and  $\lambda_3$ in comparison to
$\lambda_1$, all form factors are multiplied by the appropriate
lattice momenta. The lattice momentum $K(p)$ is used here to
reduce lattice artefacts and ensure the continuum limit is approached
more rapidly; note that all momentum variables are equivalent in the
continuum limit and that no variable is intrinsically preferred over
any other away from the continuum limit.

For all ensembles, regardless of quark mass, volume and lattice
spacing, we find that $\lambda_1$ has the biggest contribution to the quark-gluon vertex. The second and third in the hierarchy are the 
$\lambda_2$ and $\lambda_3$ form factors, respectively. There is no IR
divergence in the soft gluon kinematics and hence we expect
$p^2\lambda_2$ and $p\lambda_3$ to vanish in the deep infrared.
Lattice calculations of $\lambda_2$ and $\lambda_3$ in the deep
infrared region are numerically extremely challenging; nevertheless in
all the plots in  Fig.~\ref{fig:allLambdas}  we see that
their contributions approach zero as $p\to0$.

\section{Conclusions and outlook}
\label{sec:conclude}
We have performed the first ever study of the quark--gluon vertex in Landau gauge lattice QCD with $N_f=2$ dynamical fermions.  The study has been carried out in the soft gluon limit (gluon momentum $q=0$) which gives access to three out of the non-transverse form factors, $(\lambda_1, \lambda_2, \lambda_3)$. 

The quark--gluon vertex is an essential element of QCD, yet its
complete non-perturbative form is still unknown. This
calculation may be used to calibrate any approximation made for the quark--gluon vertex. Furthermore it can be used to validate proposed non-perturbative models of the quark--gluon vertex.    

We find that the leading form factor $\lambda_1$ is significantly
enhanced in the infrared, and that this enhancement is stronger than
in the quenched approximation and increases as the chiral limit is
approached.  The indications are that the enhancement is further
strengthened as the continuum limit is approached.  No significant
finite volume effects are found.

The subleading vector form factor $\lambda_2$ also exhibits an
infrared strength commensurate with, if somewhat smaller than
$\lambda_1$.  This form factor has only a mild dependence on the
number of quark flavors, but the enhancement appears to increase as
the continuum, infinite-volume and chiral limits are approached.

The infrared strength of the scalar form factor $\lambda_3$ is considerably larger than in the quenched approximation and also appears to increase as the continuum limit is approached.  No
significant volume effect is found, but the strength is slightly
increased for smaller quark masses.  The latter is in contrast to
naive expectations as this form factor violates chiral symmetry and is
therefore sensitive to the 
explicit chiral symmetry breaking from the quark mass, but suggests
that it is primarily governed by the dynamical chiral symmetry
breaking.

In all three cases, the non-perturbative effects are found to be
orders of magnitude larger than the corresponding one-loop
perturbative contributions to these form factors. In all form factors
the lightest masses cause the larger IR enhancement and it decreases
as the mass increases. The dimensionless $\lambda_2$ and $\lambda_3$ form
factors approach zero in the infrared, indicating no kinematic
singularities in the soft gluon kinematics.

In the future we plan to extend this study to general kinematics which will allow us to determine all form factors of the transverse-projected vertex.  It is also of interest to compute
the vertex in general covariant gauges.   Apart from addressing gauge dependence, this would also allow us to disentangle the longitudinal and transverse components, which is not possible in Landau gauge.

\begin{acknowledgments}
The gauge fixing and calculations of the fermion propagators were
performed on the HLRN supercomputing facilities in Berlin and Hanover.
JIS has been supported by Science Foundation Ireland grant
11/RFP.1/PHY/1362. AS acknowledges support by the DFG as member of the
projects SFB/TRR55 and GRK1523. OO and PJS acknowledge support from 
FCT (Portugal) Projects No. UID/FIS/04564/2019 and UID/FIS/04564/2020.
PJS acknowledges financial support from FCT (Portugal) under Contract
No. CEECIND/00488/2017.
JIS expresses his deep appreciation for the hospitality of the Centre
for the Subatomic Structure of Matter,
University of Adelaide, where much of this work was carried out, as
well as the Galileo Galilei Institute, Florence, where this work was
brought to completion. We thank A.G. Williams for stimulating discussions and
support in the early stages of the project.
AK thanks Prof. A.W. Thomas for supporting this work.
\end{acknowledgments}
\bibliography{qgv_references}

\begin{thebibliography}{62}%
\makeatletter
\providecommand \@ifxundefined [1]{%
 \@ifx{#1\undefined}
}%
\providecommand \@ifnum [1]{%
 \ifnum #1\expandafter \@firstoftwo
 \else \expandafter \@secondoftwo
 \fi
}%
\providecommand \@ifx [1]{%
 \ifx #1\expandafter \@firstoftwo
 \else \expandafter \@secondoftwo
 \fi
}%
\providecommand \natexlab [1]{#1}%
\providecommand \enquote  [1]{``#1''}%
\providecommand \bibnamefont  [1]{#1}%
\providecommand \bibfnamefont [1]{#1}%
\providecommand \citenamefont [1]{#1}%
\providecommand \href@noop [0]{\@secondoftwo}%
\providecommand \href [0]{\begingroup \@sanitize@url \@href}%
\providecommand \@href[1]{\@@startlink{#1}\@@href}%
\providecommand \@@href[1]{\endgroup#1\@@endlink}%
\providecommand \@sanitize@url [0]{\catcode `\\12\catcode `\$12\catcode
  `\&12\catcode `\#12\catcode `\^12\catcode `\_12\catcode `\%12\relax}%
\providecommand \@@startlink[1]{}%
\providecommand \@@endlink[0]{}%
\providecommand \url  [0]{\begingroup\@sanitize@url \@url }%
\providecommand \@url [1]{\endgroup\@href {#1}{\urlprefix }}%
\providecommand \urlprefix  [0]{URL }%
\providecommand \Eprint [0]{\href }%
\providecommand \doibase [0]{http://dx.doi.org/}%
\providecommand \selectlanguage [0]{\@gobble}%
\providecommand \bibinfo  [0]{\@secondoftwo}%
\providecommand \bibfield  [0]{\@secondoftwo}%
\providecommand \translation [1]{[#1]}%
\providecommand \BibitemOpen [0]{}%
\providecommand \bibitemStop [0]{}%
\providecommand \bibitemNoStop [0]{.\EOS\space}%
\providecommand \EOS [0]{\spacefactor3000\relax}%
\providecommand \BibitemShut  [1]{\csname bibitem#1\endcsname}%
\let\auto@bib@innerbib\@empty
\bibitem [{\citenamefont {K{\i}z{\i}lers{\"u}}\ \emph
  {et~al.}(2013)\citenamefont {K{\i}z{\i}lers{\"u}}, \citenamefont {Sizer},\
  and\ \citenamefont {Williams}}]{Kizilersu:2013ksw}%
  \BibitemOpen
  \bibfield  {author} {\bibinfo {author} {\bibfnamefont {Ay{\c s}e}\
  \bibnamefont {K{\i}z{\i}lers{\"u}}}, \bibinfo {author} {\bibfnamefont {Tom}\
  \bibnamefont {Sizer}}, \ and\ \bibinfo {author} {\bibfnamefont {Anthony~G.}\
  \bibnamefont {Williams}},\ }\bibfield  {title} {\enquote {\bibinfo {title}
  {Strongly-coupled unquenched {QED$_4$} propagators using {S}chwinger--{D}yson
  equations},}\ }\href {\doibase 10.1103/PhysRevD.88.045008} {\bibfield
  {journal} {\bibinfo  {journal} {Phys. Rev.}\ }\textbf {\bibinfo {volume}
  {D88}},\ \bibinfo {pages} {045008} (\bibinfo {year} {2013})},\ \Eprint
  {http://arxiv.org/abs/1310.3160} {arXiv:1310.3160 [hep-ph]} \BibitemShut
  {NoStop}%
\bibitem [{\citenamefont {Maris}\ \emph {et~al.}(1998)\citenamefont {Maris},
  \citenamefont {Roberts},\ and\ \citenamefont {Tandy}}]{Maris:1997hd}%
  \BibitemOpen
  \bibfield  {author} {\bibinfo {author} {\bibfnamefont {Pieter}\ \bibnamefont
  {Maris}}, \bibinfo {author} {\bibfnamefont {Craig~D.}\ \bibnamefont
  {Roberts}}, \ and\ \bibinfo {author} {\bibfnamefont {Peter~C.}\ \bibnamefont
  {Tandy}},\ }\bibfield  {title} {\enquote {\bibinfo {title} {{Pion mass and
  decay constant}},}\ }\href {\doibase 10.1016/S0370-2693(97)01535-9}
  {\bibfield  {journal} {\bibinfo  {journal} {Phys. Lett. B}\ }\textbf
  {\bibinfo {volume} {420}},\ \bibinfo {pages} {267--273} (\bibinfo {year}
  {1998})},\ \Eprint {http://arxiv.org/abs/nucl-th/9707003}
  {arXiv:nucl-th/9707003} \BibitemShut {NoStop}%
\bibitem [{\citenamefont {Maris}\ and\ \citenamefont
  {Tandy}(1999)}]{Maris:1999nt}%
  \BibitemOpen
  \bibfield  {author} {\bibinfo {author} {\bibfnamefont {Pieter}\ \bibnamefont
  {Maris}}\ and\ \bibinfo {author} {\bibfnamefont {Peter~C.}\ \bibnamefont
  {Tandy}},\ }\bibfield  {title} {\enquote {\bibinfo {title} {{Bethe-Salpeter
  study of vector meson masses and decay constants}},}\ }\href {\doibase
  10.1103/PhysRevC.60.055214} {\bibfield  {journal} {\bibinfo  {journal} {Phys.
  Rev. C}\ }\textbf {\bibinfo {volume} {60}},\ \bibinfo {pages} {055214}
  (\bibinfo {year} {1999})},\ \Eprint {http://arxiv.org/abs/nucl-th/9905056}
  {arXiv:nucl-th/9905056} \BibitemShut {NoStop}%
\bibitem [{\citenamefont {Maris}\ and\ \citenamefont
  {Tandy}(2000)}]{Maris:2000sk}%
  \BibitemOpen
  \bibfield  {author} {\bibinfo {author} {\bibfnamefont {Pieter}\ \bibnamefont
  {Maris}}\ and\ \bibinfo {author} {\bibfnamefont {Peter~C.}\ \bibnamefont
  {Tandy}},\ }\bibfield  {title} {\enquote {\bibinfo {title} {The {$\pi, K^+$},
  and {$K^0$} electromagnetic form-factors},}\ }\href {\doibase
  10.1103/PhysRevC.62.055204} {\bibfield  {journal} {\bibinfo  {journal} {Phys.
  Rev. C}\ }\textbf {\bibinfo {volume} {62}},\ \bibinfo {pages} {055204}
  (\bibinfo {year} {2000})},\ \Eprint {http://arxiv.org/abs/nucl-th/0005015}
  {arXiv:nucl-th/0005015} \BibitemShut {NoStop}%
\bibitem [{\citenamefont {Chang}\ \emph
  {et~al.}(2013{\natexlab{a}})\citenamefont {Chang}, \citenamefont {Clo\"et},
  \citenamefont {Roberts}, \citenamefont {Schmidt},\ and\ \citenamefont
  {Tandy}}]{Chang:2013nia}%
  \BibitemOpen
  \bibfield  {author} {\bibinfo {author} {\bibfnamefont {L.}~\bibnamefont
  {Chang}}, \bibinfo {author} {\bibfnamefont {I.C.}\ \bibnamefont {Clo\"et}},
  \bibinfo {author} {\bibfnamefont {C.D.}\ \bibnamefont {Roberts}}, \bibinfo
  {author} {\bibfnamefont {S.M.}\ \bibnamefont {Schmidt}}, \ and\ \bibinfo
  {author} {\bibfnamefont {P.C.}\ \bibnamefont {Tandy}},\ }\bibfield  {title}
  {\enquote {\bibinfo {title} {{Pion electromagnetic form factor at spacelike
  momenta}},}\ }\href {\doibase 10.1103/PhysRevLett.111.141802} {\bibfield
  {journal} {\bibinfo  {journal} {Phys. Rev. Lett.}\ }\textbf {\bibinfo
  {volume} {111}},\ \bibinfo {pages} {141802} (\bibinfo {year}
  {2013}{\natexlab{a}})},\ \Eprint {http://arxiv.org/abs/1307.0026}
  {arXiv:1307.0026 [nucl-th]} \BibitemShut {NoStop}%
\bibitem [{\citenamefont {Bashir}\ \emph {et~al.}(2012)\citenamefont {Bashir},
  \citenamefont {Bermudez}, \citenamefont {Chang},\ and\ \citenamefont
  {Roberts}}]{Bashir:2011dp}%
  \BibitemOpen
  \bibfield  {author} {\bibinfo {author} {\bibfnamefont {A.}~\bibnamefont
  {Bashir}}, \bibinfo {author} {\bibfnamefont {R.}~\bibnamefont {Bermudez}},
  \bibinfo {author} {\bibfnamefont {L.}~\bibnamefont {Chang}}, \ and\ \bibinfo
  {author} {\bibfnamefont {C.D.}\ \bibnamefont {Roberts}},\ }\bibfield  {title}
  {\enquote {\bibinfo {title} {{Dynamical chiral symmetry breaking and the
  fermion--gauge-boson vertex}},}\ }\href {\doibase 10.1103/PhysRevC.85.045205}
  {\bibfield  {journal} {\bibinfo  {journal} {Phys. Rev. C}\ }\textbf {\bibinfo
  {volume} {85}},\ \bibinfo {pages} {045205} (\bibinfo {year} {2012})},\
  \Eprint {http://arxiv.org/abs/1112.4847} {arXiv:1112.4847 [nucl-th]}
  \BibitemShut {NoStop}%
\bibitem [{\citenamefont {Chang}\ \emph
  {et~al.}(2013{\natexlab{b}})\citenamefont {Chang}, \citenamefont {Cloet},
  \citenamefont {Cobos-Martinez}, \citenamefont {Roberts}, \citenamefont
  {Schmidt},\ and\ \citenamefont {Tandy}}]{Chang:2013pq}%
  \BibitemOpen
  \bibfield  {author} {\bibinfo {author} {\bibfnamefont {Lei}\ \bibnamefont
  {Chang}}, \bibinfo {author} {\bibfnamefont {I.C.}\ \bibnamefont {Cloet}},
  \bibinfo {author} {\bibfnamefont {J.J.}\ \bibnamefont {Cobos-Martinez}},
  \bibinfo {author} {\bibfnamefont {C.D.}\ \bibnamefont {Roberts}}, \bibinfo
  {author} {\bibfnamefont {S.M.}\ \bibnamefont {Schmidt}}, \ and\ \bibinfo
  {author} {\bibfnamefont {P.C.}\ \bibnamefont {Tandy}},\ }\bibfield  {title}
  {\enquote {\bibinfo {title} {{Imaging dynamical chiral symmetry breaking:
  pion wave function on the light front}},}\ }\href {\doibase
  10.1103/PhysRevLett.110.132001} {\bibfield  {journal} {\bibinfo  {journal}
  {Phys. Rev. Lett.}\ }\textbf {\bibinfo {volume} {110}},\ \bibinfo {pages}
  {132001} (\bibinfo {year} {2013}{\natexlab{b}})},\ \Eprint
  {http://arxiv.org/abs/1301.0324} {arXiv:1301.0324 [nucl-th]} \BibitemShut
  {NoStop}%
\bibitem [{\citenamefont {Chang}\ and\ \citenamefont
  {Roberts}(2009)}]{Chang:2009zb}%
  \BibitemOpen
  \bibfield  {author} {\bibinfo {author} {\bibfnamefont {Lei}\ \bibnamefont
  {Chang}}\ and\ \bibinfo {author} {\bibfnamefont {Craig~D.}\ \bibnamefont
  {Roberts}},\ }\bibfield  {title} {\enquote {\bibinfo {title} {{Sketching the
  Bethe-Salpeter kernel}},}\ }\href {\doibase 10.1103/PhysRevLett.103.081601}
  {\bibfield  {journal} {\bibinfo  {journal} {Phys. Rev. Lett.}\ }\textbf
  {\bibinfo {volume} {103}},\ \bibinfo {pages} {081601} (\bibinfo {year}
  {2009})},\ \Eprint {http://arxiv.org/abs/0903.5461} {arXiv:0903.5461
  [nucl-th]} \BibitemShut {NoStop}%
\bibitem [{\citenamefont {Maas}(2013)}]{Maas:2011se}%
  \BibitemOpen
  \bibfield  {author} {\bibinfo {author} {\bibfnamefont {Axel}\ \bibnamefont
  {Maas}},\ }\bibfield  {title} {\enquote {\bibinfo {title} {{Describing gauge
  bosons at zero and finite temperature}},}\ }\href {\doibase
  10.1016/j.physrep.2012.11.002} {\bibfield  {journal} {\bibinfo  {journal}
  {Phys. Rept.}\ }\textbf {\bibinfo {volume} {524}},\ \bibinfo {pages}
  {203--300} (\bibinfo {year} {2013})},\ \Eprint
  {http://arxiv.org/abs/1106.3942} {arXiv:1106.3942 [hep-ph]} \BibitemShut
  {NoStop}%
\bibitem [{\citenamefont {Sternbeck}\ and\ \citenamefont {von
  Smekal}(2010)}]{Sternbeck:2008mv}%
  \BibitemOpen
  \bibfield  {author} {\bibinfo {author} {\bibfnamefont {Andre}\ \bibnamefont
  {Sternbeck}}\ and\ \bibinfo {author} {\bibfnamefont {Lorenz}\ \bibnamefont
  {von Smekal}},\ }\bibfield  {title} {\enquote {\bibinfo {title} {Infrared
  exponents and the strong-coupling limit in lattice {Landau} gauge},}\ }\href
  {\doibase 10.1140/epjc/s10052-010-1381-8} {\bibfield  {journal} {\bibinfo
  {journal} {Eur. Phys. J. C}\ }\textbf {\bibinfo {volume} {68}},\ \bibinfo
  {pages} {487--503} (\bibinfo {year} {2010})},\ \Eprint
  {http://arxiv.org/abs/0811.4300} {arXiv:0811.4300 [hep-lat]} \BibitemShut
  {NoStop}%
\bibitem [{\citenamefont {Sternbeck}\ and\ \citenamefont
  {M\"uller-Preussker}(2013)}]{Sternbeck:2012mf}%
  \BibitemOpen
  \bibfield  {author} {\bibinfo {author} {\bibfnamefont {Andr\'e}\ \bibnamefont
  {Sternbeck}}\ and\ \bibinfo {author} {\bibfnamefont {Michael}\ \bibnamefont
  {M\"uller-Preussker}},\ }\bibfield  {title} {\enquote {\bibinfo {title}
  {Lattice evidence for the family of decoupling solutions of {L}andau gauge
  {Yang--Mills} theory},}\ }\href {\doibase 10.1016/j.physletb.2013.08.017}
  {\bibfield  {journal} {\bibinfo  {journal} {Phys. Lett. B}\ }\textbf
  {\bibinfo {volume} {726}},\ \bibinfo {pages} {396--403} (\bibinfo {year}
  {2013})},\ \Eprint {http://arxiv.org/abs/1211.3057} {arXiv:1211.3057
  [hep-lat]} \BibitemShut {NoStop}%
\bibitem [{\citenamefont {Oliveira}\ \emph {et~al.}(2017)\citenamefont
  {Oliveira}, \citenamefont {Duarte}, \citenamefont {Dudal},\ and\
  \citenamefont {Silva}}]{Oliveira:2016stx}%
  \BibitemOpen
  \bibfield  {author} {\bibinfo {author} {\bibfnamefont {O.}~\bibnamefont
  {Oliveira}}, \bibinfo {author} {\bibfnamefont {A.G.}\ \bibnamefont {Duarte}},
  \bibinfo {author} {\bibfnamefont {D.}~\bibnamefont {Dudal}}, \ and\ \bibinfo
  {author} {\bibfnamefont {P.J.}\ \bibnamefont {Silva}},\ }\bibfield  {title}
  {\enquote {\bibinfo {title} {Gluon and ghost dynamics from lattice {QCD}},}\
  }\href {\doibase 10.1007/s00601-017-1269-3} {\bibfield  {journal} {\bibinfo
  {journal} {Few Body Syst.}\ }\textbf {\bibinfo {volume} {58}},\ \bibinfo
  {pages} {99} (\bibinfo {year} {2017})},\ \Eprint
  {http://arxiv.org/abs/1611.03642} {arXiv:1611.03642 [hep-lat]} \BibitemShut
  {NoStop}%
\bibitem [{\citenamefont {Duarte}\ \emph {et~al.}(2016)\citenamefont {Duarte},
  \citenamefont {Oliveira},\ and\ \citenamefont {Silva}}]{Duarte:2016iko}%
  \BibitemOpen
  \bibfield  {author} {\bibinfo {author} {\bibfnamefont {Anthony~G.}\
  \bibnamefont {Duarte}}, \bibinfo {author} {\bibfnamefont {Orlando}\
  \bibnamefont {Oliveira}}, \ and\ \bibinfo {author} {\bibfnamefont {Paulo~J.}\
  \bibnamefont {Silva}},\ }\bibfield  {title} {\enquote {\bibinfo {title}
  {{Lattice Gluon and Ghost Propagators, and the Strong Coupling in Pure
  {SU(3)} {Y}ang--{M}ills Theory: {F}inite Lattice Spacing and Volume
  Effects}},}\ }\href {\doibase 10.1103/PhysRevD.94.014502} {\bibfield
  {journal} {\bibinfo  {journal} {Phys. Rev. D}\ }\textbf {\bibinfo {volume}
  {94}},\ \bibinfo {pages} {014502} (\bibinfo {year} {2016})},\ \Eprint
  {http://arxiv.org/abs/1605.00594} {arXiv:1605.00594 [hep-lat]} \BibitemShut
  {NoStop}%
\bibitem [{\citenamefont {Parappilly}\ \emph {et~al.}(2006)\citenamefont
  {Parappilly}, \citenamefont {Bowman}, \citenamefont {Heller}, \citenamefont
  {Leinweber}, \citenamefont {Williams},\ and\ \citenamefont
  {Zhang}}]{Parappilly:2005ei}%
  \BibitemOpen
  \bibfield  {author} {\bibinfo {author} {\bibfnamefont {Maria~B.}\
  \bibnamefont {Parappilly}}, \bibinfo {author} {\bibfnamefont {Patrick~O.}\
  \bibnamefont {Bowman}}, \bibinfo {author} {\bibfnamefont {Urs~M.}\
  \bibnamefont {Heller}}, \bibinfo {author} {\bibfnamefont {Derek~B.}\
  \bibnamefont {Leinweber}}, \bibinfo {author} {\bibfnamefont {Anthony~G.}\
  \bibnamefont {Williams}}, \ and\ \bibinfo {author} {\bibfnamefont {J.B}\
  \bibnamefont {Zhang}},\ }\bibfield  {title} {\enquote {\bibinfo {title}
  {Scaling behavior of quark propagator in full {QCD}},}\ }\href {\doibase
  10.1103/PhysRevD.73.054504} {\bibfield  {journal} {\bibinfo  {journal} {Phys.
  Rev. D}\ }\textbf {\bibinfo {volume} {73}},\ \bibinfo {pages} {054504}
  (\bibinfo {year} {2006})},\ \Eprint {http://arxiv.org/abs/hep-lat/0511007}
  {arXiv:hep-lat/0511007} \BibitemShut {NoStop}%
\bibitem [{\citenamefont {Kamleh}\ \emph {et~al.}(2007)\citenamefont {Kamleh},
  \citenamefont {Bowman}, \citenamefont {Leinweber}, \citenamefont {Williams},\
  and\ \citenamefont {Zhang}}]{Kamleh:2007ud}%
  \BibitemOpen
  \bibfield  {author} {\bibinfo {author} {\bibfnamefont {Waseem}\ \bibnamefont
  {Kamleh}}, \bibinfo {author} {\bibfnamefont {Patrick~O.}\ \bibnamefont
  {Bowman}}, \bibinfo {author} {\bibfnamefont {Derek~B.}\ \bibnamefont
  {Leinweber}}, \bibinfo {author} {\bibfnamefont {Anthony~G.}\ \bibnamefont
  {Williams}}, \ and\ \bibinfo {author} {\bibfnamefont {Jianbo}\ \bibnamefont
  {Zhang}},\ }\bibfield  {title} {\enquote {\bibinfo {title} {{Unquenching
  effects in the quark and gluon propagator}},}\ }\href {\doibase
  10.1103/PhysRevD.76.094501} {\bibfield  {journal} {\bibinfo  {journal} {Phys.
  Rev. D}\ }\textbf {\bibinfo {volume} {76}},\ \bibinfo {pages} {094501}
  (\bibinfo {year} {2007})},\ \Eprint {http://arxiv.org/abs/0705.4129}
  {arXiv:0705.4129 [hep-lat]} \BibitemShut {NoStop}%
\bibitem [{\citenamefont {Oliveira}\ \emph {et~al.}(2019)\citenamefont
  {Oliveira}, \citenamefont {Silva}, \citenamefont {Skullerud},\ and\
  \citenamefont {Sternbeck}}]{Oliveira:2018lln}%
  \BibitemOpen
  \bibfield  {author} {\bibinfo {author} {\bibfnamefont {Orlando}\ \bibnamefont
  {Oliveira}}, \bibinfo {author} {\bibfnamefont {Paulo~J.}\ \bibnamefont
  {Silva}}, \bibinfo {author} {\bibfnamefont {Jon-Ivar}\ \bibnamefont
  {Skullerud}}, \ and\ \bibinfo {author} {\bibfnamefont {Andr{\'e}}\
  \bibnamefont {Sternbeck}},\ }\bibfield  {title} {\enquote {\bibinfo {title}
  {Quark propagator with two flavors of {O(a)}-improved {W}ilson fermions},}\
  }\href {\doibase 10.1103/PhysRevD.99.094506} {\bibfield  {journal} {\bibinfo
  {journal} {Phys. Rev.}\ }\textbf {\bibinfo {volume} {D99}},\ \bibinfo {pages}
  {094506} (\bibinfo {year} {2019})},\ \Eprint
  {http://arxiv.org/abs/1809.02541} {arXiv:1809.02541 [hep-lat]} \BibitemShut
  {NoStop}%
\bibitem [{\citenamefont {Fischer}(2006)}]{Fischer:2006ub}%
  \BibitemOpen
  \bibfield  {author} {\bibinfo {author} {\bibfnamefont {Christian~S.}\
  \bibnamefont {Fischer}},\ }\bibfield  {title} {\enquote {\bibinfo {title}
  {Infrared properties of {QCD} from {D}yson--{S}chwinger equations},}\ }\href
  {\doibase 10.1088/0954-3899/32/8/R02} {\bibfield  {journal} {\bibinfo
  {journal} {J. Phys. G}\ }\textbf {\bibinfo {volume} {32}},\ \bibinfo {pages}
  {R253--R291} (\bibinfo {year} {2006})},\ \Eprint
  {http://arxiv.org/abs/hep-ph/0605173} {arXiv:hep-ph/0605173} \BibitemShut
  {NoStop}%
\bibitem [{\citenamefont {Binosi}\ and\ \citenamefont
  {Papavassiliou}(2009)}]{Binosi:2009qm}%
  \BibitemOpen
  \bibfield  {author} {\bibinfo {author} {\bibfnamefont {Daniele}\ \bibnamefont
  {Binosi}}\ and\ \bibinfo {author} {\bibfnamefont {Joannis}\ \bibnamefont
  {Papavassiliou}},\ }\bibfield  {title} {\enquote {\bibinfo {title} {Pinch
  technique: Theory and applications},}\ }\href {\doibase
  10.1016/j.physrep.2009.05.001} {\bibfield  {journal} {\bibinfo  {journal}
  {Phys. Rept.}\ }\textbf {\bibinfo {volume} {479}},\ \bibinfo {pages} {1--152}
  (\bibinfo {year} {2009})},\ \Eprint {http://arxiv.org/abs/0909.2536}
  {arXiv:0909.2536 [hep-ph]} \BibitemShut {NoStop}%
\bibitem [{\citenamefont {Braun}\ \emph {et~al.}(2016)\citenamefont {Braun},
  \citenamefont {Fister}, \citenamefont {Pawlowski},\ and\ \citenamefont
  {Rennecke}}]{Braun:2014ata}%
  \BibitemOpen
  \bibfield  {author} {\bibinfo {author} {\bibfnamefont {Jens}\ \bibnamefont
  {Braun}}, \bibinfo {author} {\bibfnamefont {Leonard}\ \bibnamefont {Fister}},
  \bibinfo {author} {\bibfnamefont {Jan~M.}\ \bibnamefont {Pawlowski}}, \ and\
  \bibinfo {author} {\bibfnamefont {Fabian}\ \bibnamefont {Rennecke}},\
  }\bibfield  {title} {\enquote {\bibinfo {title} {From quarks and gluons to
  hadrons: Chiral symmetry breaking in dynamical {QCD}},}\ }\href {\doibase
  10.1103/PhysRevD.94.034016} {\bibfield  {journal} {\bibinfo  {journal} {Phys.
  Rev. D}\ }\textbf {\bibinfo {volume} {94}},\ \bibinfo {pages} {034016}
  (\bibinfo {year} {2016})},\ \Eprint {http://arxiv.org/abs/1412.1045}
  {arXiv:1412.1045 [hep-ph]} \BibitemShut {NoStop}%
\bibitem [{\citenamefont {Cyrol}\ \emph {et~al.}(2018)\citenamefont {Cyrol},
  \citenamefont {Mitter}, \citenamefont {Pawlowski},\ and\ \citenamefont
  {Strodthoff}}]{Cyrol:2017ewj}%
  \BibitemOpen
  \bibfield  {author} {\bibinfo {author} {\bibfnamefont {Anton~K.}\
  \bibnamefont {Cyrol}}, \bibinfo {author} {\bibfnamefont {Mario}\ \bibnamefont
  {Mitter}}, \bibinfo {author} {\bibfnamefont {Jan~M.}\ \bibnamefont
  {Pawlowski}}, \ and\ \bibinfo {author} {\bibfnamefont {Nils}\ \bibnamefont
  {Strodthoff}},\ }\bibfield  {title} {\enquote {\bibinfo {title}
  {Nonperturbative quark, gluon, and meson correlators of unquenched {QCD}},}\
  }\href {\doibase 10.1103/PhysRevD.97.054006} {\bibfield  {journal} {\bibinfo
  {journal} {Phys. Rev.}\ }\textbf {\bibinfo {volume} {D97}},\ \bibinfo {pages}
  {054006} (\bibinfo {year} {2018})},\ \Eprint
  {http://arxiv.org/abs/1706.06326} {arXiv:1706.06326 [hep-ph]} \BibitemShut
  {NoStop}%
\bibitem [{\citenamefont {Aguilar}\ \emph {et~al.}(2020)\citenamefont
  {Aguilar}, \citenamefont {De~Soto}, \citenamefont {Ferreira}, \citenamefont
  {Papavassiliou}, \citenamefont {Rodr\'\i{}guez-Quintero},\ and\ \citenamefont
  {Zafeiropoulos}}]{Aguilar:2019uob}%
  \BibitemOpen
  \bibfield  {author} {\bibinfo {author} {\bibfnamefont {A.~C.}\ \bibnamefont
  {Aguilar}}, \bibinfo {author} {\bibfnamefont {F.}~\bibnamefont {De~Soto}},
  \bibinfo {author} {\bibfnamefont {M.~N.}\ \bibnamefont {Ferreira}}, \bibinfo
  {author} {\bibfnamefont {J.}~\bibnamefont {Papavassiliou}}, \bibinfo {author}
  {\bibfnamefont {J.}~\bibnamefont {Rodr\'\i{}guez-Quintero}}, \ and\ \bibinfo
  {author} {\bibfnamefont {S.}~\bibnamefont {Zafeiropoulos}},\ }\bibfield
  {title} {\enquote {\bibinfo {title} {Gluon propagator and three-gluon vertex
  with dynamical quarks},}\ }\href {\doibase 10.1140/epjc/s10052-020-7741-0}
  {\bibfield  {journal} {\bibinfo  {journal} {Eur. Phys. J. C}\ }\textbf
  {\bibinfo {volume} {80}},\ \bibinfo {pages} {154} (\bibinfo {year} {2020})},\
  \Eprint {http://arxiv.org/abs/1912.12086} {arXiv:1912.12086 [hep-ph]}
  \BibitemShut {NoStop}%
\bibitem [{\citenamefont {Bhagwat}\ and\ \citenamefont
  {Tandy}(2004)}]{Bhagwat:2004kj}%
  \BibitemOpen
  \bibfield  {author} {\bibinfo {author} {\bibfnamefont {M.~S.}\ \bibnamefont
  {Bhagwat}}\ and\ \bibinfo {author} {\bibfnamefont {P.~C.}\ \bibnamefont
  {Tandy}},\ }\bibfield  {title} {\enquote {\bibinfo {title} {Quark-gluon
  vertex model and lattice-{QCD} data},}\ }\href@noop {} {\bibfield  {journal}
  {\bibinfo  {journal} {Phys. Rev.}\ }\textbf {\bibinfo {volume} {D70}},\
  \bibinfo {pages} {094039} (\bibinfo {year} {2004})},\ \Eprint
  {http://arxiv.org/abs/hep-ph/0407163} {hep-ph/0407163} \BibitemShut {NoStop}%
\bibitem [{\citenamefont {K{\i}z{\i}lers{\"u}}\ and\ \citenamefont
  {Pennington}(2009)}]{Kizilersu:2009kg}%
  \BibitemOpen
  \bibfield  {author} {\bibinfo {author} {\bibfnamefont {Ay{\c s}e}\
  \bibnamefont {K{\i}z{\i}lers{\"u}}}\ and\ \bibinfo {author} {\bibfnamefont
  {Mike~R.}\ \bibnamefont {Pennington}},\ }\bibfield  {title} {\enquote
  {\bibinfo {title} {Building the full fermion-photon vertex of {QED} by
  imposing multiplicative renormalizability of the {S}chwinger-{D}yson
  equations for the fermion and photon propagators},}\ }\href {\doibase
  10.1103/PhysRevD.79.125020} {\bibfield  {journal} {\bibinfo  {journal} {Phys.
  Rev.}\ }\textbf {\bibinfo {volume} {D79}},\ \bibinfo {pages} {125020}
  (\bibinfo {year} {2009})},\ \Eprint {http://arxiv.org/abs/0904.3483}
  {arXiv:0904.3483 [hep-th]} \BibitemShut {NoStop}%
\bibitem [{\citenamefont {Chang}\ \emph {et~al.}(2011)\citenamefont {Chang},
  \citenamefont {Liu},\ and\ \citenamefont {Roberts}}]{Chang:2010hb}%
  \BibitemOpen
  \bibfield  {author} {\bibinfo {author} {\bibfnamefont {Lei}\ \bibnamefont
  {Chang}}, \bibinfo {author} {\bibfnamefont {Yu-Xin}\ \bibnamefont {Liu}}, \
  and\ \bibinfo {author} {\bibfnamefont {Craig~D.}\ \bibnamefont {Roberts}},\
  }\bibfield  {title} {\enquote {\bibinfo {title} {Dressed-quark anomalous
  magnetic moments},}\ }\href {\doibase 10.1103/PhysRevLett.106.072001}
  {\bibfield  {journal} {\bibinfo  {journal} {Phys. Rev. Lett.}\ }\textbf
  {\bibinfo {volume} {106}},\ \bibinfo {pages} {072001} (\bibinfo {year}
  {2011})},\ \Eprint {http://arxiv.org/abs/1009.3458} {arXiv:1009.3458
  [nucl-th]} \BibitemShut {NoStop}%
\bibitem [{\citenamefont {Mitter}\ \emph {et~al.}(2015)\citenamefont {Mitter},
  \citenamefont {Pawlowski},\ and\ \citenamefont
  {Strodthoff}}]{Mitter:2014wpa}%
  \BibitemOpen
  \bibfield  {author} {\bibinfo {author} {\bibfnamefont {Mario}\ \bibnamefont
  {Mitter}}, \bibinfo {author} {\bibfnamefont {Jan~M.}\ \bibnamefont
  {Pawlowski}}, \ and\ \bibinfo {author} {\bibfnamefont {Nils}\ \bibnamefont
  {Strodthoff}},\ }\bibfield  {title} {\enquote {\bibinfo {title} {Chiral
  symmetry breaking in continuum {QCD}},}\ }\href {\doibase
  10.1103/PhysRevD.91.054035} {\bibfield  {journal} {\bibinfo  {journal} {Phys.
  Rev. D}\ }\textbf {\bibinfo {volume} {91}},\ \bibinfo {pages} {054035}
  (\bibinfo {year} {2015})},\ \Eprint {http://arxiv.org/abs/1411.7978}
  {arXiv:1411.7978 [hep-ph]} \BibitemShut {NoStop}%
\bibitem [{\citenamefont {Aguilar}\ \emph {et~al.}(2014)\citenamefont
  {Aguilar}, \citenamefont {Binosi}, \citenamefont {Iba\~nez},\ and\
  \citenamefont {Papavassiliou}}]{Aguilar:2014lha}%
  \BibitemOpen
  \bibfield  {author} {\bibinfo {author} {\bibfnamefont {A.~C.}\ \bibnamefont
  {Aguilar}}, \bibinfo {author} {\bibfnamefont {D.}~\bibnamefont {Binosi}},
  \bibinfo {author} {\bibfnamefont {D.}~\bibnamefont {Iba\~nez}}, \ and\
  \bibinfo {author} {\bibfnamefont {J.}~\bibnamefont {Papavassiliou}},\
  }\bibfield  {title} {\enquote {\bibinfo {title} {New method for determining
  the quark-gluon vertex},}\ }\href {\doibase 10.1103/PhysRevD.90.065027}
  {\bibfield  {journal} {\bibinfo  {journal} {Phys. Rev.}\ }\textbf {\bibinfo
  {volume} {D90}},\ \bibinfo {pages} {065027} (\bibinfo {year} {2014})},\
  \Eprint {http://arxiv.org/abs/1405.3506} {arXiv:1405.3506 [hep-ph]}
  \BibitemShut {NoStop}%
\bibitem [{\citenamefont {Aguilar}\ \emph {et~al.}(2017)\citenamefont
  {Aguilar}, \citenamefont {Cardona}, \citenamefont {Ferreira},\ and\
  \citenamefont {Papavassiliou}}]{Aguilar:2016lbe}%
  \BibitemOpen
  \bibfield  {author} {\bibinfo {author} {\bibfnamefont {A.~C.}\ \bibnamefont
  {Aguilar}}, \bibinfo {author} {\bibfnamefont {J.~C.}\ \bibnamefont
  {Cardona}}, \bibinfo {author} {\bibfnamefont {M.~N.}\ \bibnamefont
  {Ferreira}}, \ and\ \bibinfo {author} {\bibfnamefont {J.}~\bibnamefont
  {Papavassiliou}},\ }\bibfield  {title} {\enquote {\bibinfo {title}
  {Non-abelian {B}all--{C}hiu vertex for arbitrary euclidean momenta},}\ }\href
  {\doibase 10.1103/PhysRevD.96.014029} {\bibfield  {journal} {\bibinfo
  {journal} {Phys. Rev.}\ }\textbf {\bibinfo {volume} {D96}},\ \bibinfo {pages}
  {014029} (\bibinfo {year} {2017})},\ \Eprint
  {http://arxiv.org/abs/1610.06158} {arXiv:1610.06158 [hep-ph]} \BibitemShut
  {NoStop}%
\bibitem [{\citenamefont {Binosi}\ \emph {et~al.}(2017)\citenamefont {Binosi},
  \citenamefont {Chang}, \citenamefont {Papavassiliou}, \citenamefont {Qin},\
  and\ \citenamefont {Roberts}}]{Binosi:2016wcx}%
  \BibitemOpen
  \bibfield  {author} {\bibinfo {author} {\bibfnamefont {Daniele}\ \bibnamefont
  {Binosi}}, \bibinfo {author} {\bibfnamefont {Lei}\ \bibnamefont {Chang}},
  \bibinfo {author} {\bibfnamefont {Joannis}\ \bibnamefont {Papavassiliou}},
  \bibinfo {author} {\bibfnamefont {Si-Xue}\ \bibnamefont {Qin}}, \ and\
  \bibinfo {author} {\bibfnamefont {Craig~D.}\ \bibnamefont {Roberts}},\
  }\bibfield  {title} {\enquote {\bibinfo {title} {Natural constraints on the
  gluon-quark vertex},}\ }\href {\doibase 10.1103/PhysRevD.95.031501}
  {\bibfield  {journal} {\bibinfo  {journal} {Phys. Rev.}\ }\textbf {\bibinfo
  {volume} {D95}},\ \bibinfo {pages} {031501(R)} (\bibinfo {year} {2017})},\
  \Eprint {http://arxiv.org/abs/1609.02568} {arXiv:1609.02568 [nucl-th]}
  \BibitemShut {NoStop}%
\bibitem [{\citenamefont {Bermudez}\ \emph {et~al.}(2017)\citenamefont
  {Bermudez}, \citenamefont {Albino}, \citenamefont {Guti\'errez-Guerrero},
  \citenamefont {Tejeda-Yeomans},\ and\ \citenamefont
  {Bashir}}]{Bermudez:2017bpx}%
  \BibitemOpen
  \bibfield  {author} {\bibinfo {author} {\bibfnamefont {R.}~\bibnamefont
  {Bermudez}}, \bibinfo {author} {\bibfnamefont {L.}~\bibnamefont {Albino}},
  \bibinfo {author} {\bibfnamefont {L.~X.}\ \bibnamefont
  {Guti\'errez-Guerrero}}, \bibinfo {author} {\bibfnamefont {M.~E.}\
  \bibnamefont {Tejeda-Yeomans}}, \ and\ \bibinfo {author} {\bibfnamefont
  {A.}~\bibnamefont {Bashir}},\ }\bibfield  {title} {\enquote {\bibinfo {title}
  {Quark-gluon vertex: A perturbation theory primer and beyond},}\ }\href
  {\doibase 10.1103/PhysRevD.95.034041} {\bibfield  {journal} {\bibinfo
  {journal} {Phys. Rev.}\ }\textbf {\bibinfo {volume} {D95}},\ \bibinfo {pages}
  {034041} (\bibinfo {year} {2017})},\ \Eprint
  {http://arxiv.org/abs/1702.04437} {arXiv:1702.04437 [hep-ph]} \BibitemShut
  {NoStop}%
\bibitem [{\citenamefont {K\i{}z\i{}lers\"u}\ \emph {et~al.}(2015)\citenamefont
  {K\i{}z\i{}lers\"u}, \citenamefont {Sizer}, \citenamefont {Pennington},
  \citenamefont {Williams},\ and\ \citenamefont
  {Williams}}]{Kizilersu:2014ela}%
  \BibitemOpen
  \bibfield  {author} {\bibinfo {author} {\bibfnamefont {Ay\c{s}e}\
  \bibnamefont {K\i{}z\i{}lers\"u}}, \bibinfo {author} {\bibfnamefont {Tom}\
  \bibnamefont {Sizer}}, \bibinfo {author} {\bibfnamefont {Michael~R.}\
  \bibnamefont {Pennington}}, \bibinfo {author} {\bibfnamefont {Anthony~G.}\
  \bibnamefont {Williams}}, \ and\ \bibinfo {author} {\bibfnamefont {Richard}\
  \bibnamefont {Williams}},\ }\bibfield  {title} {\enquote {\bibinfo {title}
  {Dynamical mass generation in unquenched {QED} using the {Dyson--Schwinger}
  equations},}\ }\href {\doibase 10.1103/PhysRevD.91.065015} {\bibfield
  {journal} {\bibinfo  {journal} {Phys. Rev. D}\ }\textbf {\bibinfo {volume}
  {91}},\ \bibinfo {pages} {065015} (\bibinfo {year} {2015})},\ \Eprint
  {http://arxiv.org/abs/1409.5979} {arXiv:1409.5979 [hep-ph]} \BibitemShut
  {NoStop}%
\bibitem [{\citenamefont {Williams}(2015)}]{Williams:2014iea}%
  \BibitemOpen
  \bibfield  {author} {\bibinfo {author} {\bibfnamefont {Richard}\ \bibnamefont
  {Williams}},\ }\bibfield  {title} {\enquote {\bibinfo {title} {{The
  quark-gluon vertex in Landau gauge bound-state studies}},}\ }\href {\doibase
  10.1140/epja/i2015-15057-4} {\bibfield  {journal} {\bibinfo  {journal} {Eur.
  Phys. J. A}\ }\textbf {\bibinfo {volume} {51}},\ \bibinfo {pages} {57}
  (\bibinfo {year} {2015})},\ \Eprint {http://arxiv.org/abs/1404.2545}
  {arXiv:1404.2545 [hep-ph]} \BibitemShut {NoStop}%
\bibitem [{\citenamefont {Oliveira}\ \emph {et~al.}(2018)\citenamefont
  {Oliveira}, \citenamefont {Frederico}, \citenamefont {de~Paula},\ and\
  \citenamefont {de~Melo}}]{Oliveira:2018fkj}%
  \BibitemOpen
  \bibfield  {author} {\bibinfo {author} {\bibfnamefont {Orlando}\ \bibnamefont
  {Oliveira}}, \bibinfo {author} {\bibfnamefont {T.}~\bibnamefont {Frederico}},
  \bibinfo {author} {\bibfnamefont {W.}~\bibnamefont {de~Paula}}, \ and\
  \bibinfo {author} {\bibfnamefont {J.P.B.C.}\ \bibnamefont {de~Melo}},\
  }\bibfield  {title} {\enquote {\bibinfo {title} {Exploring the quark-gluon
  vertex with {S}lavnov--{T}aylor identities and lattice simulations},}\ }\href
  {\doibase 10.1140/epjc/s10052-018-6037-0} {\bibfield  {journal} {\bibinfo
  {journal} {Eur. Phys. J. C}\ }\textbf {\bibinfo {volume} {78}},\ \bibinfo
  {pages} {553} (\bibinfo {year} {2018})},\ \Eprint
  {http://arxiv.org/abs/1807.00675} {arXiv:1807.00675 [hep-ph]} \BibitemShut
  {NoStop}%
\bibitem [{\citenamefont {Oliveira}\ \emph {et~al.}(2020)\citenamefont
  {Oliveira}, \citenamefont {Frederico},\ and\ \citenamefont
  {de~Paula}}]{Oliveira:2020yac}%
  \BibitemOpen
  \bibfield  {author} {\bibinfo {author} {\bibfnamefont {Orlando}\ \bibnamefont
  {Oliveira}}, \bibinfo {author} {\bibfnamefont {Tobias}\ \bibnamefont
  {Frederico}}, \ and\ \bibinfo {author} {\bibfnamefont {Wayne}\ \bibnamefont
  {de~Paula}},\ }\bibfield  {title} {\enquote {\bibinfo {title} {{The
  soft-gluon limit and the infrared enhancement of the quark-gluon vertex}},}\
  }\href {\doibase 10.1140/epjc/s10052-020-8037-0} {\bibfield  {journal}
  {\bibinfo  {journal} {Eur. Phys. J. C}\ }\textbf {\bibinfo {volume} {80}},\
  \bibinfo {pages} {484} (\bibinfo {year} {2020})},\ \Eprint
  {http://arxiv.org/abs/2006.04982} {arXiv:2006.04982 [hep-ph]} \BibitemShut
  {NoStop}%
\bibitem [{\citenamefont {Serna}\ \emph {et~al.}(2019)\citenamefont {Serna},
  \citenamefont {Chen},\ and\ \citenamefont {El-Bennich}}]{Serna:2018dwk}%
  \BibitemOpen
  \bibfield  {author} {\bibinfo {author} {\bibfnamefont {Fernando~E.}\
  \bibnamefont {Serna}}, \bibinfo {author} {\bibfnamefont {Chen}\ \bibnamefont
  {Chen}}, \ and\ \bibinfo {author} {\bibfnamefont {Bruno}\ \bibnamefont
  {El-Bennich}},\ }\bibfield  {title} {\enquote {\bibinfo {title} {{Interplay
  of dynamical and explicit chiral symmetry breaking effects on a quark}},}\
  }\href {\doibase 10.1103/PhysRevD.99.094027} {\bibfield  {journal} {\bibinfo
  {journal} {Phys. Rev. D}\ }\textbf {\bibinfo {volume} {99}},\ \bibinfo
  {pages} {094027} (\bibinfo {year} {2019})},\ \Eprint
  {http://arxiv.org/abs/1812.01096} {arXiv:1812.01096 [hep-ph]} \BibitemShut
  {NoStop}%
\bibitem [{\citenamefont {Gao}\ \emph {et~al.}(2021)\citenamefont {Gao},
  \citenamefont {Papavassiliou},\ and\ \citenamefont
  {Pawlowski}}]{Gao:2021wun}%
  \BibitemOpen
  \bibfield  {author} {\bibinfo {author} {\bibfnamefont {Fei}\ \bibnamefont
  {Gao}}, \bibinfo {author} {\bibfnamefont {Joannis}\ \bibnamefont
  {Papavassiliou}}, \ and\ \bibinfo {author} {\bibfnamefont {Jan~M.}\
  \bibnamefont {Pawlowski}},\ }\bibfield  {title} {\enquote {\bibinfo {title}
  {Fully coupled functional equations for the quark sector of {QCD}},}\ }\href
  {\doibase 10.1103/PhysRevD.103.094013} {\bibfield  {journal} {\bibinfo
  {journal} {Phys. Rev. D}\ }\textbf {\bibinfo {volume} {103}},\ \bibinfo
  {pages} {094013} (\bibinfo {year} {2021})},\ \Eprint
  {http://arxiv.org/abs/2102.13053} {arXiv:2102.13053 [hep-ph]} \BibitemShut
  {NoStop}%
\bibitem [{\citenamefont {Williams}\ \emph {et~al.}(2016)\citenamefont
  {Williams}, \citenamefont {Fischer},\ and\ \citenamefont
  {Heupel}}]{Williams:2015cvx}%
  \BibitemOpen
  \bibfield  {author} {\bibinfo {author} {\bibfnamefont {Richard}\ \bibnamefont
  {Williams}}, \bibinfo {author} {\bibfnamefont {Christian~S.}\ \bibnamefont
  {Fischer}}, \ and\ \bibinfo {author} {\bibfnamefont {Walter}\ \bibnamefont
  {Heupel}},\ }\bibfield  {title} {\enquote {\bibinfo {title} {Light mesons in
  {QCD} and unquenching effects from the {3PI} effective action},}\ }\href
  {\doibase 10.1103/PhysRevD.93.034026} {\bibfield  {journal} {\bibinfo
  {journal} {Phys. Rev. D}\ }\textbf {\bibinfo {volume} {93}},\ \bibinfo
  {pages} {034026} (\bibinfo {year} {2016})},\ \Eprint
  {http://arxiv.org/abs/1512.00455} {arXiv:1512.00455 [hep-ph]} \BibitemShut
  {NoStop}%
\bibitem [{\citenamefont {Sultan}\ \emph {et~al.}(2021)\citenamefont {Sultan},
  \citenamefont {Akram}, \citenamefont {Masud},\ and\ \citenamefont
  {Raya}}]{Sultan:2018qpx}%
  \BibitemOpen
  \bibfield  {author} {\bibinfo {author} {\bibfnamefont {M.~Atif}\ \bibnamefont
  {Sultan}}, \bibinfo {author} {\bibfnamefont {Faisal}\ \bibnamefont {Akram}},
  \bibinfo {author} {\bibfnamefont {Bilal}\ \bibnamefont {Masud}}, \ and\
  \bibinfo {author} {\bibfnamefont {Kh\'epani}\ \bibnamefont {Raya}},\
  }\bibfield  {title} {\enquote {\bibinfo {title} {Effect of the quark-gluon
  vertex on dynamical chiral symmetry breaking},}\ }\href {\doibase
  10.1103/PhysRevD.103.054036} {\bibfield  {journal} {\bibinfo  {journal}
  {Phys. Rev. D}\ }\textbf {\bibinfo {volume} {103}},\ \bibinfo {pages}
  {054036} (\bibinfo {year} {2021})},\ \Eprint
  {http://arxiv.org/abs/1810.01396} {arXiv:1810.01396 [nucl-th]} \BibitemShut
  {NoStop}%
\bibitem [{\citenamefont {Aguilar}\ \emph {et~al.}(2018)\citenamefont
  {Aguilar}, \citenamefont {Cardona}, \citenamefont {Ferreira},\ and\
  \citenamefont {Papavassiliou}}]{Aguilar:2018epe}%
  \BibitemOpen
  \bibfield  {author} {\bibinfo {author} {\bibfnamefont {A.~C.}\ \bibnamefont
  {Aguilar}}, \bibinfo {author} {\bibfnamefont {J.~C.}\ \bibnamefont
  {Cardona}}, \bibinfo {author} {\bibfnamefont {M.~N.}\ \bibnamefont
  {Ferreira}}, \ and\ \bibinfo {author} {\bibfnamefont {J.}~\bibnamefont
  {Papavassiliou}},\ }\bibfield  {title} {\enquote {\bibinfo {title} {Quark gap
  equation with non-abelian {B}all--{C}hiu vertex},}\ }\href {\doibase
  10.1103/PhysRevD.98.014002} {\bibfield  {journal} {\bibinfo  {journal} {Phys.
  Rev.}\ }\textbf {\bibinfo {volume} {D98}},\ \bibinfo {pages} {014002}
  (\bibinfo {year} {2018})},\ \Eprint {http://arxiv.org/abs/1804.04229}
  {arXiv:1804.04229 [hep-ph]} \BibitemShut {NoStop}%
\bibitem [{\citenamefont {He}\ \emph {et~al.}(2000)\citenamefont {He},
  \citenamefont {Khanna},\ and\ \citenamefont {Takahashi}}]{He:2000we}%
  \BibitemOpen
  \bibfield  {author} {\bibinfo {author} {\bibfnamefont {Han-Xin}\ \bibnamefont
  {He}}, \bibinfo {author} {\bibfnamefont {F.~C.}\ \bibnamefont {Khanna}}, \
  and\ \bibinfo {author} {\bibfnamefont {Y.}~\bibnamefont {Takahashi}},\
  }\bibfield  {title} {\enquote {\bibinfo {title} {Transverse
  {W}ard--{T}akahashi identity for the fermion boson vertex in gauge
  theories},}\ }\href {\doibase 10.1016/S0370-2693(00)00353-1} {\bibfield
  {journal} {\bibinfo  {journal} {Phys. Lett.}\ }\textbf {\bibinfo {volume}
  {B480}},\ \bibinfo {pages} {222--228} (\bibinfo {year} {2000})}\BibitemShut
  {NoStop}%
\bibitem [{\citenamefont {Kondo}(1997)}]{Kondo:1996xn}%
  \BibitemOpen
  \bibfield  {author} {\bibinfo {author} {\bibfnamefont {Kei-Ichi}\
  \bibnamefont {Kondo}},\ }\bibfield  {title} {\enquote {\bibinfo {title}
  {Transverse {Ward--Takahashi} identity, anomaly and {Schwinger--Dyson}
  equation},}\ }\href {\doibase 10.1142/S0217751X97002978} {\bibfield
  {journal} {\bibinfo  {journal} {Int. J. Mod. Phys.}\ }\textbf {\bibinfo
  {volume} {A12}},\ \bibinfo {pages} {5651--5686} (\bibinfo {year} {1997})},\
  \Eprint {http://arxiv.org/abs/hep-th/9608100} {arXiv:hep-th/9608100 [hep-th]}
  \BibitemShut {NoStop}%
\bibitem [{\citenamefont {Pennington}\ and\ \citenamefont
  {Williams}(2006)}]{Pennington:2005mw}%
  \BibitemOpen
  \bibfield  {author} {\bibinfo {author} {\bibfnamefont {M.~R.}\ \bibnamefont
  {Pennington}}\ and\ \bibinfo {author} {\bibfnamefont {R.}~\bibnamefont
  {Williams}},\ }\bibfield  {title} {\enquote {\bibinfo {title} {Checking the
  transverse {W}ard-{T}akahashi relation at one loop order in 4-dimensions},}\
  }\href {\doibase 10.1088/0954-3899/32/11/014} {\bibfield  {journal} {\bibinfo
   {journal} {J. Phys.}\ }\textbf {\bibinfo {volume} {G32}},\ \bibinfo {pages}
  {2219--2234} (\bibinfo {year} {2006})},\ \Eprint
  {http://arxiv.org/abs/hep-ph/0511254} {arXiv:hep-ph/0511254 [hep-ph]}
  \BibitemShut {NoStop}%
\bibitem [{\citenamefont {He}(2009)}]{He:2009sj}%
  \BibitemOpen
  \bibfield  {author} {\bibinfo {author} {\bibfnamefont {Han-xin}\ \bibnamefont
  {He}},\ }\bibfield  {title} {\enquote {\bibinfo {title} {Transverse symmetry
  transformations and the quark-gluon vertex function in {QCD}},}\ }\href
  {\doibase 10.1103/PhysRevD.80.016004} {\bibfield  {journal} {\bibinfo
  {journal} {Phys. Rev.}\ }\textbf {\bibinfo {volume} {D80}},\ \bibinfo {pages}
  {016004} (\bibinfo {year} {2009})},\ \Eprint {http://arxiv.org/abs/0906.2834}
  {arXiv:0906.2834 [hep-ph]} \BibitemShut {NoStop}%
\bibitem [{\citenamefont {Skullerud}\ and\ \citenamefont
  {K{\i}z{\i}lers{\"u}}(2002)}]{Skullerud:2002ge}%
  \BibitemOpen
  \bibfield  {author} {\bibinfo {author} {\bibfnamefont {Jon-Ivar}\
  \bibnamefont {Skullerud}}\ and\ \bibinfo {author} {\bibfnamefont {Ayse}\
  \bibnamefont {K{\i}z{\i}lers{\"u}}},\ }\bibfield  {title} {\enquote {\bibinfo
  {title} {Quark-gluon vertex from lattice {QCD}},}\ }\href@noop {} {\bibfield
  {journal} {\bibinfo  {journal} {JHEP}\ }\textbf {\bibinfo {volume} {09}},\
  \bibinfo {pages} {013} (\bibinfo {year} {2002})},\ \Eprint
  {http://arxiv.org/abs/hep-ph/0205318} {hep-ph/0205318} \BibitemShut {NoStop}%
\bibitem [{\citenamefont {Skullerud}\ \emph {et~al.}(2003)\citenamefont
  {Skullerud}, \citenamefont {Bowman}, \citenamefont {K{\i}z{\i}lers{\"u}},
  \citenamefont {Leinweber},\ and\ \citenamefont
  {Williams}}]{Skullerud:2003qu}%
  \BibitemOpen
  \bibfield  {author} {\bibinfo {author} {\bibfnamefont {Jon-Ivar}\
  \bibnamefont {Skullerud}}, \bibinfo {author} {\bibfnamefont {Patrick~O.}\
  \bibnamefont {Bowman}}, \bibinfo {author} {\bibfnamefont {Ay{\c s}e}\
  \bibnamefont {K{\i}z{\i}lers{\"u}}}, \bibinfo {author} {\bibfnamefont
  {Derek~B.}\ \bibnamefont {Leinweber}}, \ and\ \bibinfo {author}
  {\bibfnamefont {Anthony~G.}\ \bibnamefont {Williams}},\ }\bibfield  {title}
  {\enquote {\bibinfo {title} {Nonperturbative structure of the quark gluon
  vertex},}\ }\href@noop {} {\bibfield  {journal} {\bibinfo  {journal} {JHEP}\
  }\textbf {\bibinfo {volume} {04}},\ \bibinfo {pages} {047} (\bibinfo {year}
  {2003})},\ \Eprint {http://arxiv.org/abs/hep-ph/0303176} {hep-ph/0303176}
  \BibitemShut {NoStop}%
\bibitem [{\citenamefont {K{\i}z{\i}lers{\"u}}\ \emph
  {et~al.}(2007)\citenamefont {K{\i}z{\i}lers{\"u}}, \citenamefont {Leinweber},
  \citenamefont {Skullerud},\ and\ \citenamefont
  {Williams}}]{Kizilersu:2006et}%
  \BibitemOpen
  \bibfield  {author} {\bibinfo {author} {\bibfnamefont {Ay{\c s}e}\
  \bibnamefont {K{\i}z{\i}lers{\"u}}}, \bibinfo {author} {\bibfnamefont
  {Derek~B.}\ \bibnamefont {Leinweber}}, \bibinfo {author} {\bibfnamefont
  {Jon-Ivar}\ \bibnamefont {Skullerud}}, \ and\ \bibinfo {author}
  {\bibfnamefont {Anthony~G.}\ \bibnamefont {Williams}},\ }\bibfield  {title}
  {\enquote {\bibinfo {title} {Quark-gluon vertex in general kinematics},}\
  }\href {\doibase 10.1140/epjc/s10052-007-0250-6} {\bibfield  {journal}
  {\bibinfo  {journal} {Eur. Phys. J. C}\ }\textbf {\bibinfo {volume} {50}},\
  \bibinfo {pages} {871--875} (\bibinfo {year} {2007})},\ \Eprint
  {http://arxiv.org/abs/hep-lat/0610078} {arXiv:hep-lat/0610078 [hep-lat]}
  \BibitemShut {NoStop}%
\bibitem [{\citenamefont {Pel\'aez}\ \emph {et~al.}(2015)\citenamefont
  {Pel\'aez}, \citenamefont {Tissier},\ and\ \citenamefont
  {Wschebor}}]{Pelaez:2015tba}%
  \BibitemOpen
  \bibfield  {author} {\bibinfo {author} {\bibfnamefont {Marcela}\ \bibnamefont
  {Pel\'aez}}, \bibinfo {author} {\bibfnamefont {Matthieu}\ \bibnamefont
  {Tissier}}, \ and\ \bibinfo {author} {\bibfnamefont {Nicol\'as}\ \bibnamefont
  {Wschebor}},\ }\bibfield  {title} {\enquote {\bibinfo {title} {{Quark-gluon
  vertex from the Landau gauge Curci-Ferrari model}},}\ }\href {\doibase
  10.1103/PhysRevD.92.045012} {\bibfield  {journal} {\bibinfo  {journal} {Phys.
  Rev. D}\ }\textbf {\bibinfo {volume} {92}},\ \bibinfo {pages} {045012}
  (\bibinfo {year} {2015})},\ \Eprint {http://arxiv.org/abs/1504.05157}
  {arXiv:1504.05157 [hep-th]} \BibitemShut {NoStop}%
\bibitem [{\citenamefont {Oliveira}\ \emph {et~al.}(2016)\citenamefont
  {Oliveira}, \citenamefont {K{\i}z{\i}lers{\"u}}, \citenamefont {Silva},
  \citenamefont {Skullerud}, \citenamefont {Sternbeck},\ and\ \citenamefont
  {Williams}}]{Oliveira:2016muq}%
  \BibitemOpen
  \bibfield  {author} {\bibinfo {author} {\bibfnamefont {Orlando}\ \bibnamefont
  {Oliveira}}, \bibinfo {author} {\bibfnamefont {Ay{\c s}e}\ \bibnamefont
  {K{\i}z{\i}lers{\"u}}}, \bibinfo {author} {\bibfnamefont {Paulo~J.}\
  \bibnamefont {Silva}}, \bibinfo {author} {\bibfnamefont {Jon-Ivar}\
  \bibnamefont {Skullerud}}, \bibinfo {author} {\bibfnamefont {Andre}\
  \bibnamefont {Sternbeck}}, \ and\ \bibinfo {author} {\bibfnamefont
  {Anthony~G.}\ \bibnamefont {Williams}},\ }\bibfield  {title} {\enquote
  {\bibinfo {title} {Lattice {L}andau gauge quark propagator and the
  quark-gluon vertex},}\ }\bibfield  {booktitle} {\emph {\bibinfo {booktitle}
  {{Proceedings, International Meeting Excited QCD 2016: Costa da Caparica,
  Portugal, March 6-12, 2016}}},\ }\href {\doibase 10.5506/APhysPolBSupp.9.363}
  {\bibfield  {journal} {\bibinfo  {journal} {Acta Phys. Polon. Supp.}\
  }\textbf {\bibinfo {volume} {9}},\ \bibinfo {pages} {363--368} (\bibinfo
  {year} {2016})},\ \Eprint {http://arxiv.org/abs/1605.09632} {arXiv:1605.09632
  [hep-lat]} \BibitemShut {NoStop}%
\bibitem [{\citenamefont {Sternbeck}\ \emph {et~al.}(2017)\citenamefont
  {Sternbeck}, \citenamefont {Balduf}, \citenamefont {K{\i}z{\i}lers{\"u}},
  \citenamefont {Oliveira}, \citenamefont {Silva}, \citenamefont {Skullerud},\
  and\ \citenamefont {Williams}}]{Sternbeck:2017ntv}%
  \BibitemOpen
  \bibfield  {author} {\bibinfo {author} {\bibfnamefont {Andre}\ \bibnamefont
  {Sternbeck}}, \bibinfo {author} {\bibfnamefont {Paul-Hermann}\ \bibnamefont
  {Balduf}}, \bibinfo {author} {\bibfnamefont {Ay{\c s}e}\ \bibnamefont
  {K{\i}z{\i}lers{\"u}}}, \bibinfo {author} {\bibfnamefont {Orlando}\
  \bibnamefont {Oliveira}}, \bibinfo {author} {\bibfnamefont {Paulo~J.}\
  \bibnamefont {Silva}}, \bibinfo {author} {\bibfnamefont {Jon-Ivar}\
  \bibnamefont {Skullerud}}, \ and\ \bibinfo {author} {\bibfnamefont
  {Anthony~G.}\ \bibnamefont {Williams}},\ }\bibfield  {title} {\enquote
  {\bibinfo {title} {Triple-gluon and quark-gluon vertex from lattice {QCD} in
  {L}andau gauge},}\ }\bibfield  {booktitle} {\emph {\bibinfo {booktitle}
  {{Proceedings, 34th International Symposium on Lattice Field Theory (Lattice
  2016): Southampton, UK, July 24-30, 2016}}},\ }\href {\doibase
  10.22323/1.256.0349} {\bibfield  {journal} {\bibinfo  {journal} {PoS}\
  }\textbf {\bibinfo {volume} {LATTICE2016}},\ \bibinfo {pages} {349} (\bibinfo
  {year} {2017})},\ \Eprint {http://arxiv.org/abs/1702.00612} {arXiv:1702.00612
  [hep-lat]} \BibitemShut {NoStop}%
\bibitem [{\citenamefont {Skullerud}\ and\ \citenamefont
  {Williams}(2001)}]{Skullerud:2000un}%
  \BibitemOpen
  \bibfield  {author} {\bibinfo {author} {\bibfnamefont {Jon~Ivar}\
  \bibnamefont {Skullerud}}\ and\ \bibinfo {author} {\bibfnamefont
  {Anthony~G.}\ \bibnamefont {Williams}},\ }\bibfield  {title} {\enquote
  {\bibinfo {title} {Quark propagator in the {L}andau gauge},}\ }\href@noop {}
  {\bibfield  {journal} {\bibinfo  {journal} {Phys. Rev.}\ }\textbf {\bibinfo
  {volume} {D63}},\ \bibinfo {pages} {054508} (\bibinfo {year} {2001})},\
  \Eprint {http://arxiv.org/abs/hep-lat/0007028} {hep-lat/0007028} \BibitemShut
  {NoStop}%
\bibitem [{\citenamefont {Skullerud}\ \emph {et~al.}(2001)\citenamefont
  {Skullerud}, \citenamefont {Leinweber},\ and\ \citenamefont
  {Williams}}]{Skullerud:2001aw}%
  \BibitemOpen
  \bibfield  {author} {\bibinfo {author} {\bibfnamefont {Jon-Ivar}\
  \bibnamefont {Skullerud}}, \bibinfo {author} {\bibfnamefont {Derek~B.}\
  \bibnamefont {Leinweber}}, \ and\ \bibinfo {author} {\bibfnamefont
  {Anthony~G.}\ \bibnamefont {Williams}},\ }\bibfield  {title} {\enquote
  {\bibinfo {title} {Nonperturbative improvement and tree-level correction of
  the quark propagator},}\ }\href@noop {} {\bibfield  {journal} {\bibinfo
  {journal} {Phys. Rev.}\ }\textbf {\bibinfo {volume} {D64}},\ \bibinfo {pages}
  {074508} (\bibinfo {year} {2001})},\ \Eprint
  {http://arxiv.org/abs/hep-lat/0102013} {hep-lat/0102013} \BibitemShut
  {NoStop}%
\bibitem [{\citenamefont {Ball}\ and\ \citenamefont
  {Chiu}(1980)}]{Ball:1980ay}%
  \BibitemOpen
  \bibfield  {author} {\bibinfo {author} {\bibfnamefont {James~S.}\
  \bibnamefont {Ball}}\ and\ \bibinfo {author} {\bibfnamefont {Ting-Wai}\
  \bibnamefont {Chiu}},\ }\bibfield  {title} {\enquote {\bibinfo {title}
  {Analytic properties of the vertex function in gauge theories. 1.}}\ }\href
  {\doibase 10.1103/PhysRevD.22.2542} {\bibfield  {journal} {\bibinfo
  {journal} {Phys. Rev. D}\ }\textbf {\bibinfo {volume} {22}},\ \bibinfo
  {pages} {2542} (\bibinfo {year} {1980})}\BibitemShut {NoStop}%
\bibitem [{\citenamefont {K{\i}z{\i}lers{\"u}}\ \emph
  {et~al.}(1995)\citenamefont {K{\i}z{\i}lers{\"u}}, \citenamefont {Reenders},\
  and\ \citenamefont {Pennington}}]{Kizilersu:1995iz}%
  \BibitemOpen
  \bibfield  {author} {\bibinfo {author} {\bibfnamefont {A.}~\bibnamefont
  {K{\i}z{\i}lers{\"u}}}, \bibinfo {author} {\bibfnamefont {M.}~\bibnamefont
  {Reenders}}, \ and\ \bibinfo {author} {\bibfnamefont {M.R.}\ \bibnamefont
  {Pennington}},\ }\bibfield  {title} {\enquote {\bibinfo {title} {One loop
  {QED} vertex in any covariant gauge: Its complete analytic form},}\ }\href
  {\doibase 10.1103/PhysRevD.52.1242} {\bibfield  {journal} {\bibinfo
  {journal} {Phys. Rev. D}\ }\textbf {\bibinfo {volume} {52}},\ \bibinfo
  {pages} {1242--1259} (\bibinfo {year} {1995})},\ \Eprint
  {http://arxiv.org/abs/hep-ph/9503238} {arXiv:hep-ph/9503238} \BibitemShut
  {NoStop}%
\bibitem [{\citenamefont {Sheikholeslami}\ and\ \citenamefont
  {Wohlert}(1985)}]{Sheikholeslami:1985ij}%
  \BibitemOpen
  \bibfield  {author} {\bibinfo {author} {\bibfnamefont {B.}~\bibnamefont
  {Sheikholeslami}}\ and\ \bibinfo {author} {\bibfnamefont {R.}~\bibnamefont
  {Wohlert}},\ }\bibfield  {title} {\enquote {\bibinfo {title} {Improved
  continuum limit lattice action for {QCD} with {W}ilson fermions},}\
  }\href@noop {} {\bibfield  {journal} {\bibinfo  {journal} {Nucl. Phys.}\
  }\textbf {\bibinfo {volume} {B259}},\ \bibinfo {pages} {572} (\bibinfo {year}
  {1985})}\BibitemShut {NoStop}%
\bibitem [{\citenamefont {Capitani}\ \emph {et~al.}(2001)\citenamefont
  {Capitani} \emph {et~al.}}]{Capitani:2000xi}%
  \BibitemOpen
  \bibfield  {author} {\bibinfo {author} {\bibfnamefont {S.}~\bibnamefont
  {Capitani}} \emph {et~al.},\ }\bibfield  {title} {\enquote {\bibinfo {title}
  {Renormalisation and off-shell improvement in lattice perturbation theory},}\
  }\href@noop {} {\bibfield  {journal} {\bibinfo  {journal} {Nucl. Phys.}\
  }\textbf {\bibinfo {volume} {B593}},\ \bibinfo {pages} {183--228} (\bibinfo
  {year} {2001})},\ \Eprint {http://arxiv.org/abs/hep-lat/0007004}
  {hep-lat/0007004} \BibitemShut {NoStop}%
\bibitem [{\citenamefont {Bali}\ \emph {et~al.}(2013)\citenamefont {Bali} \emph
  {et~al.}}]{Bali:2012qs}%
  \BibitemOpen
  \bibfield  {author} {\bibinfo {author} {\bibfnamefont {G.~S.}\ \bibnamefont
  {Bali}} \emph {et~al.},\ }\bibfield  {title} {\enquote {\bibinfo {title}
  {Nucleon mass and sigma term from lattice {QCD} with two light fermion
  flavors},}\ }\href {\doibase 10.1016/j.nuclphysb.2012.08.009} {\bibfield
  {journal} {\bibinfo  {journal} {Nucl. Phys.}\ }\textbf {\bibinfo {volume}
  {B866}},\ \bibinfo {pages} {1--25} (\bibinfo {year} {2013})},\ \Eprint
  {http://arxiv.org/abs/1206.7034} {arXiv:1206.7034 [hep-lat]} \BibitemShut
  {NoStop}%
\bibitem [{\citenamefont {Bali}\ \emph {et~al.}(2014)\citenamefont {Bali},
  \citenamefont {Collins}, \citenamefont {Gl{\"a}{\ss}le}, \citenamefont
  {G{\"o}ckeler}, \citenamefont {Najjar}, \citenamefont {R{\"o}dl},
  \citenamefont {Sch{\"a}fer}, \citenamefont {Schiel}, \citenamefont
  {Sternbeck},\ and\ \citenamefont {S{\"o}ldner}}]{Bali:2014gha}%
  \BibitemOpen
  \bibfield  {author} {\bibinfo {author} {\bibfnamefont {Gunnar~S.}\
  \bibnamefont {Bali}}, \bibinfo {author} {\bibfnamefont {Sara}\ \bibnamefont
  {Collins}}, \bibinfo {author} {\bibfnamefont {Benjamin}\ \bibnamefont
  {Gl{\"a}{\ss}le}}, \bibinfo {author} {\bibfnamefont {Meinulf}\ \bibnamefont
  {G{\"o}ckeler}}, \bibinfo {author} {\bibfnamefont {Johannes}\ \bibnamefont
  {Najjar}}, \bibinfo {author} {\bibfnamefont {Rudolf~H.}\ \bibnamefont
  {R{\"o}dl}}, \bibinfo {author} {\bibfnamefont {Andreas}\ \bibnamefont
  {Sch{\"a}fer}}, \bibinfo {author} {\bibfnamefont {Rainer~W.}\ \bibnamefont
  {Schiel}}, \bibinfo {author} {\bibfnamefont {Andr{\'e}}\ \bibnamefont
  {Sternbeck}}, \ and\ \bibinfo {author} {\bibfnamefont {Wolfgang}\
  \bibnamefont {S{\"o}ldner}},\ }\bibfield  {title} {\enquote {\bibinfo {title}
  {The moment {$\langle x\rangle_{u-d}$} of the nucleon from {$N_f=2$} lattice
  {QCD} down to nearly physical quark masses},}\ }\href {\doibase
  10.1103/PhysRevD.90.074510} {\bibfield  {journal} {\bibinfo  {journal} {Phys.
  Rev.}\ }\textbf {\bibinfo {volume} {D90}},\ \bibinfo {pages} {074510}
  (\bibinfo {year} {2014})},\ \Eprint {http://arxiv.org/abs/1408.6850}
  {arXiv:1408.6850 [hep-lat]} \BibitemShut {NoStop}%
\bibitem [{\citenamefont {Bali}\ \emph {et~al.}(2015)\citenamefont {Bali},
  \citenamefont {Collins}, \citenamefont {Gl{\"a}{\ss}le}, \citenamefont
  {G{\"o}ckeler}, \citenamefont {Najjar}, \citenamefont {R{\"o}dl},
  \citenamefont {Sch{\"a}fer}, \citenamefont {Schiel}, \citenamefont
  {S{\"o}ldner},\ and\ \citenamefont {Sternbeck}}]{Bali:2014nma}%
  \BibitemOpen
  \bibfield  {author} {\bibinfo {author} {\bibfnamefont {Gunnar~S.}\
  \bibnamefont {Bali}}, \bibinfo {author} {\bibfnamefont {Sara}\ \bibnamefont
  {Collins}}, \bibinfo {author} {\bibfnamefont {Benjamin}\ \bibnamefont
  {Gl{\"a}{\ss}le}}, \bibinfo {author} {\bibfnamefont {Meinulf}\ \bibnamefont
  {G{\"o}ckeler}}, \bibinfo {author} {\bibfnamefont {Johannes}\ \bibnamefont
  {Najjar}}, \bibinfo {author} {\bibfnamefont {Rudolf~H.}\ \bibnamefont
  {R{\"o}dl}}, \bibinfo {author} {\bibfnamefont {Andreas}\ \bibnamefont
  {Sch{\"a}fer}}, \bibinfo {author} {\bibfnamefont {Rainer~W.}\ \bibnamefont
  {Schiel}}, \bibinfo {author} {\bibfnamefont {Wolfgang}\ \bibnamefont
  {S{\"o}ldner}}, \ and\ \bibinfo {author} {\bibfnamefont {Andr{\'e}}\
  \bibnamefont {Sternbeck}},\ }\bibfield  {title} {\enquote {\bibinfo {title}
  {Nucleon isovector couplings from {$N_f=2$} lattice {QCD}},}\ }\href
  {\doibase 10.1103/PhysRevD.91.054501} {\bibfield  {journal} {\bibinfo
  {journal} {Phys. Rev.}\ }\textbf {\bibinfo {volume} {D91}},\ \bibinfo {pages}
  {054501} (\bibinfo {year} {2015})},\ \Eprint {http://arxiv.org/abs/1412.7336}
  {arXiv:1412.7336 [hep-lat]} \BibitemShut {NoStop}%
\bibitem [{Note1()}]{Note1}%
  \BibitemOpen
  \bibinfo {note} {$\lambda _4=0$ in this kinematics because of charge
  conjugation symmetry, which dictates that $\lambda _4(p^2,q^2,k^2)=-\lambda
  _4(k^2,q^2,p^2)$.}\BibitemShut {Stop}%
\bibitem [{\citenamefont {Heatlie}\ \emph {et~al.}(1991)\citenamefont
  {Heatlie}, \citenamefont {Martinelli}, \citenamefont {Pittori}, \citenamefont
  {Rossi},\ and\ \citenamefont {Sachrajda}}]{Heatlie:1991kg}%
  \BibitemOpen
  \bibfield  {author} {\bibinfo {author} {\bibfnamefont {G.}~\bibnamefont
  {Heatlie}}, \bibinfo {author} {\bibfnamefont {G.}~\bibnamefont {Martinelli}},
  \bibinfo {author} {\bibfnamefont {C.}~\bibnamefont {Pittori}}, \bibinfo
  {author} {\bibfnamefont {G.~C.}\ \bibnamefont {Rossi}}, \ and\ \bibinfo
  {author} {\bibfnamefont {C.~T.}\ \bibnamefont {Sachrajda}},\ }\bibfield
  {title} {\enquote {\bibinfo {title} {The improvement of hadronic matrix
  elements in lattice {QCD}},}\ }\href@noop {} {\bibfield  {journal} {\bibinfo
  {journal} {Nucl. Phys.}\ }\textbf {\bibinfo {volume} {B352}},\ \bibinfo
  {pages} {266--288} (\bibinfo {year} {1991})}\BibitemShut {NoStop}%
\bibitem [{\citenamefont {Capitani}\ and\ \citenamefont
  {Rossi}(1995)}]{Capitani:1995qn}%
  \BibitemOpen
  \bibfield  {author} {\bibinfo {author} {\bibfnamefont {Stefano}\ \bibnamefont
  {Capitani}}\ and\ \bibinfo {author} {\bibfnamefont {Giancarlo}\ \bibnamefont
  {Rossi}},\ }\bibfield  {title} {\enquote {\bibinfo {title} {Deep inelastic
  scattering in improved lattice {QCD}. 1. the first moment of structure
  functions},}\ }\href@noop {} {\bibfield  {journal} {\bibinfo  {journal}
  {Nucl. Phys.}\ }\textbf {\bibinfo {volume} {B433}},\ \bibinfo {pages}
  {351--389} (\bibinfo {year} {1995})},\ \Eprint
  {http://arxiv.org/abs/hep-lat/9401014} {hep-lat/9401014} \BibitemShut
  {NoStop}%
\bibitem [{\citenamefont {Bethke}(2009)}]{Bethke:2009jm}%
  \BibitemOpen
  \bibfield  {author} {\bibinfo {author} {\bibfnamefont {Siegfried}\
  \bibnamefont {Bethke}},\ }\bibfield  {title} {\enquote {\bibinfo {title} {The
  2009 world average of $\alpha_s$},}\ }\href {\doibase
  10.1140/epjc/s10052-009-1173-1} {\bibfield  {journal} {\bibinfo  {journal}
  {Eur. Phys. J. C}\ }\textbf {\bibinfo {volume} {64}},\ \bibinfo {pages}
  {689--703} (\bibinfo {year} {2009})},\ \Eprint
  {http://arxiv.org/abs/0908.1135} {arXiv:0908.1135 [hep-ph]} \BibitemShut
  {NoStop}%
\bibitem [{\citenamefont {Bethke}\ \emph {et~al.}(2016)\citenamefont {Bethke},
  \citenamefont {Dissertori},\ and\ \citenamefont {Salam}}]{Bethke:2015etp}%
  \BibitemOpen
  \bibfield  {author} {\bibinfo {author} {\bibfnamefont {Siegfried}\
  \bibnamefont {Bethke}}, \bibinfo {author} {\bibfnamefont {Gunther}\
  \bibnamefont {Dissertori}}, \ and\ \bibinfo {author} {\bibfnamefont
  {Gavin~P.}\ \bibnamefont {Salam}},\ }\bibfield  {title} {\enquote {\bibinfo
  {title} {World summary of $\alpha_s$ (2015)},}\ }\href {\doibase
  10.1051/epjconf/201612007005} {\bibfield  {journal} {\bibinfo  {journal} {EPJ
  Web Conf.}\ }\textbf {\bibinfo {volume} {120}},\ \bibinfo {pages} {07005}
  (\bibinfo {year} {2016})}\BibitemShut {NoStop}%
\end{thebibliography}%

\appendix
\section{Tree-level lattice expressions}
\label{app:tree-level}


In the following we will make use of the lattice
momentum variables defined by
\begin{align}
\widetilde{K}_{\nu}(p)  & = \frac{1}{2a}\,\sin{2p_{\nu} a}\,,  \\
K_{\nu}(p)  & = \frac{1}{a}\,\sin{p_{\nu} a}\,,  \\
Q_{\nu}(p)  & =  \frac{2}{a}\sin\left(\frac{ {p_{\nu} a} }{2} \right) \,, \\
C_{\nu}(p)  & = \cos(p_{\nu}a)\,, \\
{\overline C}_{\nu}(p) & = \cos\left( \frac{p_{\nu}a}{2} \right) \,.
\end{align}
In terms of these variables, 
the tree-level quark--gluon vertex in \eqqref{eq:vtx-tree1} can be rewritten as
\begin{align}
\overline{\Lambda}_{0,\nu}^{(0)}(p,q,k) &= (-ig_0)\, \left\{  
                \bm{ \gamma}_\nu a{\overline{C}}_\nu(p+k) 
               -\frac{i}{2}\, aQ_\nu(p+k)\,\bm{I} 
-i \frac{\csw}{2} \, a{\overline{C}}_\nu(q)\,\sum_{\lambda} \bm{\sigma_{\nu\lambda}}  \,a K_\lambda(q)  \right\}\,,
\label{eqa:vtx-tree}
 \end{align}
for the case of
the `unimproved' propagator $S_0(x,y)=\braket{\psi(x)\psibar(y)}$.
The improved vertex obtained using the rotated propagator
\eqref{eq:rotprop} is given at tree level by
\begin{align}
  \overline{\Lambda}^{(0)}_{R,\nu}(p,q,k)
  &= (1+b_q am)\,[S_R^{(0)}(p)]^{-1}\,S_0^{(0)}(p)\,
  \left[ \Lambda_{0,\nu}^{a(0)}(p,q,k)\right]\,
  S_0^{(0)}(k)\,[S_R^{(0)}(k)]^{-1}\,,
\label{eqa:tree-imp} 
\end{align}
where $S_0^\z(p)$ is the tree-level, unimproved, dimensionless Wilson quark
propagator, while $S_R^\z(p)$ is the level expression for the
improved propagator \eqref{eq:rotprop}, given by \cite{Skullerud:2001aw}
\begin{align}
S_0^\z(p) &= \frac{1}{D}\, \Big[-ia {\not\!\!K}(p) + ma+\half \, a^2\qsp\Big]\,, 
\label{eqa:unimpS}\\
S_R^\z(p)^{-1} &= \frac{D}{(1+ma/2)\,D_R}\,\left(ia \Kslash(p)\,A_R+B_R\right) \,,
\label{eqa:rotS}\\
 \mbox{where} \hspace*{4cm} &\nonumber \\
D &=  a^2\ksp + \Big(ma+\half\, a^2 \qsp\Big)^2\,, \\
D_R &= a^2K^2(p)\,A_R^2+B_R^2 \,,\\
A_R & = 1+\frac{1}{2}\,am + \frac{1}{4}a^2Q^2(p) - \frac{1}{16}a^2K^2(p) \,,\\
B_R & =  \left( am + \frac{1}{2}a^2Q^2(p)  \right)\,\left( 1-\frac{a^2K^2(p)}{16}\right) - \frac{1}{2}\, a^2K^2(p) \,.
\end{align}
Making use of the tree level unimproved \eqref{eqa:unimpS} and the rotated-improved \eqref{eqa:rotS} propagators in \eqqref{eqa:tree-imp}, the tree level quark-gluon vertex can be rewritten as
\begin{align}
\overline{\Lambda}^{(0)}_{R,\nu}(p,q) &= \frac{(1+b_q am)}{(1+am/2)^2\,D_R(p)\,D_R(k)} \nonumber\\
& \times \left[ -ia {\not\!\!K}(p) A_V^R(p)+B_V^R(p) \right]\,
\left[ \Lambda_{0,\nu}^{a(0)}(p,q,k)\right]\,
\left[ -ia {\not\!\!K}(k) A_V^R(k)+B_V^R(k) \right]\ \,,
\label{eqa:tree-imp-full} 
\end{align}
\noindent
where
\begin{align}
A_V^R(p) &= 2c_q D(p) \,,\\
B_V^R(p) & = (1-c_q^2 a^2 K^2(p))\,D^2(p)\,, \\
D_R^2(p) & = \left(1+\frac{a^2K^2(p)}{16}\right)^2\,D(p)\,.
\label{eqa:tree-imp1} 
\end{align}
and after further manipulation we arrive at the expression
\begin{align}
  \overline{\Lambda}^{(0)}_{R,\nu}(p,q,k)
  &= \frac{(1+b_q am)}{(1+am/2)^2}\,\frac{D(p)\,D(k)}{D_R(p)\,D_R(k)} \notag\\
  & \times \left[ 2ic_q a{\not\!\!K}(p)+1-c_q^2a^2K^2(p) \right]\,
      \left[ \Lambda_{0,\nu}^{a(0)}(p,q,k)\right]\,
      \left[ 2ic_q a{\not\!\!K}(k)+1-c_q^2a^2K^2(k) \right]   \,. 
  \label{eqa:tree-imp2} 
\end{align}
As discussed above [see \eqqref{eq:transverse_vertex}], for $q\neq0$ we can only access the transverse vertex on the lattice in the Landau gauge. Hence, using the lattice equivalent of the transverse projector, $P_{\mu\nu}^T = \sum_\nu \left( \delta_{\mu\nu}-\frac{K_\mu(q)\,K_\nu(q)}{K^2(q)} \right)$, the tree level quark-gluon vertex in general kinematics is given by
\begin{align}
  \overline\Lambda^{(0)}_{R,\mu}(p,q,k)
  &= P_{\mu\nu}^T(q) \,\Lambda^{(0)}_{R,\nu}(p,q,k) \nonumber\\
  & = \frac{(1+b_q am)}{(1+am/2)^2}\,
  \frac{1}{ \left(1+c_q^2a^2K^2(p)\right)^2\,
    \left(1+c_q^2a^2K^2(k)\right)^2\ } \nonumber\\
   \times & \sum_\nu \left( \delta_{\mu\nu}-\frac{K_\mu(q)\,K_\nu(q)}{K^2(q)} \right)\, \left[ 2ic_q a{\not\!\!K}(p)+1-c_q^2a^2K^2(p) \right]\,
   \left[ \Lambda_{0,\nu}^{a(0)}(p,q,k)\right]\,
   \left[ 2ic_q a{\not\!\!K}(k)+1-c_q^2a^2K^2(k) \right]          \,.         
  \label{eqa:tree-imp4} 
\end{align}

The general form of the tree-level clover-rotated vertex \eqref{eqa:tree-imp4} is highly complicated.  However, it simplifies in the soft-gluon kinematics which is the case we are studying in this paper. In this kinematics the gluon momentum $q=0$ and both the quark momenta are equal, $k=p$. In this case the `unimproved' tree-level vertex \eqref{eqa:vtx-tree} becomes
\begin{equation}
  \overline{\Lambda}_{0,\nu}^{(0)}(p,0,p)
  = -ig_0\,  \Big( \bm{ \gamma}_\nu {\overline{C}}_\nu(2p)-\frac{i}{2}\, aQ_\nu(2p)\,\bm{I} \Big)
= -ig_0\,\Big(\bm{ \gamma}_\nu C_\nu(p)- i\,aK_\nu(p)\,\bm{I} \Big)\,,
\end{equation}
and the tree-level clover-rotated vertex \eqref{eqa:tree-imp4} reduces to
\begin{align}
  \overline\Lambda^{(0)}_{R,\mu}(p,0,p)
  &=  (-ig_0)F(p)
  \left[ 2ic_q a{\not\!\!K}(p)+1-c_q^2a^2K^2(p) \right]\,
  \left[ \Lambda_{0,\mu}^{a(0)}(p,0)\right]\,
  \left[ 2ic_q a{\not\!\!K}(p)+1-c_q^2a^2K^2(p) \right] \,,
  \label{eqa:tree-imp5} 
\end{align}
where the common prefactor $F(p)$ is given by
\begin{equation}
  F(p) = \frac{(1+b_q am)}{(1+am/2)^2}\,
  \frac{1}{ \left(1+c_q^2a^2K^2(p)\right)^4}\,.
\end{equation}
After rearranging \eqqref{eqa:tree-imp5} to match the form of
\eqqref{eq:sqk_cont_vertex}, the final shape of the tree-level
clover-rotated quark-gluon vertex in the soft gluon kinematics is 
\begin{align}
  \overline\Lambda^{(0)}_{R,\mu}(p,0,p)
  &=  (-ig_0)\,\frac{(1+b_q am)}{(1+am/2)^2}\,\frac{1}{ \big(1+ c_q^2 a^2 K^2(p) \big)^4}  \nonumber\\
  \times &  \bigg\{  
         {\boldsymbol{ \gamma_\mu}}\, \Big[ \big(1+c_q^2 a^2K^2(p)\big)^2
           \,C_\mu(p) \Big]  \nonumber\\
         &  \,\, \bm{-4a^2K_\mu\Kslash(p)}\,
         \Big[- c_q\big(1-c_q^2a^2K^2(p)\big) + 2c_q^2 C_\mu(p)
            \Big]  \nonumber\\
         & \,\,  \bm{-2i aK_\mu}\, \Big[ -2c_q^2 a^2K^2(p)
   +\frac{1}{2}\big(1-c_q^2 a^2K^2(p) \big) -2c_q
   \big(1-c_q^2a^2K^2(p)\big)\,C_\mu(p) \Big]
 \bigg\}         \,.    
  \label{eqa:tree-imp6} 
\end{align}
Making a comparison between this expression and the continuum vertex
\eqqref{eq:sqk_cont_vertex}, we can identify the tree-level expressions
for the form factors in the soft-gluon kinematics:
\begin{align}
  \lambda_1^{(0)} + \overline\lambda_{1(\mu)}^{(0)}
  & =  F(p)\,\Big[ \big(1+c_q^2 a^2 K^2(p) \big)^2 C_\mu(p)
    \Big]\bigg|_{\substack{p_\mu=0}}
   =  \frac{(1+b_q am)}{(1+am/2)^2}\,\frac{1}{ \big(1+ c_q^2 a^2 K^2(p) \big)^2}\,, 
 \label{eqa:lam1-corr} \\
 \lambda_2^{(0)}+ \overline\lambda_{2(\mu)}^{(0)} & = 
 a^2F(p)\,\Big[-c_q\big(1-c_q^2a^2K^2(p) \big) + 2c_q^2 C_\mu(p)\Big]\,,
 \label{eqa:lam2-corr} \\
\lambda_3^{(0)} + \overline\lambda_{3(\mu)}^{(0)} &= aF(p)\,
\Big[ -2c_q^2 a^2K^2(p) +\frac{1}{2}\big(1-c_q^2 a^2K^2(p) \big)^2
  - 2c_q \big(1-c_q^2a^2K^2(p)\big)\,C_\mu(p)\Big] \,.
\label{eqa:lam3-corr}   
\end{align}
We note in particular that there are two separate tensor structures
appearing in the tree-level vertex \eqref{eqa:tree-imp6} which in the
continuum become equal to $L_{2,\mu}$, and likewise for $L_{3,\mu}$.
In Eqs~\eqref{eqa:lam2-corr} and \eqref{eqa:lam3-corr} these are
associated with separate form factors $\lambda_i^\z$ and
$\overline{\lambda}_i^\z$ respectively.

Inspecting \eqqref{eqa:tree-imp6} and comparing this with the
continuum expressions
\eqref{eq:l1-cont-noncov}, \eqref{eq:l2-cont-noncov},
\eqref{eq:l3-cont-covariant}, we can write down the lattice
equivalents of these expressions, namely
\begin{align}
 \lambda_1(p^2,0,p^2)  &=  \frac{1}{(-ig_0)}\, \left\{  \,\,\Big[\trd(\gamma_\alpha \overline\Lambda_\mu)\Big]\,    \bigg|_{\substack{\alpha=\mu\\p_\mu=0}}   \right\}  \nonumber\\
 & =  \frac{{\bm{\mbox{Im}}}}{g_0}\, \left\{  \,\,\Big[\trd(\gamma_\alpha \overline\Lambda_\mu)\Big]\,    \bigg|_{\substack{\alpha=\mu\\p_\mu=0}}   \right\} \, ,  
 \label{eqa:l1-lat-noncov} \\
 \lambda_2 (p^2,0,p^2)&=  \frac{1}{(-ig_0)}\, \left\{    - \frac{1}{4\,K(p)^2}\,\frac{K_{\alpha}(p)K_{\mu}(p)}{K(p)^2}\, \Big[ \trd(\gamma_\alpha \overline\Lambda_\mu) \, \Big|_{\alpha \neq\mu} \Big]  \right\} 
    \nonumber\\[2mm]
  &=  \frac{{\bm{\mbox{Im}}}}{g_0}\, \left\{    - \frac{1}{4\,K(p)^2}\,\frac{K_{\alpha}(p)K_{\mu}(p)}{K(p)^2}\, \Big[ \trd(\gamma_\alpha \overline\Lambda_\mu) \, \Big|_{\alpha \neq\mu} \Big]  \right\} 
  \label{eqa:l2-lat-noncov} \,, \\[2mm]
 \lambda_3(p^2,0,p^2) &= \frac{1}{(-ig_0)}\, \left\{ \frac{i}{2}\, \frac{K_\mu(p)}{K^2(p)} \trd(I\, \overline\Lambda_{\mu}) \right\}  \nonumber\\[2mm]
  &= \frac{{\bm{\mbox{Re}}}}{(-g_0)}\, \left\{ \frac{1}{2}\, \frac{K_\mu(p)}{K^2(p)} \trd(I\, \overline\Lambda_{\mu}) \right\}
\label{eqa:l3-lat-noncov} \,.
\end{align}
Although \eqqref{eqa:l3-lat-noncov} is based on the covariant
expression \eqref{eq:l3-cont-covariant}, we will in practice not sum over
$\mu$ in \eqqref{eqa:l3-lat-noncov}, just as there is no sum over
$\mu$ and $\alpha$ in Eqs \eqref{eqa:l1-lat-noncov} and
\eqref{eqa:l2-lat-noncov}.  Combining this with the tree-level form
factors, Eqs~\eqref{eqa:lam1-corr}, \eqref{eqa:lam2-corr},
\eqref{eqa:lam3-corr} we arrive at the expressions we will use to
determine the tree-level corrected form factors $\lambda_i$ in the
soft gluon kinematics,
%
\begin{align}
\lambda_1(p^2,0,p^2)  
& =  \frac{{\bm{\mbox{Im}}}}{g_0}\,
\left\{  \,\,\Big[\trd(\gamma_\alpha \overline\Lambda_\mu)\Big]\,
\bigg|_{\substack{\alpha=\mu\\p_\mu=0}}   \right\} \bigg/
\lambda_1^{(0)}\, ,  
 \label{eqa:l1-lat-noncov-corr} \\[2mm]
 \lambda_2 (p^2,0,p^2)
 &=  \frac{{\bm{\mbox{Im}}}}{g_0}\,
 \left\{-\frac{1}{4\,K(p)^2}\,\frac{K_{\alpha}(p)K_{\mu}(p)}{K(p)^2}\,
 \Big[\trd(\gamma_\alpha\overline\Lambda_\mu)\,\Big|_{\alpha \neq\mu}\Big]\right\} 
  -\left(\lambda_2^\z +\overline\lambda_{2(\mu)}^\z\right)  
  \label{eqa:l2-lat-noncov-corr} \,, \\[2mm]
 \lambda_3(p^2,0,p^2) 
  &= \frac{{\bm{\mbox{Re}}}}{(-g_0)}\, \left\{ \frac{1}{2}\, \frac{K_\mu(p)}{K^2(p)} \trd(I\, \overline\Lambda_{\mu}) \right\}
  -\left(\lambda_3^\z +\overline\lambda_{3(\mu)}^\z\right) 
\label{eqa:l3-lat-noncov-corr} \,.
\end{align}
These ($\lambda_1,\lambda_2,\lambda_3$)  are the complete form factors required to determine the quark-gluon vertex in the soft gluon limits and 
Eqs (\ref{eqa:l1-lat-noncov-corr},\ref{eqa:l2-lat-noncov-corr},\ref{eqa:l3-lat-noncov-corr}) define the exact procedures to calculate them on the lattice.

For completeness and future reference, we also include the tree-level
lattice equivalents of the covariant expressions
(\ref{eq:l1-cont-covariant}--\ref{eq:l3-cont-covariant}),
\begin{align}
  \sum_\mu\trd\big(\gm\overline{\Lambda}^{R(0)}_\mu(p,0)\big)
  &= F(p)\Bigg\{\bigg[\Big(1-c_q^2K^2(p)\Big)^2\! + 4c_q^2K^2(p)\bigg]
  \Big(4-\frac{1}{2}Q^2(p)\Big) \notag\\
 &\quad  + 4c_q\Big(1-c_q^2K^2(p)\Big)K^2(p)
  - 8c_q^2K(p)\!\cdot\!\Kt(p)\Bigg\} \label{eqa:tr-gmu}\\
  \sum_{\alpha\mu}K_\alpha(p)K_\mu(p)
  \trd\big(\ga\overline{\Lambda}^{R(0)}_\mu(p,0)\big)
  &= F(p)\Bigg\{\bigg[\Big(1-c_q^2K^2(p)\Big)^2\! - 4c_q^2K^2(p)\bigg]
  K(p)\!\cdot\!\Kt(p) \notag \\
  &\quad + c_q\Big(1-c_q^2K^2(p)\Big)\big(K^2(p)\big)^2\Bigg\}
  \label{eqa:qa-qmu-tr-ga}\\
  \sum_{\mu}\frac{K_\mu(p)}{K^2(p)}
  \trd\big(\overline{\Lambda}^{R(0)}_\mu(p,0)\big)
  &= -\frac{iF(p)}{2}\bigg[\Big(1-c_q^2K^2(p)\Big)^2\!- 4c_q^2K^2(p)
    -4c_q\Big(1-c_q^2K^2(p)\Big)\frac{K(p)\!\cdot\Kt(p)}{K^2(p)}\bigg]
  \label{eqa:l3-lat-covariant} \\
   \lambda^\z_{1,\text{cov}}
   = \frac{1}{3}\bigg[\eqref{eqa:tr-gmu}
     - \frac{\eqref{eqa:qa-qmu-tr-ga}}{K^2(p)}\bigg]
  &= \frac{F(p)}{3}\bigg[4-\frac{1}{2}Q^2(p)
    -\frac{K(p)\!\cdot\!\Kt(p)}{K^2(p)}\bigg]
  \bigg[\Big(1-c_q^2K^2(p)\Big)^2 + 4c_q^2K^2(p)\bigg]
  \label{eqa:l1-lat-covariant} \\
  \lambda^\z_{2,\text{cov}}= \frac{1}{12K^2(p)}\bigg[
  \eqref{eqa:tr-gmu} - 4\frac{\eqref{eqa:qa-qmu-tr-ga}}{K^2(p)}\bigg]
  &= F(p)\Bigg\{\frac{\Big(1-c_q^2K^2(p)\Big)^2}{3K^2(p)}
  \bigg[1-\frac{K(p)\!\cdot\!\Kt(p)}{K^2(p)}
    -\frac{1}{8}Q^2(p)\bigg] \notag\\
 &\quad - c_q\Big(1-c_q^2K^2(p)\Big)
  + \frac{4c_q^2}{3}\bigg[1+\frac{K(p)\!\cdot\!\Kt(p)}{2K^2(p)}
    - \frac{1}{8}Q^2(p)\bigg]\Bigg\}
  \label{eqa:l2-lat-covariant1} 
\end{align}
\end{document}